%% file: main.tex
\documentclass{article}
\usepackage[utf8]{inputenc}

\usepackage[a4paper, total={6in, 10in}]{geometry}
\usepackage{graphicx}
\usepackage{caption}
\usepackage{subcaption}
\usepackage{pict2e}
\usepackage{amsmath}
\usepackage{amssymb}
\usepackage{bm}
\usepackage{hyperref}
\usepackage[capitalize]{cleveref}

\usepackage{tikz,pgfplots}
\usepgfplotslibrary{patchplots}
\usepgfplotslibrary{colormaps}
\pgfplotsset{compat=1.17}
\DeclareUnicodeCharacter{2212}{−}
\usepgfplotslibrary{groupplots,dateplot}

\usetikzlibrary{shapes, arrows, positioning, calc, fadings}
\usetikzlibrary{patterns} 
\usetikzlibrary{spy} 
\tikzfading[name=fade right, left color=transparent!0, right color=transparent!100]
\tikzfading[name=fade left, left color=transparent!200, right color=transparent!0]
\tikzstyle{terminator} = [rectangle, draw, text centered, rounded corners, minimum height=2em]
\tikzstyle{process} = [rectangle, draw, text centered, minimum height=2em]
\tikzstyle{decision} = [diamond, draw, text centered, minimum height=1em]
\tikzstyle{data}=[trapezium, draw, text centered, trapezium left angle=60, trapezium right angle=120, minimum height=2em]
\tikzstyle{connector} = [draw, -latex']

\newcommand{\tripledot}{%
\tikz[baseline=-0.2ex]{ \draw[black,fill=black] (0,0) circle (.1ex); \draw[black,fill=black] (0,0.6ex) circle (.1ex); \draw[black,fill=black] (0,1.2ex) circle (.1ex) }%
}

\usepackage{authblk}
\title{
Topology Optimization of self-contacting structures
}
\author[1]{Andreas Henrik Frederiksen\thanks{andfr@dtu.dk}}
\author[1]{Ole Sigmund}
\author[1]{Konstantinos Poulios}
\affil[1]{Department of Civil and Mechanical Engineering, 
Technical University of Denmark, Koppels Allé
Building 404
DK- 2800 Kgs. Lyngby
Denmark}
\date{May 11, 2023}

\begin{document}

\maketitle

\section*{Abstract}
Inclusion of contact in mechanical designs opens a large range of design possibilities, this includes classical designs with contact, such as gears, couplings, switches, clamps etc. However, incorporation of contact in topology optimization is challenging, as classical contact models are not readily applicable when the boundaries are not defined. This paper aims to address the limitations of contact in topology optimization by extending the third medium contact method for topology optimization problems with internal contact.

When the objective is to maximize a given contact load for a specified displacement, instabilities may arise as an optimum is approached.
In order to alleviate stability problems as well as provide robustness of the optimized designs, a tangent stiffness requirement is introduced to the design objective.
To avoid a non-physical exploitation of the third medium in optimized designs, small features are penalized by evaluating the volume constraint on a dilated design.
The present work incorporates well-established methods in topology optimization including Helmholtz PDE filtering, threshold projection, Solid Isotropic Material Interpolation with Penalization, and the Method of Moving Asymptotes.
Three examples are used to illustrate how the approach exploits internal contact in the topology optimization of structures subjected to large deformations.

\section{Introduction}
Topology optimization has found multiple applications ranging from micro and macro-scale problems to multiphysics problems such as fluid flow, heat transfer, and nanophotonics. 
Non-linearities are nothing new to topology optimization, and geometrical and material non-linearity for large strain topology optimization has been applied successfully in the design of structures with large deformations \cite{Jung2004TopologyStructures,Bruns2001TopologyMechanisms,Bruns2002NumericalSnap-through,Buhl2000StiffnessOptimization,Wang2011OnOptimization,Luo2015TopologyTechnique,Lahuerta2013TowardsDeformation}.
When subjected to large deformations such structures may engage in contact, and including such contact in the topology optimization process extends the solution space, making more advanced designs possible.
This may be of special interest in the design of compliant mechanisms or soft robotics where human-made designs already utilize contact. 
However, until recently there has been no viable approach to include internal contact in topology optimization due to the seeming paradox: how to assign contact forces when the boundaries are not defined?

A few approaches to include internal contact in topology optimization exist. One such approach \cite{Kumar2021OnMechanisms} has been applied successfully in the design of shape morphing compliant mechanisms. 
This approach employs an explicit treatment of contact surfaces - solid and void phases are determined based on a negative mask scheme, a smooth boundary is then determined by a boundary resolution and smoothing scheme. Contact between these surfaces is modelled by a Lagrange multiplier approach. A related approach, which is based on frame-structures and includes contact between different members, also exists \cite{Reddy2012SystematicMechanisms}.

Another approach, which also includes internal contact, is the level set based approach for sliding contact interfaces proposed by \cite{Lawry2015LevelInterfaces}. This approach is based on the distribution of two separate materials sharing an interface. A level set method is used to define the interface, along which the contacting bodies may slide or separate. Tractions along these interfaces are modelled through a Lagrange multiplier approach.
Lawry \& Maute illustrate the approach on the design of anchors embedded in a host material. It should be noted that the method is developed based on small displacement gradients and that there are no gaps in this approach, i.e. there is no void region between the contacting surfaces which has to be passed before contact is established.

A promising new approach to include internal self-contact in topology optimization has recently been proposed \cite{Bluhm2021InternalOptimization}. This approach is very appealing to topology optimization since it: 1) utilizes the already present void region as a contact medium, i.e. requires no additional regions or surfaces to be defined, and 2) implies a differentiable contact formulation. Specifically, the method incorporates the third medium contact method \cite{Wriggers2013AMedium} into density-based topology optimization by exploiting the void region as a contact medium.
Third medium contact is an implicit contact established by the stiffening of a highly compliant medium compressed between the contacting bodies. The method is differentiable since the third medium has a finite stiffness, albeit very low, at any level of compression with non-zero volume. This results in the contact being included in the sensitivities during optimization, thus making the third medium contact method very attractive for gradient based topology optimization.

The works by \cite{Bluhm2021InternalOptimization,Kumar2021OnMechanisms,Reddy2012SystematicMechanisms,Lawry2015LevelInterfaces} are at present the only approaches known to the authors which include internal contact in topology optimization with the aforementioned limitations. In addition to these, a range of different approaches to include external contact interfaces exist. A good overview may be found in \cite{Kristiansen2020TopologyFriction}, where the authors themselves introduce an approach to include frictional contact between members of the design domain and some external boundary. An overview of earlier work is given in a review paper by Hilding, Klarbring, and Petersson \cite{Hilding1999OptimizationContact}. The authors of \cite{Kumar2021OnMechanisms} also have prior work with a very similar approach, however, without contact between different members in the design domain, i.e. contact is limited to masks specified in the method \cite{Kumar2016SynthesisMethod}.

Usually, in fictitious domain methods, the void region will tend to invert and introduce numerical instabilities when severely deformed. This issue can be alleviated in various ways. Common approaches include 1) convergence criterion relaxation \cite{Pedersen2001TopologyMechanisms}, 2) void element removal \cite{Bruns2003AnMechanisms}, and 3) void element linearization \cite{Wang2014InterpolationProblems}. However, none of these approaches are well suited for topology optimization with third medium contact. Convergence criterion relaxation relaxes the convergence criterion for void regions, however, for third medium contact the void region is vital as it serves as a contact medium and thus should not be compromised. Void element removal is not a viable option since the void elements are needed as a contact medium. Void element linearization is not a suitable option either, as shear locking in the contact, where elements are highly compressed, will add nonphysical stiffness to the contact.
An approach more suitable when considering third medium contact has recently been introduced \cite{Bluhm2021InternalOptimization}. This method introduces a void regularization term which adds strain energy to bending and warping deformations in the void elements. The effect of the void regularization on the physical behaviour of the structure without contact is negligible since a very low degree of regularization provides sufficient stabilization.

A challenging design problem is that of a stiff coupling subjected to large deformations.
For such problems, maximization of reaction forces can lead to designs at the very limit of their stability where small displacement perturbations may cause the coupling to disengage. The present work addresses this issue by adding tangent stiffness terms to the design objective for couplings subject to large deformations, i.e. if an objective function maximizes a given point on a curve, then the slope of the curve at this point is also included in the objective function. A topology optimization approach including a dilated and an eroded density field is applied together with a staggered solution approach employing the widely used Method of Moving Asymptotes (MMA) \cite{Svanberg1987TheOptimization}. Third medium contact is included through the applied material law and Solid Isotropic Material Interpolation with Penalization (SIMP) \cite{Rozvany1992GeneralizedHomogenization}.

Section \ref{sec:modelling_approach} outlines the modelling approach by a) defining the material model including void regularization, b) defining the material interpolation scheme including the filtering and projection method employed, and c) stating the minimization problem and the solution algorithm employed to solve the problem. 
Section \ref{sec:results_and_discussion} applies the modelling approach to three problems: A) a lifting mechanism, B) a stiff coupling problem resulting in the formation of hooks, and C) a bending problem inspired by an endoscope bending section. Conclusions are included in section \ref{sec:conclusion} and the general notation employed throughout the present work is summarized in table \ref{tab:notation}.

\begin{table}[]
    \centering
    \begin{tabular}{ll}
    \hline
         $A\cdot B = A_i B_i$ & Scalar product \\
         $A:B = A_{ij} B_{ij}$ & Double contraction \\
         $A \, \tripledot \, B = A_{ijk} B_{ijk}$ & Triple contraction  \\
         $|A|=\text{det}(A)$ & Determinant of matrix $A$ \\
         $||A||=\sqrt{A:A}$ & Frobenius norm of matrix $A$ \\
         $\nabla A = \dfrac{\partial A_i}{\partial X_j}$ & Spatial gradient of a vector field $A$\\
         $\mathbb{H}A = \dfrac{\partial^2A_i}{\partial X_j \partial X_k}$ & Spatial Hessian of a vector field $A$ \\
         $\langle x \rangle = \text{max}(0,x)$ & Positive part function     \\ \hline
    \end{tabular}
    \caption{Notation conventions.}
    \label{tab:notation}
\end{table}

\section{Modelling approach}
\label{sec:modelling_approach}
\subsection{Material model}
The hyperelastic material model including void regularization presented by \cite{Bluhm2021InternalOptimization} is adopted. The model is restated below with a minor change to the scaling of the regularization term. 

Material strain energy density is expressed by an isotropic neo-Hookean law
\begin{equation}
    \overline{\Psi}(u) = \dfrac{K}{2}(\ln|F|)^2 + \dfrac{G}{2}\left(|F|^{-2/3} ||F||^2-3\right)
    \label{eq:neo_hook}
\end{equation}
where $F=I+\nabla u$ is the deformation gradient tensor, and $K$ and $G$ are bulk and shear modulus respectively, which are $K=5/3$\,MPa and $G=5/14$\,MPa for all examples treated in the present work.

An augmented strain energy density expression is obtained by adding a void regularization term to the material strain energy density
\begin{equation}
    \Psi(u) = \overline{\Psi}(u) + k_r\, \dfrac{1}{2} \mathbb{H}u \, \tripledot \, \mathbb{H}u
    \label{eq:psi_aug}
\end{equation}
where $\mathbb{H}u$ is the Hessian of the displacement field $u$, and $k_r$ is a scaling constant given by
\begin{equation}
    k_r = \Bar{k_r} L^2 K,
\end{equation}
where $L$ is a characteristic length scale of the structure. In the original form presented in \cite{Bluhm2021InternalOptimization}, the regularization term in eq.~\eqref{eq:psi_aug}, was scaled by the factor $\text{exp}(-5|F|)$, such that regularization diminishes in regions under tension. This term has been omitted in the present work to have a homogeneous regularization throughout the domain.
The constant $\Bar{k_r}$ is set to $\Bar{k_r}=10^{-6}$ for consistency with \cite{Bluhm2021InternalOptimization}. Lower values reduce possible stiffening of the void region, whereas higher values stabilize the problem.

The regularization term effectively stabilizes void regions by penalizing higher order deformations, i.e. warping and bending deformations. A value of $\Bar{k_r}$ should be chosen as small as possible to reduce the effect of the regularization on the physical behaviour of the problem, yet sufficiently large to effectively stabilize void regions.

Mechanical equilibrium can be expressed in a general weak form
\begin{equation}
    \mathcal{V}(u,q; \delta u, \delta q) = 0 \hspace{4mm} \forall \,\, \delta u, \delta q \label{eq:mech_eq_no_mat_int}
\end{equation}
where
\begin{equation}
    \mathcal{V}(u,q; \delta u, \delta q) = \int_\Omega \Psi_{,u}(u;\delta u) \, d\, \Omega + \int_{\Gamma_D} \left\{ q\cdot \delta u + (u-u_D) \cdot \delta q \right\} \, d\, \Gamma_D.
    \label{eq:virtual_work}
\end{equation}
Here, $\Omega$ is the modelling domain, $\Psi_{,u}(u;\delta u)$ is the first variation of the strain energy density with respect to $u$, $\Gamma_D$ is a boundary on the domain where a displacement $u_D$ may be prescribed, and $q$ is a vector field representing the reaction traction on the boundary $\Gamma_D$ as a result of the enforced displacement $u_D$.

\subsection{Contact through material interpolation}
In topology optimization, the stiffness of the material at any given point should depend on the material density such that solid regions represent the material model and void regions represent the absence of material. Values in between these two extremes are interpolated. In this case, the material interpolation may be introduced by multiplying the strain energy density from eq.~\eqref{eq:neo_hook} with a material interpolation function, here denoted by $\gamma(\Tilde{\rho})$. By introducing this material interpolation, the mechanical equilibrium, as introduced in the previous section, takes the form
\begin{equation}
\begin{split}
     \mathcal{R}(\Tilde{\rho},u,q; \delta u, \delta q) &= \int_\Omega \left\{ \gamma(\Tilde{\rho})\, \overline{\Psi}_{,u}(u;\delta u) + k_r\, \mathbb{H}u \, \tripledot \, \mathbb{H}\delta u \right\}\, d\, \Omega \\ 
     &+ \int_{\Gamma_D} \left\{ q\cdot \delta u + (u-u_D) \cdot \delta q \right\} \, d\, \Gamma_D = 0 \hspace{4mm} \forall \,\, \delta u, \delta q,  
\end{split}
\label{eq:residual}
\end{equation}
where $\Tilde{\rho}$ is a density field.

The material interpolation $\gamma(\Tilde{\rho})$ is obtained by combining the SIMP scheme \cite{Rozvany1992GeneralizedHomogenization} and a smooth Heaviside projection function $\Bar{\Tilde{\rho}}(\Tilde{\rho})$, resulting in
\begin{equation}
    \gamma(\Tilde{\rho}) = \gamma_0 + (1-\gamma_0) \Bar{\Tilde{\rho}}(\Tilde{\rho})^{\,p},
    \label{eq:simp}
\end{equation}
where $p$ is a penalization parameter and $\gamma_0$ is a minimum stiffness value of $\gamma(\Tilde{\rho})$. The projection function $\Bar{\Tilde{\rho}}(\Tilde{\rho})$, which yields the physical material density, projects values of $\Tilde{\rho}$ towards 0 and 1. It is described in the following section. A penalization of $p=3$ is applied throughout the present work.
In solid regions $\gamma(1)=1$ and in void regions $\gamma(0)=\gamma_0$. Thus, $\gamma_0$ expresses the difference in magnitude of the strain energy density expression from eq.~\eqref{eq:neo_hook} between solid and void regions. In the undeformed configuration this corresponds to the stiffness ratio between solid and void regions.

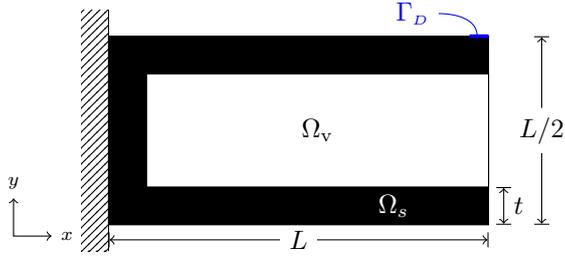
\begin{figure}
\centering
\begin{tikzpicture}
\draw [fill=black, draw=black] (0,0) rectangle (5,2.5);
\draw [fill=white!30, draw=black] (0.5,0.5) rectangle (5,2);
\fill [pattern=north east lines] (-0.35,-0.35) rectangle (0,2.85);
\draw [-] (0,-0.35) -- (0,2.85);

\node (L) at (2.5,-0.2) {\normalsize $L$};
\draw [|<-] (0,-0.2) -- (L);
\draw [->|] (L) -- (5,-0.2);

\node (H) at (5.7, 1.25) {\normalsize $L/2$};
\draw [|<-] (5.7, 0) -- (H);
\draw [->|] (H) -- (5.7,2.5);

\node (t) at (5.4, 0.25) {\normalsize $t$};
\draw [|<->|] (5.2, 0) -- (5.2,0.5);

\node [color=white] (omega_s) at (3.75, 0.25) {\normalsize $\Omega_s$};
\node (omega_v) at (2.75, 1.25) {\normalsize $\Omega_\mathrm{v}$};

\node [color=blue] (gamma_u) at (4, 2.8) {\normalsize $\Gamma_{\scriptscriptstyle D}$};
\draw [color=blue] (gamma_u.east) to[out=0, in=90] (4.85,2.5);
\draw [color=blue, line width=0.5mm] (4.75,2.5) -- (5,2.5);

\draw [<->] (-1.25,0.35) [anchor=south] node {\scriptsize $y$}  -- (-1.25,-0.15) -- (-0.75,-0.15)  [anchor=west] node {\scriptsize $x$};

\end{tikzpicture}
\caption{C-shape problem with void region $\Omega_\mathrm{v}$ and solid region $\Omega_s$. Thickness $t=0.1L$.}
\label{fig:C_shape_problem}
\end{figure}

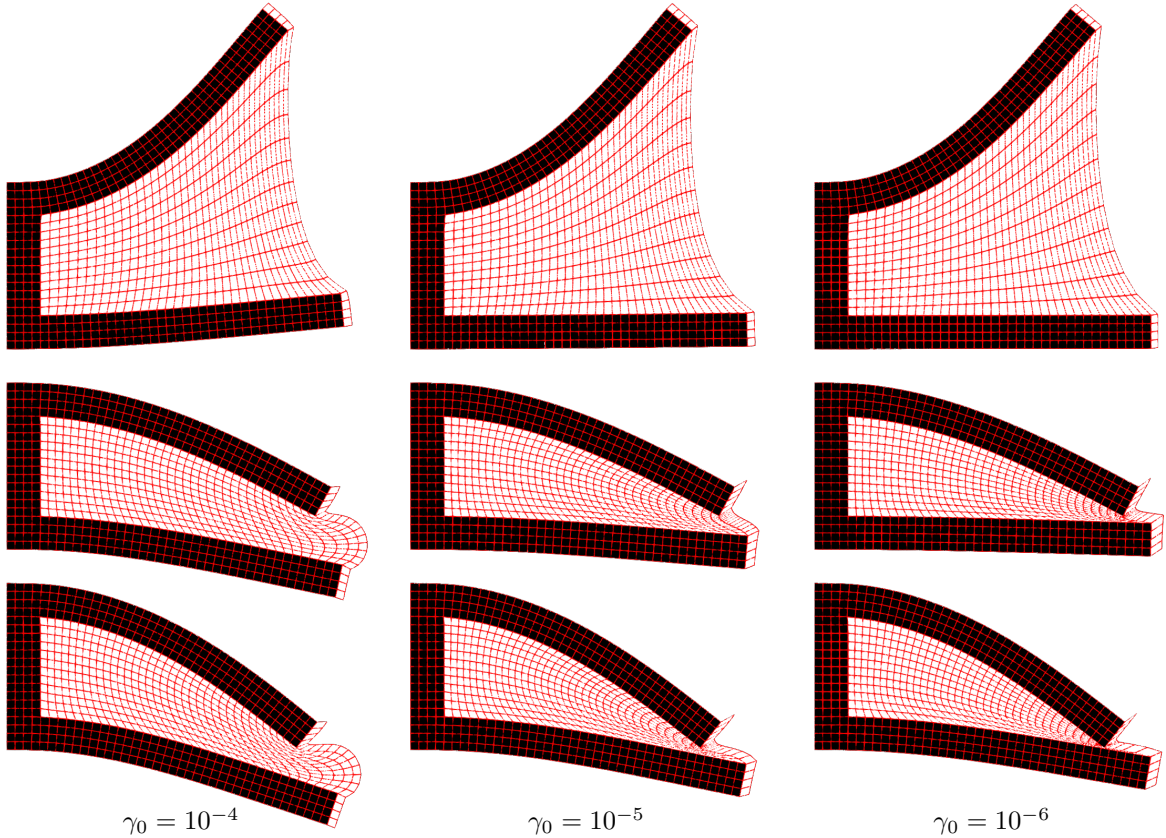
\begin{figure}[bthp]
    \centering
    \setlength{\unitlength}{0.1\textwidth}
    \begin{picture}(10,7)
       \put(0,0){\includegraphics[width=\textwidth]{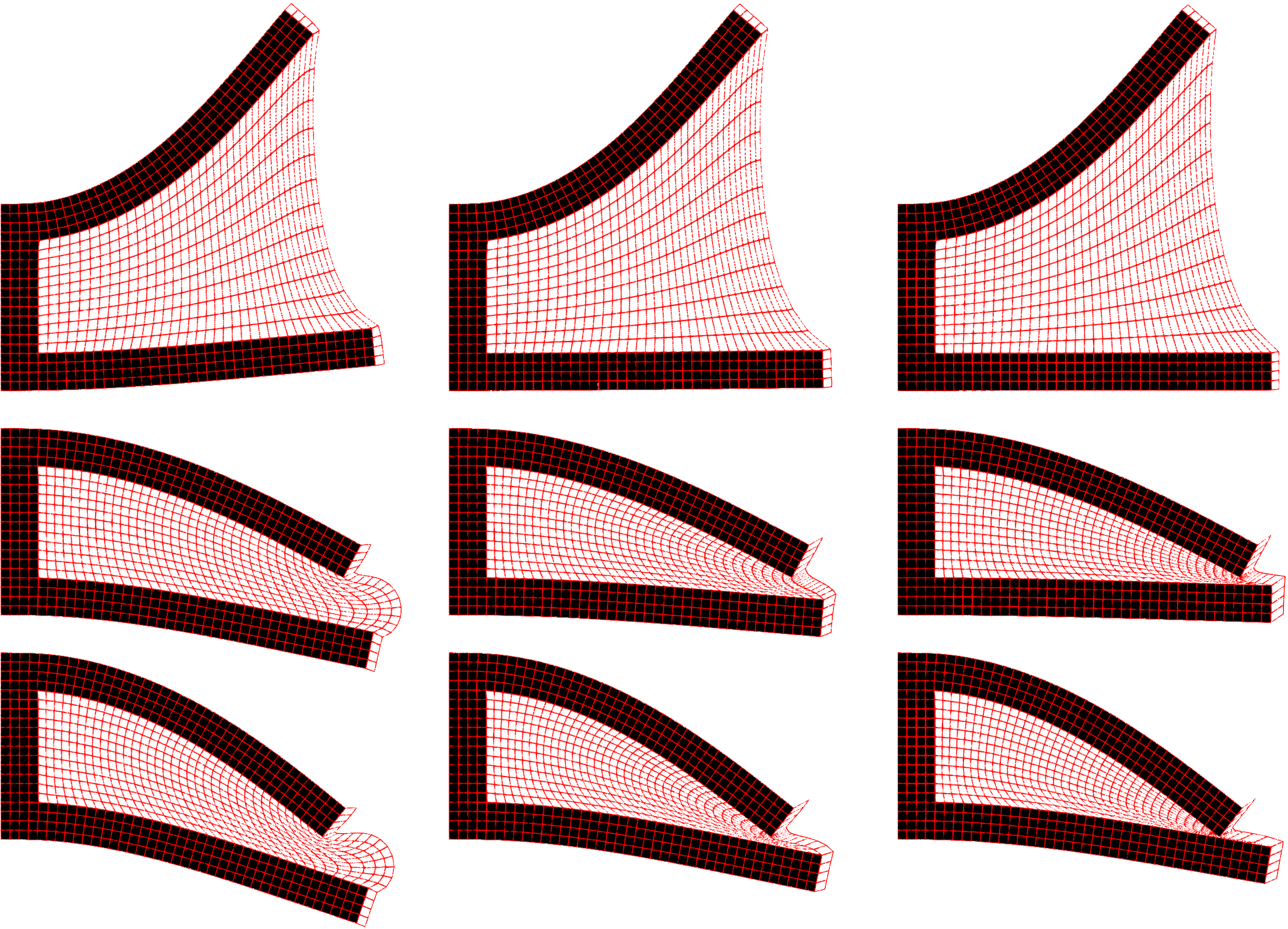}}
       \put(1,0){$\gamma_0=10^{-4}$}
       \put(4.5,0){$\gamma_0=10^{-5}$}
       \put(8,0){$\gamma_0=10^{-6}$}
    \end{picture}
    \caption{C-shape structure deformed by $u_{Dy}=0.5L$, $u_{Dy}=-0.3L$, $u_{Dy}=-0.4L$ for three different values of $\gamma_0$.}
    \label{fig:C_shape_deform}
\end{figure}

It should be emphasized that the material interpolation, i.e. the scaling of the strain energy density, naturally introduces third medium contact to the mechanical model. If the parameter $\gamma_0$ is chosen sufficiently low, the void may be deformed with close to no resistance. However, when void regions are compressed towards zero volume, the neo-Hookean material model stiffens the region and thus enables it to transfer loads. If the parameter $\gamma_0$ is chosen too high, parasitic forces will affect the design obtained as a result of the optimization. This has been a point of critique of the third medium contact method \cite{Zong2022TopologySelf-Contact}. However, choosing a sufficiently low value alleviates this issue. \cref{fig:C_shape_problem} shows the contact problem from \cite{Bluhm2021InternalOptimization} which serves as an example. The effect of $\gamma_0$ is shown in \cref{fig:C_shape_deform} for three different displacements imposed on the problem. From \cref{fig:C_shape_deform} it is evident that high values of $\gamma_0$ 1) transfer forces between non-contacting members through the void and 2) do not establish proper contact when contact is expected. 
A value of $\gamma_0=10^{-6}$ yields results with an acceptably low level of parasitic forces for the neo-Hookean material from eq. (\ref{eq:neo_hook}). Thus $\gamma_0=10^{-6}$ is used throughout the present work. This stiffness contrast value is also common in structural topology optimization in general.

\subsection*{Parametrization of the physical material density field}
\label{sec:filtering}
A well-established method in topology optimization is to work with a smooth density field~$\Tilde{\rho}$, obtained through a filtering step applied to an underlying design field $\rho$, discretized either as elementwise constant or with linear finite elements \cite{Wang2011OnOptimization,Sigmund2009ManufacturingOptimization}.

Filtering is applied to the field $\rho$ as a first step in order to control the smoothness of the material density field.
Among the available methods of filtering, Helmholtz filtering, proposed in
\cite{Lazarov2011FiltersEquations}, is a rather simple and universal approach.
Based on the design variable field $\rho$, Helmholtz filtering consists in finding a more smooth field $\Tilde{\rho}$ by solving a linear PDE
\begin{equation}
    \int_{\Omega} (\Tilde{\rho}-\rho) \, \delta \Tilde{\rho} + r^2 \nabla \Tilde{\rho} \cdot \nabla \delta \Tilde{\rho} ~ d\Omega= 0
    \label{eq:Helmholtz_weak}
\end{equation}
with natural boundary conditions on $\partial\Omega$.
In general, $\rho$ and $\Tilde{\rho}$ can be approximated in distinct discrete finite element spaces.
After discretization, eq.~\eqref{eq:Helmholtz_weak} is equivalent to
\begin{equation}
    \mathbf{M}_1 \cdot \Tilde{\bm{\rho}} + \mathbf{M}_0 \cdot \bm{\rho} = 0 \quad \Rightarrow \quad \Tilde{\bm{\rho}} = - \mathbf{M}_1^{-1} \mathbf{M}_0 \cdot \bm{\rho}
    \label{eq:M1M0}
\end{equation}
where $\bm{\rho}$ and $\Tilde{\bm{\rho}}$ are the corresponding discretized vectors, and $\mathbf{M}_0$ and $\mathbf{M}_1$ are the assembled tangent matrices from eq.~\eqref{eq:Helmholtz_weak}, for the coupling between the two variables.

As the matrices $\mathbf{M}_0$ and $\mathbf{M}_1$ remain unchanged after a design update, they only need to be computed once.
This linearity with respect to $\Tilde{\rho}$ makes it possible to prefactorize the system, as done in \cite{Lazarov2011FiltersEquations}. 
Here, an LU factorization is applied to $\mathbf{M}_1$ and its transpose. Now, rather than computing the inverse matrix in eq. (\ref{eq:M1M0}), a factorized system is solved where the factorization can be pre-completed outside the optimization loop.

The filtering step outlined above yields a smooth density field $\Tilde{\rho}$ characterized by intermediate densities. In order to 1) obtain a design dominated by solid and void regions and 2) enable sharper transitions between solid and void, the filtered density $\Tilde{\rho}$ is projected through the function
\begin{equation}
    \Bar{\Tilde{\rho}}(\Tilde{\rho}) = \dfrac{\tanh(\beta \eta) + \tanh(\beta(\Tilde{\rho}-\eta))}{\tanh(\beta \eta) + \tanh(\beta(1-\eta))}
    \label{eq:threshold}
\end{equation}
where $\beta$ is a projection parameter and $\eta$ is a threshold parameter. The projected design $\Bar{\Tilde{\rho}}(\Tilde{\rho})$ represents the physical material density, i.e. the density field used in the material interpolation from eq.~\eqref{eq:simp}. Values of $\Tilde{\rho}$ above $\eta$ are projected towards 1, and values below $\eta$ are projected towards 0. The extent to which the projected values approach 0 and 1 is controlled by the projection parameter $\beta$, high values of $\beta$ yield more forceful projections. Two thresholds are applied, $\eta_e$ and $\eta_d$ with $\eta_e>\eta_d$. The application of the thresholds, which respectively yield an eroded design $\Bar{\Tilde{\rho_e}}$ and a dilated design $\Bar{\Tilde{\rho_d}}$, is outlined in section \ref{sec:mini_prob} \cite{Wang2011OnOptimization}. Here, the aim is to penalize small features in the design, and not to control the length scale directly. Thus the choice of the thresholds is not vital, they are set to $\eta_e=0.50$ and $\eta_d=0.40$ throughout the present work. Filtering and projection is now only controlled through the filter radius $r$ and the projection parameter $\beta$. Subscripts $e$ and $d$ are added to $\Tilde{\rho}$ to indicate whether the filtered design $\Tilde{\rho}$ is to be eroded or dilated if passed through the projection function of eq.~\eqref{eq:threshold}.

\subsection{Minimization problem}
\label{sec:mini_prob}
A general formulation of the design problem is the minimization of an objective function $C(\Bar{\Tilde{\rho}},u,q)$ subject to the constraint of mechanical equilibrium, an upper bound on the total material volume in the design domain, and a box constraint on the design variable
\begin{equation}
    \begin{split}
    \min_\rho & \hspace{2mm} C(\Tilde{\rho}_e,u,q) \\
    s.t.: & \hspace{2mm} \mathcal{R}(\Tilde{\rho}_e, u,q; \delta u, \delta q) = 0 \hspace{4mm} \forall \,\, \delta u, \delta q \\
    & \hspace{2mm} \int_\Omega \Bar{\Tilde{\rho}}(\Tilde{\rho}_d) \, d\Omega \leq V^* \\
    & \hspace{2mm} 0 \leq \rho \leq 1.        
    \end{split}
    \label{eq:mini_problem}
\end{equation}
The utilization of two different design fields $\Tilde{\rho}_e$ and $\Tilde{\rho}_d$ is adapted from the robust formulation in topology optimization \cite{Wang2011OnOptimization}, where an eroded, an intermediate, and a dilated design are used to enforce a length-scale on the final design. The full robust formulation relies on a min-max formulation including all three designs. This is costly, and not strictly needed in this case as no specific length scale has to be enforced.
Thus, a different approach is applied where the objective and the mechanical equilibrium are both defined on an eroded design, while the volume constraint is defined on a dilated design. 
Letting the volume constraint be defined on a dilated design and the mechanical equilibrium be defined on an eroded design, essentially serves as a penalization of thin features, i.e. thin features will add little stiffness and be very costly in terms of the material volume they require.
By increasing the difference between $\eta_e$ and $\eta_d$ in eq.~\eqref{eq:threshold} or by increasing the filter radius $r$ in eq.~\eqref{eq:Helmholtz_weak} small features become increasingly uneconomical. This approach is distinct from the robust formulation, as it neither directly controls the length scale nor necessarily eliminates all gray-scale. Rather, it only serves as a penalization on thin features. It should be noted that since the volume constraint is defined on the dilated design only, the eroded design will always use less material volume as specified on the dilated design.

Following the approach in \cite{Bluhm2021InternalOptimization} the mechanical equilibrium constraint can be eliminated by minimizing the augmented objective function
\begin{equation}
    C^*(\Tilde{\rho}_e,u,q) = C(\Tilde{\rho}_e,u,q) + \mathcal{R}(\Tilde{\rho}_e, u,q; \lambda_u, \lambda_q), \label{eq:aug_obj}
\end{equation}
where $\lambda_u$ and $\lambda_q$ are Lagrange multipliers in the spaces of $\delta u$ and $\delta q$ respectively. Assuming an additive split of the objective function
\begin{equation}
    C(\Tilde{\rho}_e,u,q) = C_{\Tilde{\rho}}(\Tilde{\rho}_e) + C_u(u) + C_q(q)
\end{equation}
allows for a direct application of the adjoint method. The additive split results in the total variation of the objective function
\begin{equation}
    \begin{alignedat}{2}
    \delta C^* &= C_{\Tilde{\rho},\Tilde{\rho}}(\Tilde{\rho}_e; \delta \Tilde{\rho}) & &+ \mathcal{R}_{,\Tilde{\rho}}(\Tilde{\rho}_e, u,q; \lambda_u, \lambda_q ; \delta \Tilde{\rho}) \\
        &+ C_{u,u}(u; \delta u) & &+ \mathcal{R}_{,u}(\Tilde{\rho}_e, u,q; \lambda_u, \lambda_q ; \delta u) \\
        &+ C_{q,q}(q; \delta q) & &+ \mathcal{R}_{,q}(\Tilde{\rho}_e, u,q; \lambda_u, \lambda_q ; \delta q),
    \end{alignedat}
\end{equation}
which is simplified by determining the multipliers $\lambda_u$ and $\lambda_q$ satisfying the adjoint equations
\begin{alignat}{3}
    &C_{u,u}(u; \widehat{\delta u}) &&+ \mathcal{R}_{,u}(\Tilde{\rho}_e, u,q; \lambda_u, \lambda_q ; \widehat{\delta u}) &&= 0 \quad \forall \,\, \widehat{\delta u} \label{eq:adju} \quad \text{and}\\
    &C_{q,q}(q; \widehat{\delta q}) &&+ \mathcal{R}_{,q}(\Tilde{\rho}_e, u,q; \lambda_u, \lambda_q ; \widehat{\delta q}) &&= 0 \quad \forall \,\, \widehat{\delta q}, \label{eq:adjq}
\end{alignat}
where $\,\widehat{}\,$ is used to distinguish the variations in the adjoint eqs.~\eqref{eq:adju} and~\eqref{eq:adjq}, from the variations in the mechanical equilibrium eq.~\eqref{eq:residual}.
The solution of the adjoint system eliminates the variations in $\delta u$ and $\delta q$ from the variation of the objective function, thus reducing it to
\begin{equation}
    \delta C^* = C_{\Tilde{\rho},\Tilde{\rho}}(\Tilde{\rho}_e; \delta \Tilde{\rho}) + \mathcal{R}_{,\Tilde{\rho}}(\Tilde{\rho}, u,q; \lambda_u, \lambda_q ; \delta \Tilde{\rho}).
    \label{eq:optimality}
\end{equation}
This variation is stationary, i.e. $\delta C^* = 0$ at optimum. However, in the approach outlined in the following section, this stationary condition is not solved explicitly. Rather, eq.~\eqref{eq:optimality} is used to determine the sensitivities of the objective function.

In the above, variations in the design are with respect to the field $\Tilde{\rho}$, however, $\Tilde{\rho}$ depends on the field $\rho$, which in this case represents the design variable. The two fields $\Tilde{\rho}$ and $\rho$ are related through the filtering outlined in section \ref{sec:filtering}. Since the system for the filtering in section \ref{sec:filtering} is factorized, the transition from $\Tilde{\rho}$ to $\rho$ may done efficiently after discretization. This is relevant when computing the sensitivies of the augmented objective function \eqref{eq:aug_obj} and the volume constraint \eqref{eq:mini_problem} with respect to the design variable. These sensitivities are needed in order to apply MMA for the design update.
To clarify this, consider the example of $\mathcal{R}_{,\Tilde{\rho}}(\Tilde{\rho}, u,q; \lambda_u, \lambda_q ; \delta \Tilde{\rho})$ which may be written as $\partial \mathcal{R}/\partial \Tilde{\bm{\rho}} \cdot \delta \bm{\rho}$ in its discrete form. Then the discrete form of $\mathcal{R}_{,\rho}(\Tilde{\rho}, u,q; \lambda_u, \lambda_q ; \delta \rho)$ can be obtained as
\begin{equation}
    \dfrac{\partial \mathcal{R}}{\partial \bm{\rho} } \cdot \delta \bm{\rho} = \left(\dfrac{\partial \Tilde{\bm{\rho}} }{\partial \bm{\rho}}\right)^T \dfrac{\partial \mathcal{R}}{\partial \Tilde{\bm{\rho}} } \cdot \delta \bm{\rho} = -\mathbf{M}_0^{T} \mathbf{M}_1^{-T}  \dfrac{\partial \mathcal{R}}{\partial \Tilde{\bm{\rho}} } \cdot \delta \bm{\rho},
    \label{eq:sensivity_M}
\end{equation}
where, rather than computing the inverse of $\mathbf{M}_1$, the factorized system is solved. The same procedure is applied when calculating the volume constraint sensitivities.

\subsection{Solution algorithm}
\label{sec:solution_algorithm}
\begin{figure}[t]
\centering
\begin{tikzpicture}[node distance = 1.0cm]
\node [terminator] (start) {\textbf{Start}, initialize variables, $i=0$, $\beta=\beta_{ini}$};
\node [process, below of=start] (reset) {Reset variables, $u=0$, $q=0$};
\node [process, below of=reset, fill=gray!20] (mech) {Solve mechanical equilibrium ($u,q$)};
\node [process, below of=mech, fill=gray!20] (incr) {$u_D=\min(u_D+\Delta u_{incr},u_{D_0})$ on $\Gamma_D$};
\node [decision, below=0.5cm of incr, aspect=2, fill=gray!20] (full_disp) {$u_D = u_{D_0}$ on $\Gamma_D$?};
\node [process, right=0.75cm of full_disp, fill=gray!20] (step) {repeat};
\node [process, below=1.7cm of full_disp] (adj) {Solve adjoint problem ($\lambda_u$, $\lambda_q$)};
\node [process, below of=adj] (mma) {MMA, update design ($\rho$, $\Bar{\Tilde{\rho}}$)};
\node [decision, below=0.5cm of mma, aspect=2, align=center] (conv) {$\beta=\beta_{max}$ \& \\ $\max(\Delta \rho)\leq (\Delta \rho)_{max}$)?};
\node [process, align=left, node distance = 1cm] at ($(conv.center)+(6.5cm,0)$) (repeat) {i = i+1 \\ if mod(i,20) = 0: \\$\beta=\min(1.2\,\beta,\beta_{\max})$};

\node [process, above of=adj] (mech2) {Solve mechanical equilibrium ($u,q$)};

\node [terminator, below=0.5cm of conv] (end) {\textbf{End}};

\node[draw=none,anchor=south west] at (full_disp.east) (no) {No};
\node[draw=none,anchor=south east] at (full_disp.west) (yes) {Yes};
\node[draw=none,anchor=south west] at (conv.east) (no) {No};
\node[draw=none,anchor=north west] at (conv.south) (yes) {Yes};

\coordinate (fb) at ($(mech2.west)+(-1em,0)$);

\path [connector] (start) -- (reset);
\path [connector] (reset) -- (mech);
\path [connector] (mech) -- (incr);
\path [connector] (incr) -- (full_disp);
\path [connector] (full_disp) -- (step);
\path [connector] (step) |- (mech);

\draw [->] (full_disp) -| (fb) |- (adj);
\path [connector] (mech2) -- (adj);

\path [connector] (adj) -- (mma);
\path [connector] (mma) -- (conv);
\path [connector] (conv) -- (repeat);
\path [connector] (conv) -- (end);
\path [connector] (repeat) |- (mech2);
\end{tikzpicture}
\caption{Staggered solution approach.}
\label{fig:stag_solution}
\end{figure}
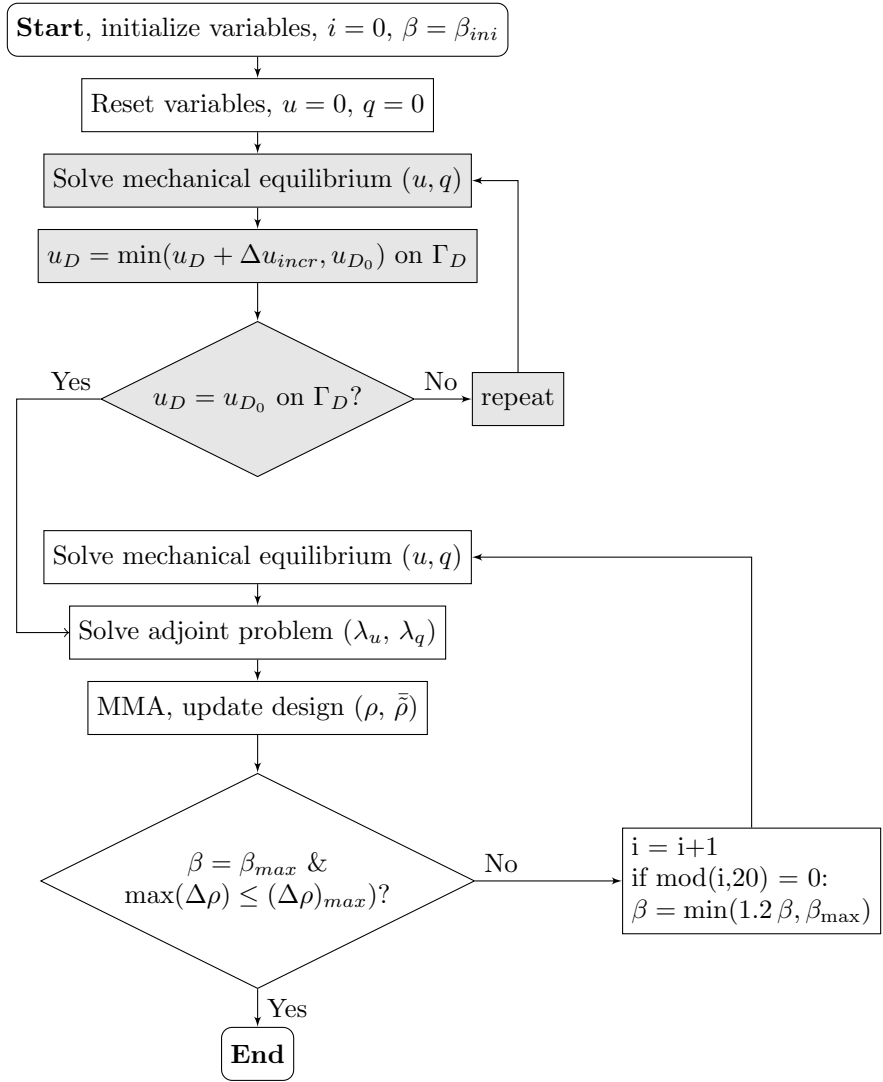

In contrast to \cite{Bluhm2021InternalOptimization} where a monolithic approach is used, here we apply a staggered solution approach where the mechanical equilibrium eq.~\eqref{eq:residual}, the adjoint problem eqs.~(\ref{eq:adju},~\ref{eq:adjq}), and the design update are solved in a consecutive order. The procedure is illustrated in \cref{fig:stag_solution}.

After an update of the mechanical equilibrium \eqref{eq:residual}, the adjoint variables $\lambda_u$ and $\lambda_q$ are determined by solving eqs. (\ref{eq:adju}) and (\ref{eq:adjq}). The adjoint variables are used to determine the value of the augmented objective function from eq. (\ref{eq:aug_obj}) and its sensitivity with respect to the design variable given in eq. (\ref{eq:optimality}). These are needed for the MMA design update.

Each design iteration is followed by an evaluation of a criterion for the increase of $\beta$. This is according to the standard robust formulation \cite{Wang2011OnOptimization} where the $\beta$-continuation yields increasingly sharp transitions between solid and void regions. The upper limit $\beta_{max}$ is chosen such that the transition between solid and void regions is sharp, yet sufficiently smooth for the contact not to depend directly on the discretization. Convergence is evaluated based on the change in the design variable $\rho$, where $(\Delta \rho)_{max}$ is the limit for the maximum change in any value of $\rho$ when the design has converged. 
It should be noted that $\beta$ is increased by a factor of 1.2 when the threshold is progressed in \cref{fig:stag_solution}. This is a rather low value for the progression parameter, however, this level is chosen in order to retain stability when the threshold is progressed.

The approach outlined above is implemented in Python where all tangent matrices and residual vectors are computed through the library GetFEM \cite{Renard2021GetFEM:Language} which performs all underlying computations in C++. Design updates are based on a Python implementation \cite{Python_MMA_Implementation} of MMA \cite{Svanberg1987TheOptimization}.

\section{Results and discussion}
\label{sec:results_and_discussion}

\subsection{Push-lift mechanism}
\label{sec:lift}
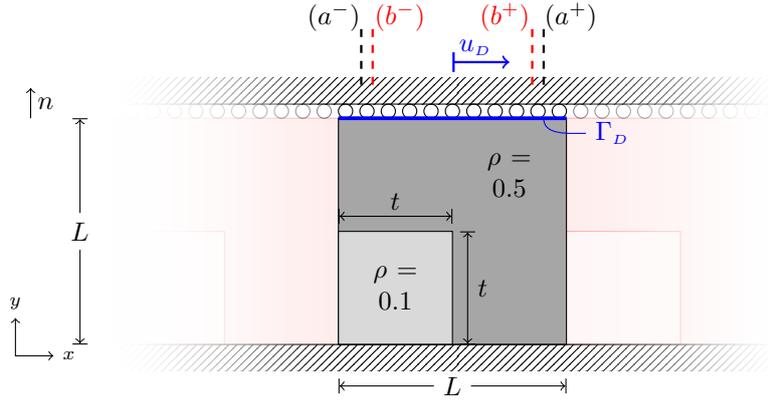
\begin{figure}[h]
\centering
\begin{tikzpicture}

\newcommand{\dy}{1.5}
\newcommand{\dx}{1.5} 
\newcommand{\LX}{3} 
\newcommand{\LY}{3} 
\newcommand{\Rcirc}{\LX/32} 
\newcommand{\dhatch}{0.35} 
\newcommand{\buff}{0.2} 

\draw [fill=red!7, draw=red, draw opacity=0.3] (-\LX,0) rectangle (0,\LY);
\draw [fill=red!3, draw=red, draw opacity=0.3] (-\LX,0) rectangle (\dx-\LX,\dy);
\draw [fill=red!7, draw=red, draw opacity=0.3] (\LX,0) rectangle (2*\LX,\LY);
\draw [fill=red!3, draw=red, draw opacity=0.3] (\LX,0) rectangle (\dx+\LX,\dy);
\draw [fill=gray!70, draw=black] (0,0) rectangle (\LX,\LY);
\draw [fill=gray!30, draw=black] (0,0) rectangle (\dx,\dy);
\fill [pattern=north east lines] (-\LX,-\dhatch) rectangle (2*\LX,0);
\fill [pattern=north east lines] (-\LX,\LY+2*\Rcirc) rectangle (2*\LX,\LY+2*\Rcirc+\dhatch);
\draw [-] (-\LX,0) -- (2*\LX,0);
\draw [-] (-\LX,\LY+2*\Rcirc) -- (2*\LX,\LY+2*\Rcirc);

\foreach \i in {0,...,10} {\draw [draw=black] (\i*3*\Rcirc+1*\Rcirc,\LY+ 1*\Rcirc) circle (\Rcirc);}
\foreach \i in {-10,...,-1} {\draw [draw=black,opacity=0.5] (\i*3*\Rcirc+1*\Rcirc,\LY+ 1*\Rcirc) circle (\Rcirc);}
\foreach \i in {11,...,20} {\draw [draw=black,opacity=0.5] (\i*3*\Rcirc+1*\Rcirc,\LY+ 1*\Rcirc) circle (\Rcirc);}

\fill [white,path fading=fade right] (-\LX-0.01,-\dhatch-0.01) rectangle (0,\LY+2*\Rcirc+\dhatch+0.01);
\fill [white,path fading=fade left] (\LX,-\dhatch-0.01) rectangle (2*\LX+0.01,\LY+2*\Rcirc+\dhatch+0.01);
\fill [white,path fading=fade right] (-\LX-0.01,-\dhatch) rectangle (-0.5\LX,\LY+2*\Rcirc+\dhatch+0.01);
\fill [white,path fading=fade left] (1.5*\LX,-\dhatch) rectangle (2*\LX+0.01,\LY+2*\Rcirc+\dhatch+0.01);

\node (L) at (\LX/2,-\dhatch-\buff) {\normalsize $L$};
\draw [|<-] (0,-\dhatch-\buff) -- (L);
\draw [->|] (L) -- (\LX,-\dhatch-\buff);
\node (H) at (0-2*\buff-\LX, \LY/2) {\normalsize $L$};
\draw [|<-] (0-2*\buff-\LX, 0) -- (H);
\draw [->|] (H) -- (0-2*\buff-\LX,\LY);

\node (gamma_u) [color=blue] at (\LX+3*\buff, \LY-\buff) {\normalsize $\Gamma_{\scriptscriptstyle D}$};
\draw [color=blue] (gamma_u.west) to [out=180, in=-90] (\LX-1.5*\buff,\LY);
\draw [line width=0.5mm,color=blue] (0,\LY) -- (\LX,\LY);

\draw [|->,line width=0.3mm,color=blue] (0.5*\LX,\LY+\Rcirc+\dhatch+1.5*\buff) -- (0.75*\LX,\LY+\Rcirc+\dhatch+1.5*\buff);
\node (u_D) [color=blue] at (0.6*\LX,\LY+\Rcirc+\dhatch+2.5*\buff) {\normalsize $u_{\scriptscriptstyle D}$};

\draw [-,dashed,line width=0.3mm,color=black] (0.5*\LX+0.4*\LX,\LY+\Rcirc+\dhatch+0*\buff) -- (0.5*\LX+0.4*\LX,\LY+\Rcirc+\dhatch+4*\buff);
\node (a) [color=black] at (0.5*\LX+0.52*\LX,\LY+\Rcirc+\dhatch+4.5*\buff) {\normalsize $(a^+)$};
\draw [-,dashed,line width=0.3mm,color=black] (0.5*\LX-0.4*\LX,\LY+\Rcirc+\dhatch+0*\buff) -- (0.5*\LX-0.4*\LX,\LY+\Rcirc+\dhatch+4*\buff);
\node (b) [color=black] at (0.5*\LX-0.52*\LX,\LY+\Rcirc+\dhatch+4.5*\buff) {\normalsize $(a^-)$};
\draw [-,dashed,line width=0.3mm,color=red] (0.5*\LX+0.35*\LX,\LY+\Rcirc+\dhatch+0*\buff) -- (0.5*\LX+0.35*\LX,\LY+\Rcirc+\dhatch+4*\buff);
\node (c) [color=red] at (0.5*\LX+0.23*\LX,\LY+\Rcirc+\dhatch+4.5*\buff) {\normalsize $(b^+)$};
\draw [-,dashed,line width=0.3mm,color=red] (0.5*\LX-0.35*\LX,\LY+\Rcirc+\dhatch+0*\buff) -- (0.5*\LX-0.35*\LX,\LY+\Rcirc+\dhatch+4*\buff);
\node (d) [color=red] at (0.5*\LX-0.23*\LX,\LY+\Rcirc+\dhatch+4.5*\buff) {\normalsize $(b^-)$};

\node[text width=1cm,align=center] (rho_1) at (0.5*\dx, 1/4*\LY) {\normalsize $\rho=$ \\ $0.1$};
\node[text width=1cm,align=center] (rho_2) at (0.75*\LX, 0.75*\LY) {\normalsize $\rho=$ \\ $0.5$};

\node (D) at (\dx/2,\dy+2*\buff) {\normalsize $t$};
\draw [|<->|] (0, \dy+\buff) -- (\dx,\dy+\buff);
\node [color=black] (t) at (\dx+2*\buff, \dx/2) {\normalsize $t$};
\draw [|<->|, color=black] (\dx+1*\buff, 0) -- (\dx+1*\buff,\dx);

\draw[->] (-3*\dhatch-\LX,\LY) -- (-3*\dhatch-\LX,\LY+2*\buff);
\node [] at (-3*\dhatch-\LX+\buff,\LY+1*\buff) {\normalsize $n$};

\draw [<->] (-1.25-\LX,0.35) [anchor=south] node {\scriptsize $y$}  -- (-1.25-\LX,-0.15) -- (-0.75-\LX,-0.15)  [anchor=west] node {\scriptsize $x$};

\end{tikzpicture}
\caption{Periodic design domain $L\times L$ for push-lift with indicated initial guess on $\rho$ for $t=0.5L$.}
\label{fig:box_problem}
\end{figure}

To illustrate how the method may be applied, we use the example shown in \cref{fig:box_problem} with periodic boundaries on the left and right side of the domain. 
In general, four horizontal translations:
\begin{alignat*}{4}
    &(a^+)\quad u_D= &&u_{max} , &&\qquad(a^-)\quad &&u_D=-u_{max} \\
    &(b^+)\quad u_D= &&u_{min} , &&\qquad(b^-)\quad &&u_D=-u_{min} \\
\end{alignat*}
are enforced on the top boundary $\Gamma_D$. As a first example, an objective function is considered that utilizes only the first two cases $(a^+)$ and $(a^-)$. This objective function maximizes the sum of the vertical reaction forces on $\Gamma_D$ in the two deformed configurations, and has the form
\begin{equation}
    C_0(\Tilde{\rho}_e,u,q) = k_q \left( -(Q_{a^+}\cdot n) - (Q_{a^-}\cdot n) \right)
    \label{eq:box_simple}
\end{equation}
where 
\begin{equation}
    Q_i = \int_{\Gamma_D} q_i\, d \Gamma_D
\end{equation}
are the reaction forces in configuration $(a^+)$ and $(a^-)$ respectively. The factor $k_q$ is a scaling constant and $n$ is the upwards unit normal vector shown in \cref{fig:box_problem}

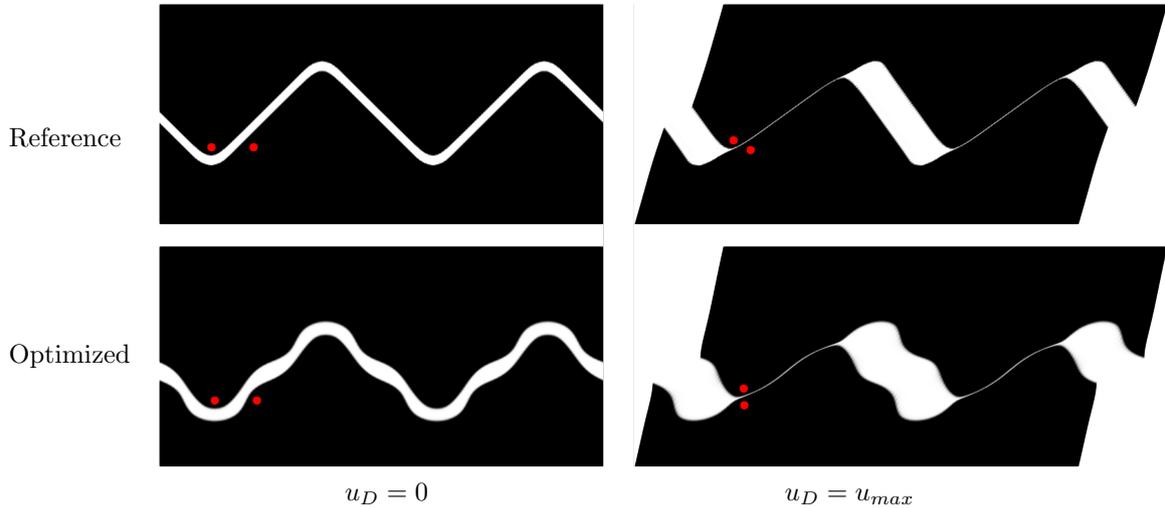
\begin{figure}
    \begin{subfigure}{\textwidth}
    \setlength{\unitlength}{0.1\textwidth}
    \begin{picture}(10,3.9)
       \put(1.3,0){\includegraphics[width=0.87\textwidth]{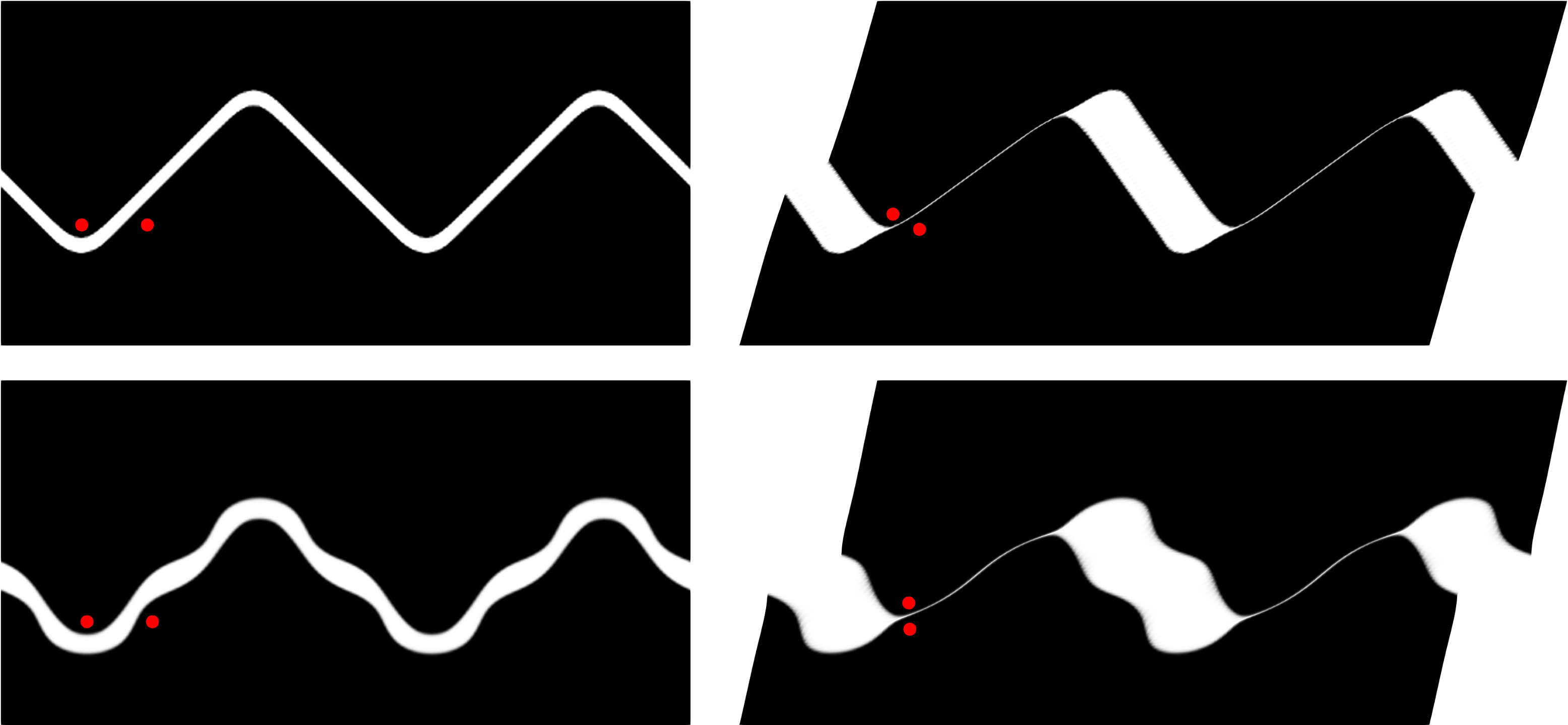}}
       \put(0.,2.8){\parbox{0.12\textwidth}{Reference}}
       \put(0.,0.9){\parbox{0.12\textwidth}{Optimized}}
       \put(2.9,-.3){$u_D=0$}
       \put(6.7,-.3){$u_D=u_{max}$}
    \end{picture}
    \end{subfigure}
    \vspace{2mm}
    \caption{Grey-scale image reference design and results for physical design ($\Bar{\Tilde{\rho}}(\Tilde{\rho}_e)$) optimized for objective function $C_0$ without volume constraint $V^*=1.0$. Red dots track displacements from undeformed to deformed configuration. Results are shown for 2 periods of the periodic domain.}
    \label{fig:box_simple}
\end{figure}

Numerical parameters for the problem are specified in \cref{tab:box_problem}. 
\begin{table}[b]
    \centering
    \begin{tabular}{llrl}
    \hline
         Domain size &$L\times L$ & $1\times 1$& mm$^2$ \\
         Enforced displacements & $(u_{min},u_{max})$ & $(0.35L,0.40L)$ & mm \\
         Mesh size & $N_x \times N_y$ & $100 \times 100$ & - \\
         Filter radius & $r$ & 0.10 & mm \\
         Threshold parameters & $(\beta_{ini},\beta_{max})$ & $(2,240)$ & -\\
         Objective weight & $k_q$ & $10^4$& - \\
         Force magnitude weight & $k_s$ & $10^3$& 1/N \\\hline
    \end{tabular}
    \caption{Parameters for push-lift mechanism.}
    \label{tab:box_problem}
\end{table}
The initial guess is as indicated by $\rho$ in \cref{fig:box_problem}, where the lower density in the $t\times t$-area serves to break symmetry. The displacement field $u$ is discretized by Q2 elements and the density fields $\rho$ and $\Tilde{\rho}$ are discretized by Q1 elements. Surface traction $q$ is discretized identically to $u$ on the boundary $\Gamma_D$. Gauss quadrature with nine Gauss-points is applied for the numerical integration on all elements.
Results without volume constraint, i.e. $V^*=1.0$, are shown in \cref{fig:box_simple}.

Despite having no volume constraint, the design shown in \cref{fig:box_simple} has a clear gap separating the structure into an upper and a lower part. The maximum threshold value $\beta_{max}$ from \cref{tab:box_problem} ensures a fast transition from solid to void, this is in order to have a physically sound contact interface. 

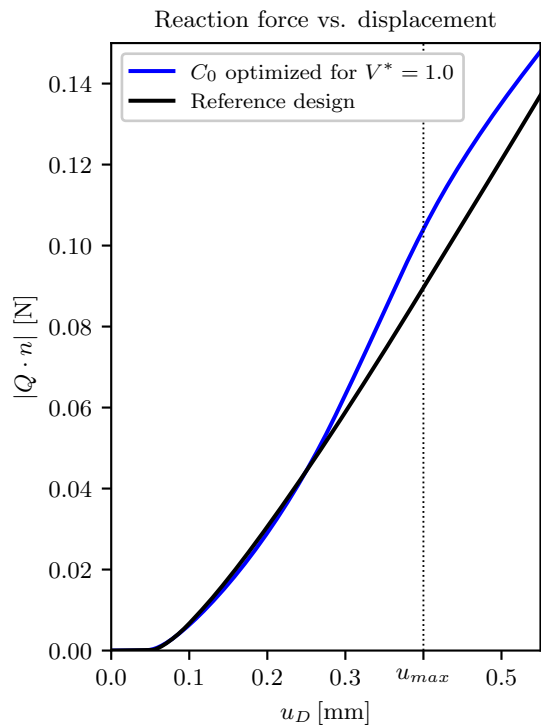
\begin{figure}[!t]
    \centering
    \input{box_force_curves_pyramid_v2.pgf}
    \caption{Force displacement curves for lifting mechanism designs. $C_0$ optimized design and reference design.}
    \label{fig:box_force_curves_pyramid}
\end{figure}

The gap width is indirectly controlled by changing the filter radius. However, the curved interfaces of the gap in the optimized design are not intuitive.
An intuitive alternative to the optimized design could be a narrow gap at a 45$^\circ$ angle as shown in \cref{fig:box_simple} as a reference design. 
A better understanding of the optimized design can be obtained by comparing the force-displacement curves for the two designs, shown in \cref{fig:box_force_curves_pyramid}.
In the deformed configuration $u_D=u_{max}$ the optimized design (blue curve) has a higher vertical load, thus also performing better with respect to the objective function $C_0$. Two red dots have been added to \cref{fig:box_simple} to motivate why this is the case. The dots are initially at the same height, with equal spacing for the optimized and the reference design. In the deformed configuration, the vertical distance between the points is higher for the optimized design than for the reference design, this indicates a higher level of compression at this point, and thus a higher reaction force in the vertical direction. Thus, the wave-like shape appears to provide an adaption of the contact surfaces that provides a larger force at large horizontal translation.

\begin{figure}
    \begin{subfigure}{\textwidth}
    \setlength{\unitlength}{0.1\textwidth}
    \begin{picture}(10,2.8)
       \put(0.7,0){\includegraphics[width=0.93\textwidth]{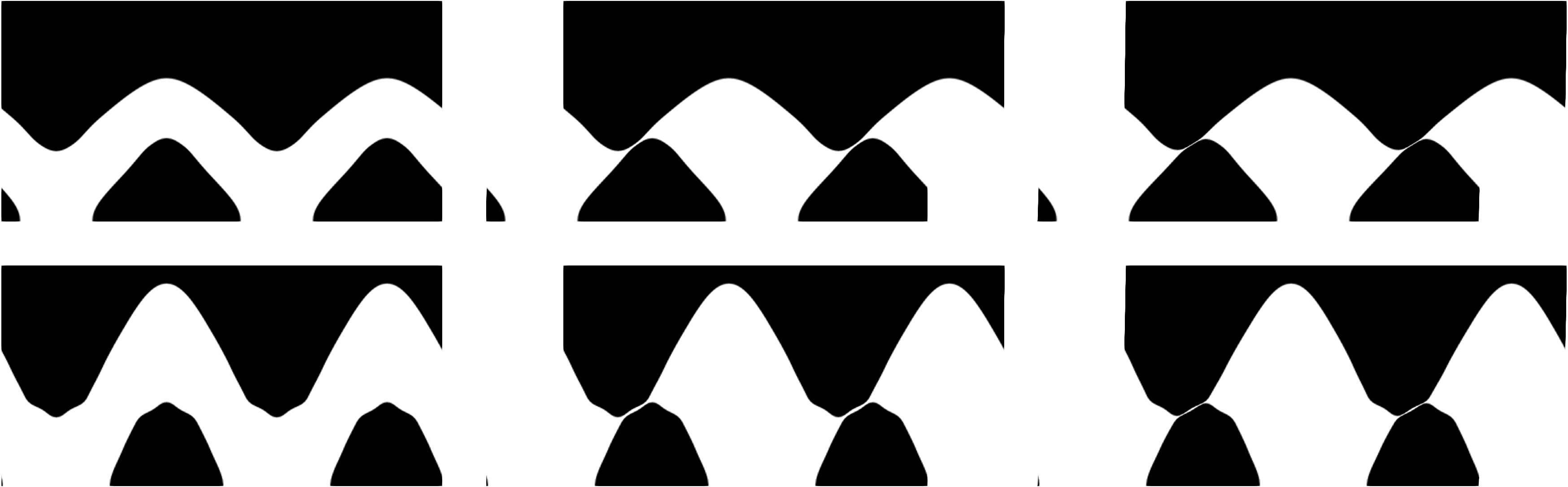}}
       \put(0.,2.0){\parbox{0.06\textwidth}{$V^*=$ $1.0$}}
       \put(0.,0.5){\parbox{0.06\textwidth}{$V^*=$ $0.6$}}
       \put(1.7,-.3){$u_D=0$}
       \put(5.0,-.3){$u_D=u_{min}$}
       \put(8.2,-.3){$u_D=u_{max}$}
    \end{picture}
    \end{subfigure}
    \vspace{2mm}
    \caption{Grey-scale image of results for physical design ($\Bar{\Tilde{\rho}}(\Tilde{\rho}_e)$) illustrated by 2 periods for objective function $C_1$ at displacements $u_D=0$, $u_D=u_{min}$, and $u_D=u_{max}$ for volume fractions $V^*=$ $1.0$ and $V^*=$ $0.6$.}
    \label{fig:boxes_shifted}
\end{figure}
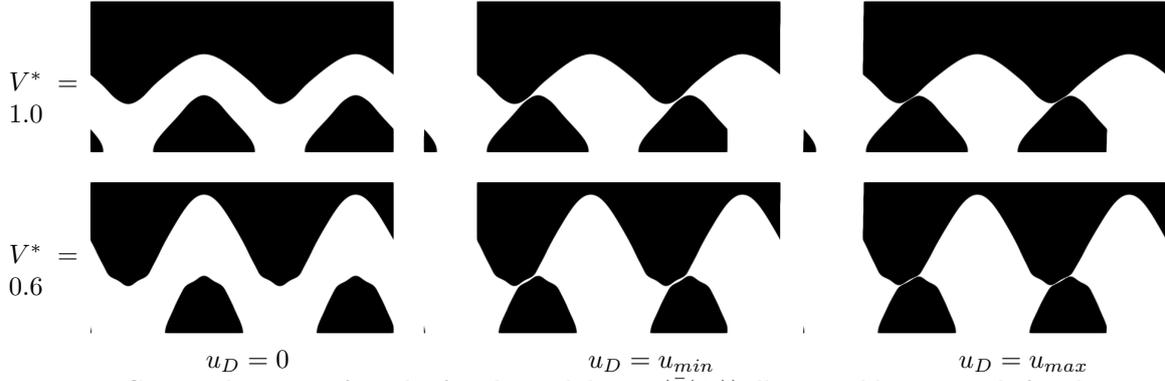

In order to gain greater control over the contact as well as the point at which contact is established, the two additional control points $(b^+)$ and $(b^-)$ may be added to the objective function
\begin{equation}
    C_1(\Tilde{\rho}_e,u,q) = k_q \left( -(Q_{a^+}\cdot n) -  (Q_{a^-}\cdot n) + k_s\,||Q_{b^+}||^2 + k_s\,||Q_{b^-}||^2 \right),
\end{equation}
where $k_s$ is a new scaling constant. The additional terms penalize any resulting force on $\Gamma_D$ at the displacement $\pm u_{min}$, thus discouraging contact for $-u_{min}<u_D<u_{min}$. Results using the objective function $C_1$ are shown in \cref{fig:boxes_shifted}.

\begin{figure}
    \centering
    \input{box_force_curves_v4.pgf}
    \caption{Force displacement curves for lifting mechanism designs.}
    \label{fig:box_force_curves}
\end{figure}
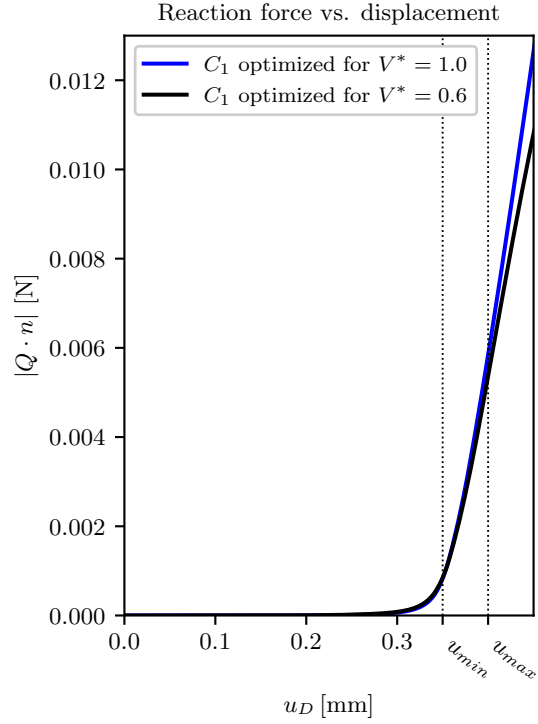

The force-displacement curves obtained by these cases are shown in \cref{fig:box_force_curves}. 
For the $C_1$ objective function there is a significant increase in the reaction force after the displacement $u_{min}$ is reached. This reflects that contact is established in the vicinity of this point. The objective function $C_1$ thus serves the intended purpose. Also, for $V^*=1.0$ the reaction force at $u_{max}$ is higher than for $V^*=0.6$, and it is slightly lower at $u_{min}$, this reflects the improvement obtained with a higher volume fraction.

\subsection{Self-engaging hooks}
\label{sec:hook}

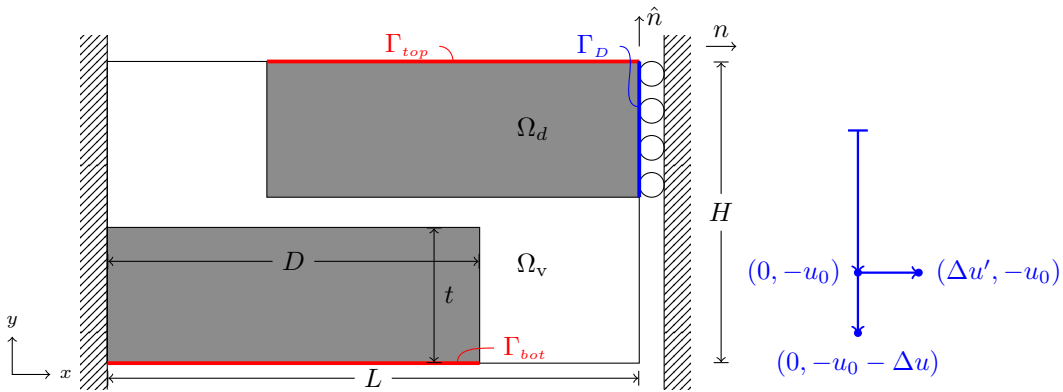
\begin{figure}[b]
\centering
\begin{tikzpicture}

\newcommand{\LX}{7} 
\newcommand{\LY}{0.57142857142*\LX} 
\newcommand{\dy}{0.45*\LY}
\newcommand{\dx}{0.7*\LX} 
\newcommand{\Rcirc}{\dy/11} 
\newcommand{\dhatch}{0.35} 
\newcommand{\buff}{0.2} 

\draw [fill=white!30, draw=black] (0,0) rectangle (\LX,\LY);
\draw [fill=gray!90, draw=black] (0,0) rectangle (\dx,\dy);
\draw [fill=gray!90, draw=black] (\LX-\dx,\LY-\dy) rectangle (\LX,\LY);
\fill [pattern=north east lines] (-\dhatch,-\dhatch) rectangle (0,\LY+\dhatch);
\fill [pattern=north east lines] (\LX+2*\Rcirc,-\dhatch) rectangle (\LX+2*\Rcirc+\dhatch,\LY+\dhatch);
\draw [-] (0,-\dhatch) -- (0,\LY+\dhatch);
\draw [-] (\LX+2*\Rcirc,-\dhatch) -- (\LX+2*\Rcirc,\LY+\dhatch);

\draw [draw=black] (\LX+\Rcirc,\LY- 1*\Rcirc) circle (\Rcirc);
\draw [draw=black] (\LX+\Rcirc,\LY- 4*\Rcirc) circle (\Rcirc);
\draw [draw=black] (\LX+\Rcirc,\LY- 7*\Rcirc) circle (\Rcirc);
\draw [draw=black] (\LX+\Rcirc,\LY-10*\Rcirc) circle (\Rcirc);

\node (L) at (\LX/2,-0.2) {\normalsize $L$};
\draw [|<-] (0,-0.2) -- (L);
\draw [->|] (L) -- (\LX,-0.2);

\node (H) at (\LX+\dhatch+2*\Rcirc+2*\buff, \LY/2) {\normalsize $H$};
\draw [|<-] (\LX+\dhatch+2*\Rcirc+2*\buff, 0) -- (H);
\draw [->|] (H) -- (\LX+\dhatch+2*\Rcirc+2*\buff,\LY);

\node (D) at (\dx/2,3/4*\dy) {\normalsize $D$};
\draw [|<-] (0,3/4*\dy) -- (D);
\draw [->|] (D) -- (\dx,3/4*\dy);

\node (omega_d) at (4/5*\LX, \LY-1/2*\dy) {\normalsize $\Omega_d$};
\node (omega_v) at (4/5*\LX, \LY-3/2*\dy) {\normalsize $\Omega_\mathrm{v}$};

\node (gamma_top) [color=red] at (\LX-1/2*\dx-3*\buff, \LY+\buff) {\normalsize $\Gamma_{\scriptscriptstyle top}$};
\draw [color=red] (gamma_top.east) to [out=0, in=90] (\LX-1/2*\dx,\LY);
\draw [line width=0.5mm,color=red] (\LX,\LY) -- (\LX-\dx,\LY);

\node (gamma_bot) [color=red] at (\dx+3*\buff, \buff) {\normalsize $\Gamma_{\scriptscriptstyle bot}$};
\draw [color=red] (gamma_bot.west) to [out=180, in=90] (9.4/10*\dx,0);
\draw [line width=0.5mm,color=red] (0,0) -- (\dx,0);

\node (gamma_u) [color=blue] at (\LX-3*\buff, \LY+\buff) {\normalsize $\Gamma_{\scriptscriptstyle D}$};
\draw [color=blue] (gamma_u.east) to [out=0, in=180] (\LX,\LY-\dy/3);
\draw [line width=0.5mm,color=blue] (\LX,\LY) -- (\LX,\LY-\dy);

\node (t) at (\dx-2*\buff, \dy/2) {\normalsize $t$};
\draw [|<->|] (\dx-3*\buff, 0) -- (\dx-3*\buff,\dy);

\draw[->] (\LX,\LY+\buff) -- (\LX,\LY+3*\buff);
\node [] at (\LX+\buff,\LY+3*\buff) {\normalsize $\hat{n}$};

\draw[->] (\LX+\dhatch+2*\Rcirc+1*\buff,\LY+\buff) -- (\LX+\dhatch+2*\Rcirc+3*\buff,\LY+\buff);
\node [] at (\LX+\dhatch+2*\Rcirc+2*\buff,\LY+2*\buff) {\normalsize $n$};

\coordinate (a0) at (\LX+\dhatch+2*\Rcirc+11*\buff,\LY-0.5*\dy);
\coordinate (a1) at (\LX+\dhatch+2*\Rcirc+11*\buff,0.3*\LY);
\coordinate (a2) at (\LX+\dhatch+2*\Rcirc+11*\buff,0.1*\LY);
\coordinate (a3) at (\LX+\dhatch+2*\Rcirc+11*\buff+0.2*\LY,0.3*\LY);
\draw[|->,line width=0.3mm ,color=blue] (a0) -- (a1);
\draw[->,line width=0.3mm,color=blue] (a1) -- (a2);
\draw[->,line width=0.3mm,color=blue] (a1) -- (a3);
\draw[blue,fill=blue] (a1) circle (.3ex);
\draw[blue,fill=blue] (a2) circle (.3ex);
\draw[blue,fill=blue] (a3) circle (.3ex);
\node [color=blue, left =0.5*\buff of a1] {\normalsize $(0,-u_0)$};
\node [color=blue, below=0.5*\buff of a2] {\normalsize $(0,-u_0-\Delta u)$};
\node [color=blue, right=0.5*\buff of a3] {\normalsize $(\Delta u',-u_0)$};

\draw [<->] (-1.25,0.35) [anchor=south] node {\scriptsize $y$}  -- (-1.25,-0.15) -- (-0.75,-0.15)  [anchor=west] node {\scriptsize $x$};

\end{tikzpicture}
\caption{Domain of self-engaging hooks problem. Initial guess $\rho=0.5$ in $\Omega_d$. Points used for enforced displacements of $\Gamma_D$ are illustrated to the right.}
\label{fig:hook_problem}
\end{figure}

Consider the domain shown in \cref{fig:hook_problem} where $\Gamma_D$ is a boundary on which a vertical displacement is prescribed.
The region $\Omega_\mathrm{v}$, which includes the boundaries $\Gamma_{top}$ and $\Gamma_{bot}$, is a prescribed void region, and $\Omega_d$ designates the design region, which is split in two parts with equal dimensions.
An objective for this problem may be to maximize the vertical reaction force on the boundary $\Gamma_D$ given a downwards displacement $u_0$. If the displacement $u_0$ is large, the reaction force on the boundary $\Gamma_D$ of the optimized design is very likely an extremum with respect to the displacement $u_0$.
In this case, any perturbation increasing the displacement may cause the structure to snap and disengage. 
Thus, the resulting design is not expected to be robust with respect to the displacement in its deformed configuration. A simple yet effective approach to obtain a more robust design is to incorporate a requirement for a positive tangent stiffness of the deformed configuration in the objective function. Here, a small perturbation $\Delta u$ is used either in the horizontal or vertical direction with respect to the initial displacement $(0,-u_0)$. 

Let $T$ and $T'$ respectively be the vertical and horizontal reaction forces on the boundary $\Gamma_D$
\begin{align}
    T  &= Q\cdot \hat{n} = \int_{\Gamma_D} q\cdot \hat{n}\, d \Gamma_D,\\
    T' &= Q\cdot n = \int_{\Gamma_D} q\cdot n\, d \Gamma_D,
\end{align}
where $\hat{n}$ is the unit tangent vector in the upwards direction and $n$ is the unit normal vector, as indicated in \cref{fig:hook_problem}.

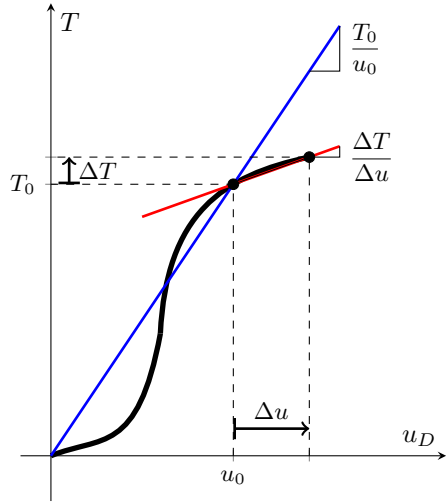
\begin{figure}
\centering
\begin{tikzpicture}
\newcommand{\xo}{6} 
\newcommand{\xp}{8.5} 
\newcommand{\yo}{6} 
\newcommand{\yp}{6.6} 
\begin{axis}[axis lines=middle,xmin=-1,xmax=13,ymin=-1,ymax=10,xlabel=$u_D$,ylabel=$T$,tick label style={font=\small,legend style={font=\small},legend pos=outer north east,},restrict y to domain=0:10,
y=0.6cm,
x=0.4cm,
xtick = {\xo,\xp},
xticklabels = {$u_{0}$,},
ytick = {\yo,\yp},
yticklabels = {$T_0$,}]
\addplot+[mark=none, line width=2pt,patch,mesh,patch type=cubic spline,faceted color=black, color=black]
coordinates {(0,0) (\xo*0.6,\yo*0.45) (1.5,0.3) (\xo*0.45,\yo*0.15)};
\addplot+[mark=none, line width=2pt,patch,mesh,patch type=cubic spline,faceted color=black, color=black]
coordinates {(\xo*0.6,\yo*0.45) (\xp,\yp) (\xo*0.72,\yo*0.8) (\xo,\yo)};
\addplot+[no marks,line width=1pt,blue,domain=-0:9.5,samples=150] {x};
\addplot+[no marks,line width=1pt,red,domain=3:9.5,samples=150] {(\yp-\yo)/(\xp-\xo)*x+(\yo-(\yp-\yo)/(\xp-\xo)*\xo)};

\addplot+[dashed,no marks,black] coordinates {(\xo,0) (\xo,\yo) (0,\yo)};
\addplot+[dashed,no marks,black] coordinates {(\xp,0) (\xp,\yp) (0,\yp)};

\addplot+[solid,no marks,black] coordinates {(\xp,\xp) (\xp+1,\xp) (\xp+1,\xp+1)};
\addplot+[solid,no marks,black] coordinates {
(\xp,{(\yp-\yo)/(\xp-\xo)*\xp+(\yo-(\yp-\yo)/(\xp-\xo)*\xo)}) 
(\xp+1,{(\yp-\yo)/(\xp-\xo)*\xp+(\yo-(\yp-\yo)/(\xp-\xo)*\xo)}) 
(\xp+1,{(\yp-\yo)/(\xp-\xo)*(\xp+1)+(\yo-(\yp-\yo)/(\xp-\xo)*\xo)})};

\addplot+[solid,|->,no marks,black,line width=1pt] coordinates {(\xo,\yo/10) (\xp,\yo/10)};
\node at (\xo+\xp/2-\xo/2,\yo/10) [anchor=south] {\small{$\Delta u$}};
\addplot+[solid,|->,no marks,black,line width=1pt] coordinates {(\xo/10,\yo) (\xo/10,\yp)};
\node at (\xo/10,\yo+\yp/2-\yo/2) [anchor=west] {\small{$\Delta T$}};

\node at (\xp+1,\xp+0.5) [anchor=west] {\small{$\dfrac{T_0}{u_{0}}$}};
\node at (\xp+1,{(\yp-\yo)/(\xp-\xo)*\xp+(\yo-(\yp-\yo)/(\xp-\xo)*\xo)}) [anchor=west] {\small{$\dfrac{\Delta T}{\Delta u}$}};

\addplot[mark=*, line width=0] coordinates {(\xo,\yo) (\xp,\yp)};

\end{axis}
\end{tikzpicture}
\caption{Representative illustration of loading history with indicated control points.}
\label{fig:hook_objective}
\end{figure}

\cref{fig:hook_objective} illustrates a qualitative loading history for the reaction force in the vertical direction $T$ on the boundary $\Gamma_D$ as a function of the applied vertical displacement $u_D$. In order to incorporate the tangent stiffness in the objective function, two control point, which are distance $\Delta u$ apart, are added, as shown in \cref{fig:hook_objective}.
A first order approximation of the aforementioned tangent stiffness can be based on these two control points
\begin{equation}
   \dfrac{\text{d}T}{\text{d}u_D} \Bigg\rvert_{u_D=u_{0}} \approx \dfrac{\Delta T}{\Delta u}.
\end{equation}
A straight-forward way to include this tangent stiffness is to penalize tangent stiffness values below a given threshold. Here we chose to penalize tangent stiffness values below the average stiffness of the structure, represented by the secant stiffness $T_0/u_0$.
The tangent stiffness can now be controlled by penalizing values of the tangent stiffness in the deformed configuration being smaller than the average stiffness, i.e. positive values of
\begin{equation}
    \dfrac{T_0}{u_{0}} - \dfrac{\Delta T}{\Delta u}.
    \label{eq:penalized_down}
\end{equation}
For improved stability, the same logic can be applied to a horizontal perturbation $\Delta u'$ on $\Gamma_D$ by penalizing tangent stiffness values below $1/10$ of the average stiffness. The perturbations are illustrated in figure \cref{fig:hook_problem}, note that the enforced displacement is negative in the $y$-direction and positive in the $x$-direction, thus the signs of the terms \eqref{eq:penalized_down} are flipped.
The objective now has the form
\begin{equation}
    C(\Tilde{\rho}_e,u,q) = k_q \left(-(T_0+\Delta T) 
    + k_s \dfrac{1}{2} \left\langle \dfrac{\Delta T}{\Delta u} - \dfrac{T_0}{u_{0}} \right\rangle^2
    + k_s \dfrac{1}{2} \left\langle -\dfrac{\Delta T'}{\Delta u'} -\dfrac{T_0}{u_{0}} \dfrac{1}{10}  \right\rangle^2 \right).
    \label{eq:hook_obj}
\end{equation}

The first term in the objective function $(T_0+\Delta T)$ is the reaction force in the deformed configuration $(0,-u_{0}-\Delta u)$. The second term in eq. (\ref{eq:hook_obj}) penalizes tangent stiffness values for downwards perturbations $\Delta u$ being smaller than the average stiffness $-T/u_{0}$. The third term penalizes transverse stiffness, i.e. the change in the reaction force $T$ with respect to a horizontal perturbation $\Delta u'$ on $\Gamma_D$. Thus, the objective function includes a tangent stiffness term for vertical as well as horizontal displacements on top of the deformed configuration. 
The factor $k_q$ scales the objective function in order to obtain an adequate scaling for the augmented objective function in eq. (\ref{eq:aug_obj}) and $k_s$ is the relative weight applied to the penalization of the tangent stiffness terms.

\begin{figure}[htbp]
\centering
    \subfloat[$u_0=2.0\,t$]{%
        \includegraphics[width=.5\linewidth]{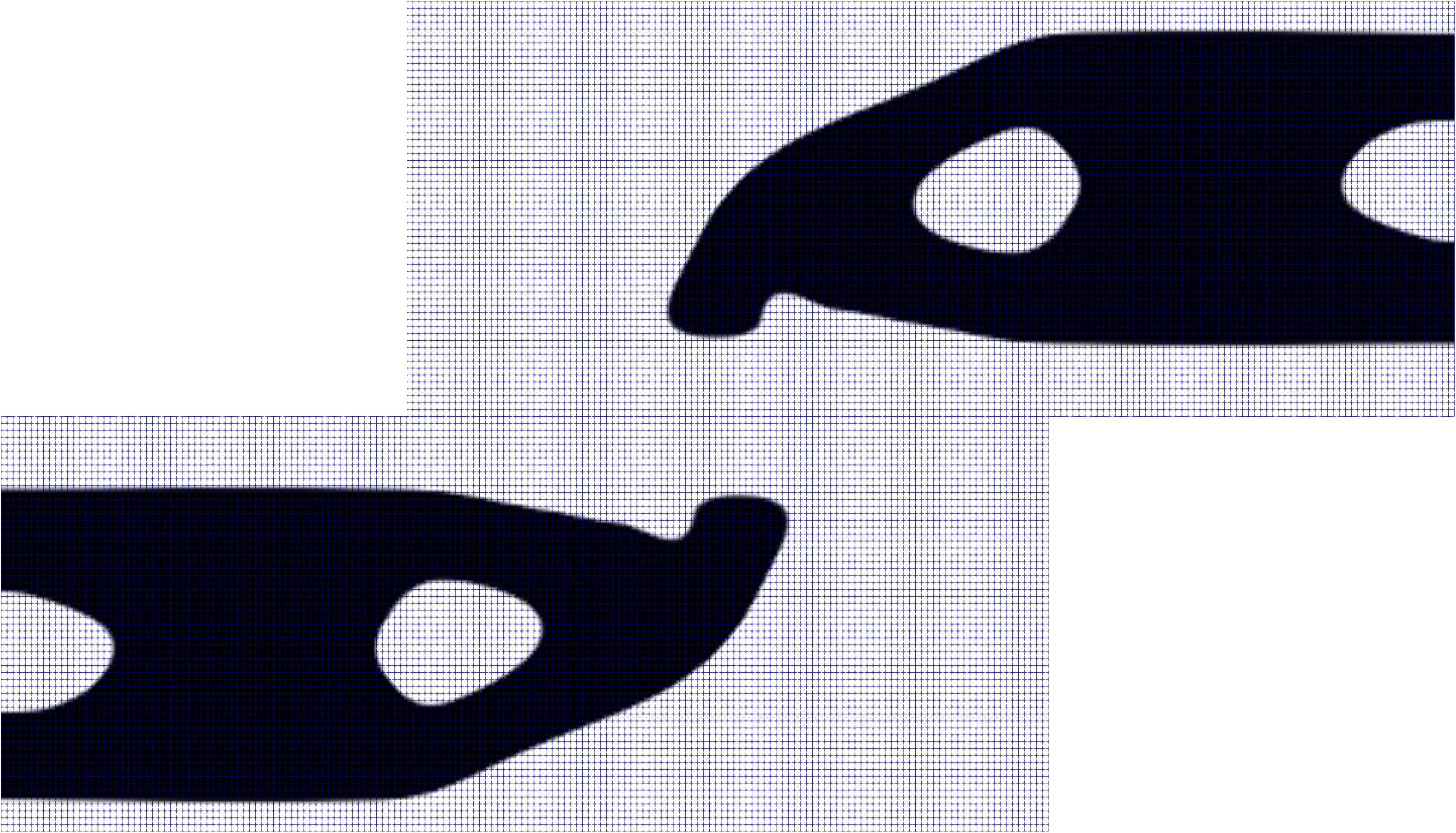}%
        \label{subfig:a}%
    }\\
    \subfloat[$u_0=2.5\,t$]{%
        \includegraphics[width=.5\linewidth]{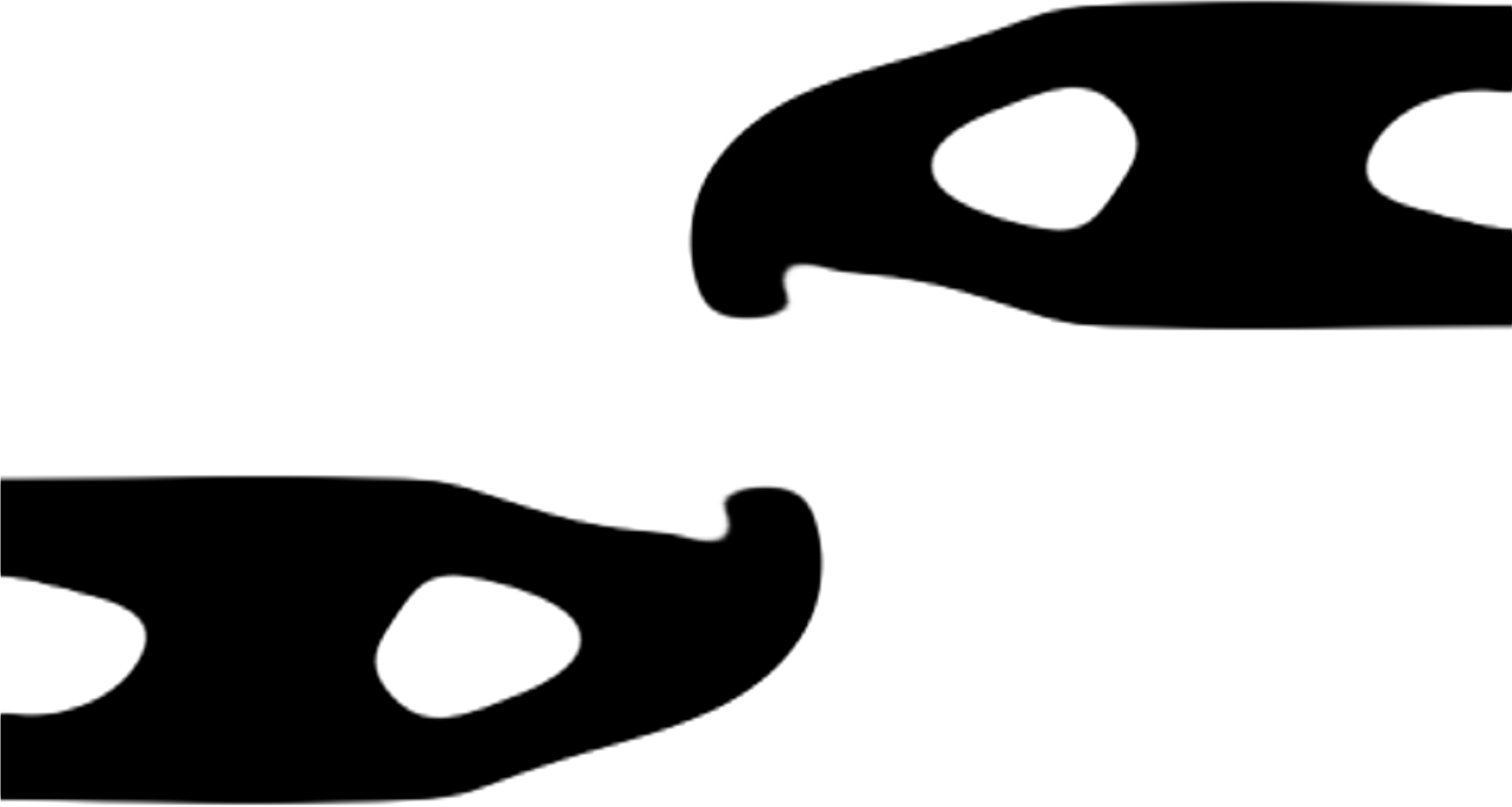}%
        \label{subfig:b}%
    }\\
    \subfloat[$u_0=3.0\,t$]{%
        \includegraphics[width=.5\linewidth]{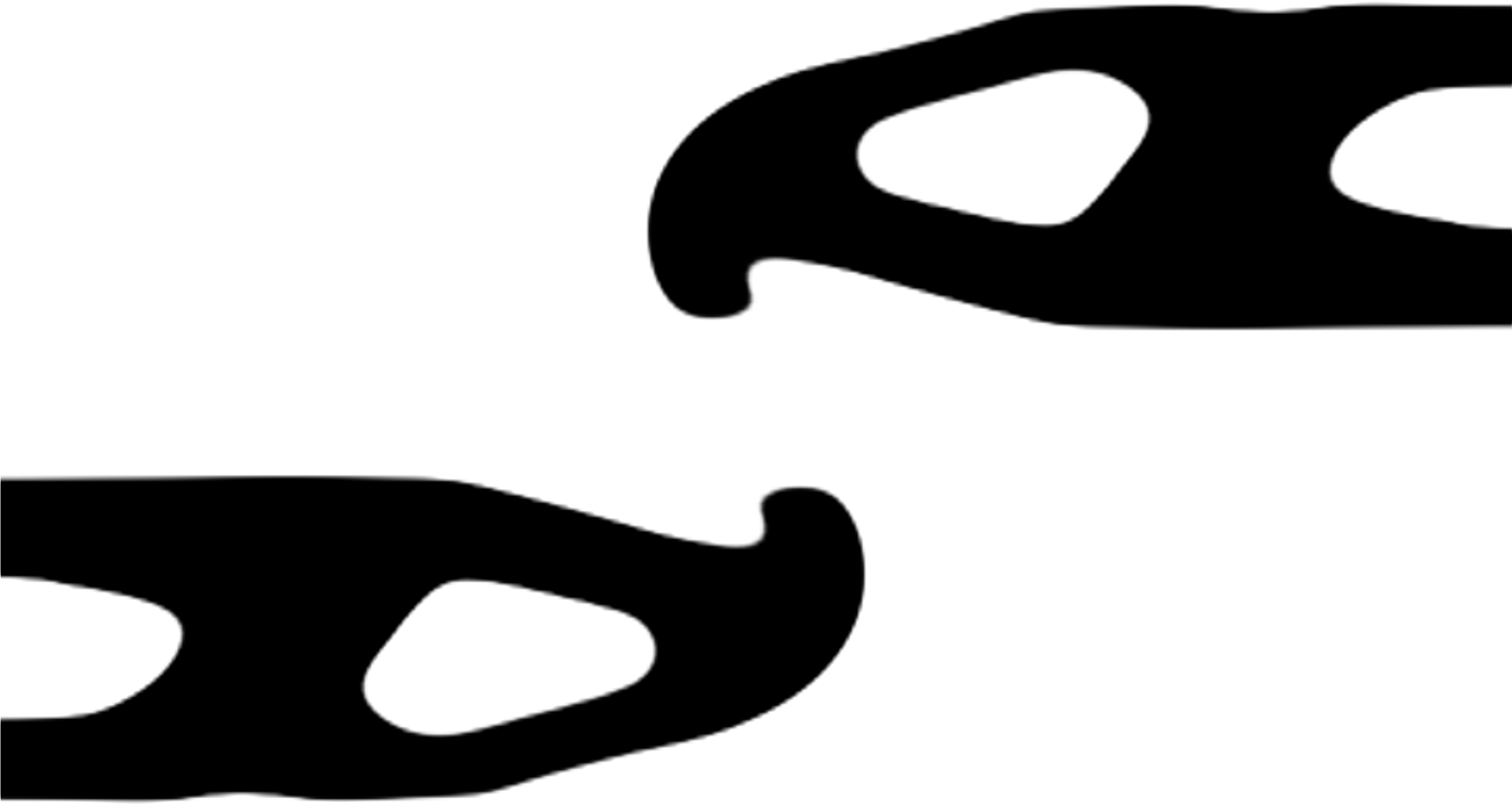}%
        \label{subfig:c}%
    }\hfill
    \caption{Optimized designs obtained for three different values of the enforced displacement $u_0$.}
    \label{fig:hooks_mult_def}
\end{figure}

\begin{figure}
\centering
\begin{tikzpicture}[scale=1.5, every node/.style={scale=1.5}]

\newcommand{\dy}{1.1}
\newcommand{\dx}{3.38} 
\newcommand{\LX}{5.2} 
\newcommand{\LY}{2.6} 
\newcommand{\Rcirc}{\dy/11} 
\newcommand{\dhatch}{0.35} 
\newcommand{\buff}{0.2} 

\newcommand{\uD}{0.95*\LY}
\newcommand{\dd}{1.6}

\fill [pattern=north east lines] (-\dhatch,-\dhatch) rectangle (0,\LY+\dhatch);
\fill [pattern=north east lines] (\LX+2*\Rcirc,-\dhatch-\uD) rectangle (\LX+2*\Rcirc+\dhatch,\LY+\dhatch);
\draw [-] (0,-\dhatch) -- (0,\LY+\dhatch);
\draw [-] (\LX+2*\Rcirc,-\dhatch-\uD) -- (\LX+2*\Rcirc,\LY+\dhatch);

\node[inner sep=0pt,anchor=south west] (picture) at (0,-1.0*\LY)
    {\includegraphics[width=\LX cm]{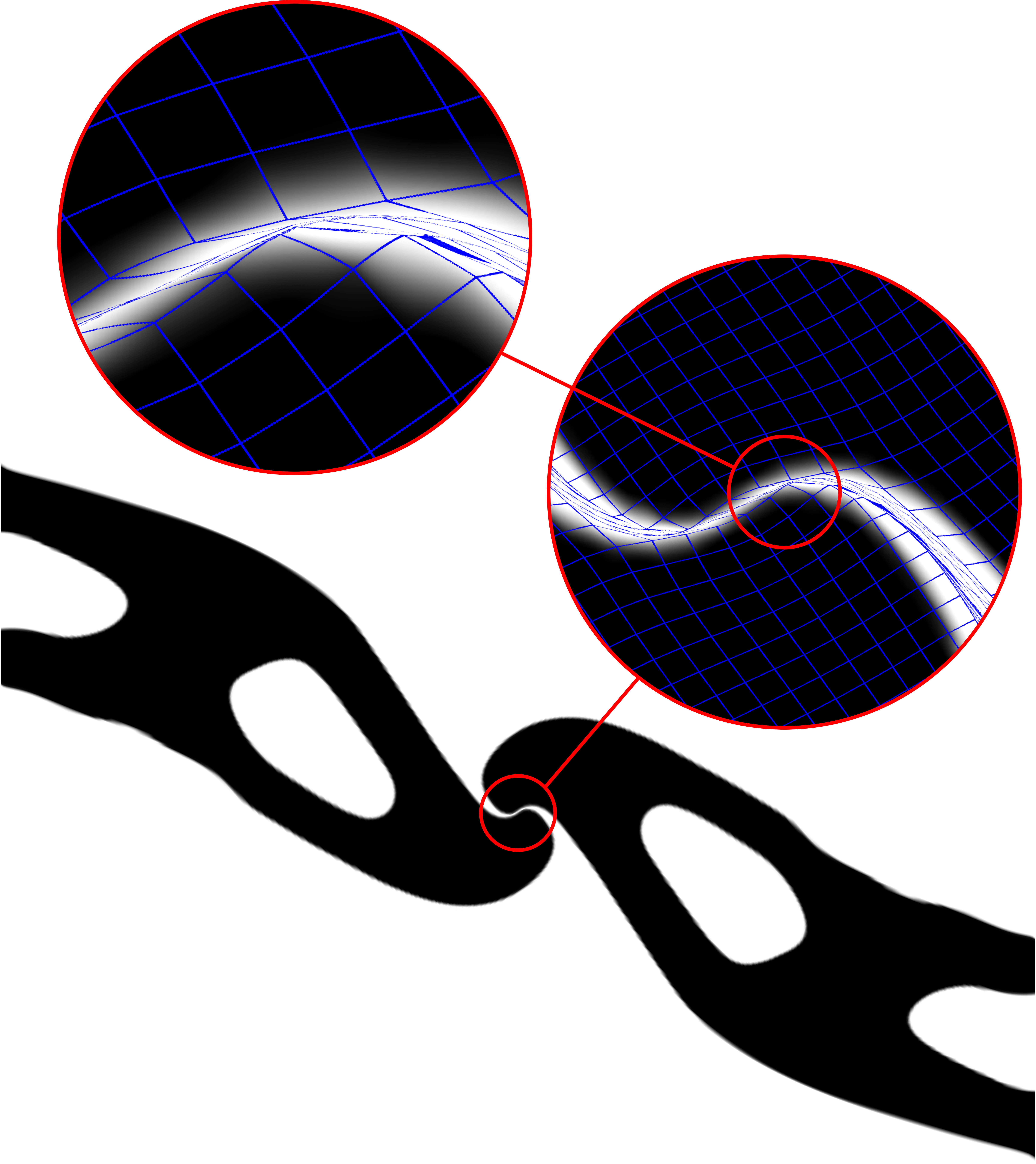}};

\draw [draw=black] (\LX+\Rcirc,\LY- 1*\Rcirc-\uD-\dd) circle (\Rcirc);
\draw [draw=black] (\LX+\Rcirc,\LY- 4*\Rcirc-\uD-\dd) circle (\Rcirc);
\draw [draw=black] (\LX+\Rcirc,\LY- 7*\Rcirc-\uD-\dd) circle (\Rcirc);
\draw [draw=black] (\LX+\Rcirc,\LY-10*\Rcirc-\uD-\dd) circle (\Rcirc);

\draw [draw=black] (\LX,\LY-\dy-\uD-\dd) -- (\LX,\LY-\uD-\dd);

\end{tikzpicture}
\caption{Hooks optimized for $u_0=3.0\,t$ at displacement in deformed configuration ($u_D=3.0\,t$).}
\label{fig:hook_deformed_design}
\end{figure}
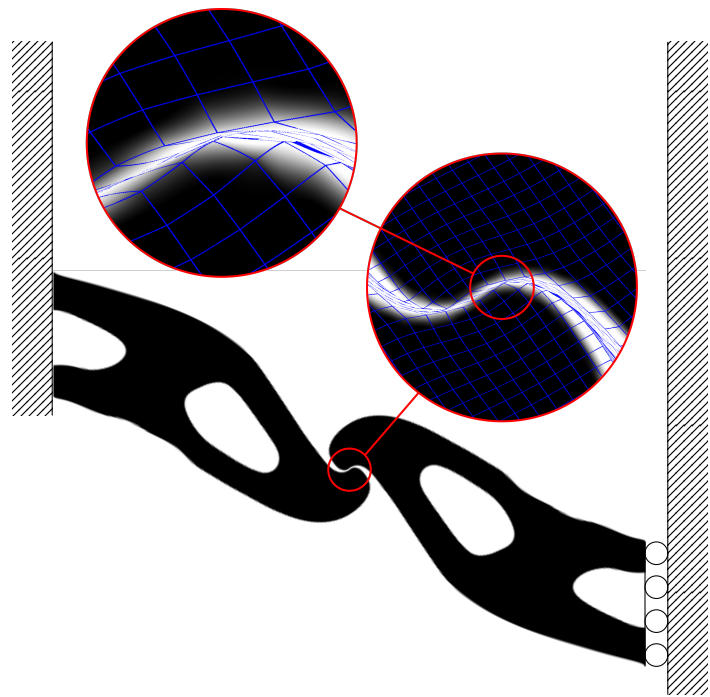

Numerical parameters for the problem are specified in \cref{tab:hook} and discretizations remain as specified in section \ref{sec:lift}.
\begin{table}[h]
    \centering
    \begin{tabular}{llrl}
    \hline
         Domain size &$L\times H$ & $56\times 32$& mm$^2$ \\
         Design region size & $t\times D$ & $0.45 H\times0.70 L$ & mm$^2$ \\
         Displacement purturbation & $\Delta u, \,\Delta u'$ & $0.01t$ & mm \\
         Mesh size & $N_x \times N_y$ & $240 \times 120$ & - \\
         Volume constraint & $V^*$ & 0.34 & -  \\
         Filter radius & $r$ & 1.60 & mm \\
         Threshold parameters & $(\beta_{ini},\beta_{max})$ & $(2,120)$ & -\\
         Objective weight & $k_q$ & $10^4$& - \\
         Tangent stiffness weight & $k_s$ & $10^4$& mm$^2$/N \\ \hline
    \end{tabular}
    \caption{Parameters for self-engaging hook problem.}
    \label{tab:hook}
\end{table}
Results for three different values of the enforced displacement $u_{0}$ are shown in \cref{fig:hooks_mult_def} with the meshed region indicated on \cref{subfig:a}.
The three designs all resemble two opposing hooks - to some extent. In the deformed configuration these hooks engage, this is shown in \cref{fig:hook_deformed_design} for the design from \cref{subfig:c}. A magnification in \cref{fig:hook_deformed_design} shows the void - which remains stable due to the regularization - being fully compressed to a state where it reflects contact.

\begin{figure}
    \centering
    \input{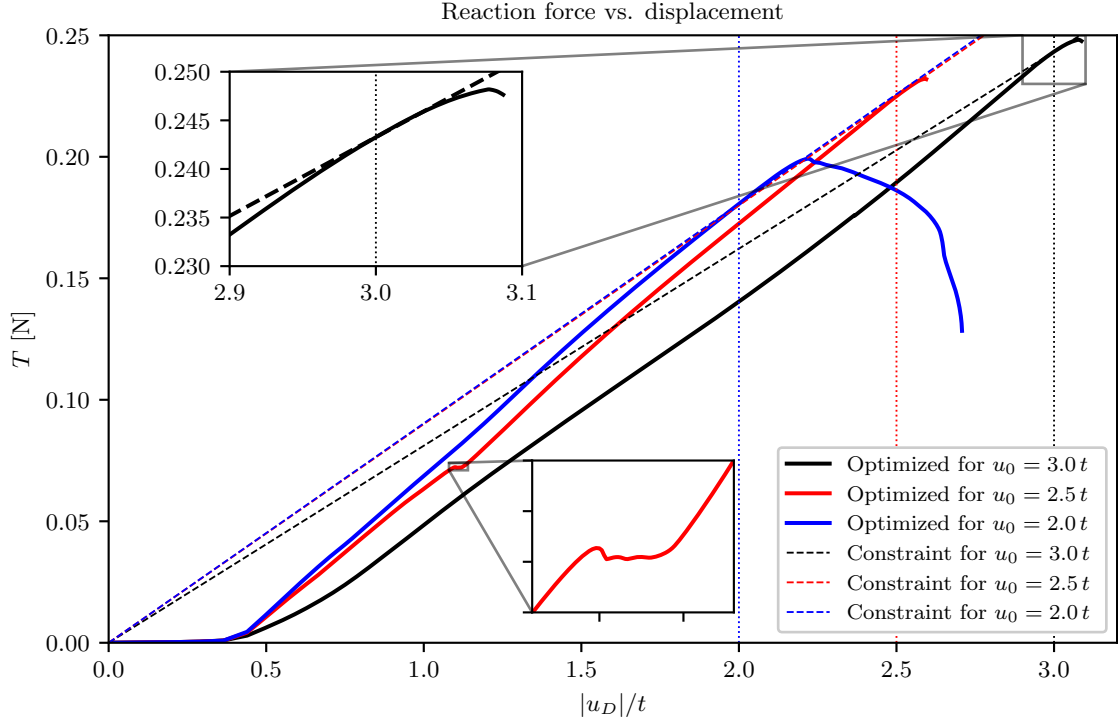}
    \caption{Force displacement curves for hook designs from \cref{fig:hook_deformed_design}.}
    \label{fig:hook_force_curves}
\end{figure}

Force-displacement curves as shown in \cref{fig:hook_force_curves} are obtained as the hooks are deformed through the prescribed loading history.
As may be expected, the design obtained for $u_{0}=2.0\,t$ has the highest force among the three designs at $u_D=2.0\,t$. Likewise for $u_{0}=2.5\,t$ at $u_D=2.5\,t$.
In the deformed configuration, as shown by the magnification for the design obtained for $u_{0}=3.0\,t$ in \cref{fig:hook_force_curves}, the tangent stiffness closely matches the average stiffness. Thus, showing that the first penalization term in the objective function eq. \eqref{eq:hook_obj} remains active until the converged state. 
A magnification of the $u_{0}=2.5\,t$ design in \cref{fig:hook_force_curves} shows a jump in the force-displacement curve. This jump is caused by the engagement of the non-overlapping tips of the hooks from \cref{subfig:b}.

\subsection{Bending mechanism}
\begin{figure}[b]
\centering

\definecolor{c0000ff}{RGB}{0,0,255}

\def \globalscale {4.000000}
\begin{tikzpicture}[y=0.80pt, x=0.80pt, yscale=-\globalscale, xscale=\globalscale, inner sep=0pt, outer sep=0pt]
  \path[draw=black,line width=0.064pt,rounded corners=0.0000cm] (2.8792,33.3831)
    rectangle (15.5182,36.1558);
\path[draw=black,line width=0.064pt,rounded corners=0.0000cm] (2.8792,38.4234)
    rectangle (15.5182,41.1961);
\path[draw=black,line width=0.064pt,rounded corners=0.0000cm] (8.4918,36.1558)
    rectangle (9.9056,38.4234);
\path[draw=black,line width=0.064pt,rounded corners=0.0000cm] (2.8792,28.3428)
    rectangle (15.5182,31.1154);
\path[draw=black,line width=0.064pt,rounded corners=0.0000cm] (8.4918,31.1155)
    rectangle (9.9056,33.3831);
\path[draw=black,line width=0.064pt,rounded corners=0.0000cm] (2.8792,23.3025)
    rectangle (15.5182,26.0751);
\path[draw=black,line width=0.064pt,rounded corners=0.0000cm] (8.4918,26.0751)
    rectangle (9.9056,28.3428);
\path[draw=black,line width=0.064pt,rounded corners=0.0000cm] (2.8792,18.2622)
    rectangle (15.5182,21.0348);
\path[draw=black,line width=0.064pt,rounded corners=0.0000cm] (8.4918,21.0348)
    rectangle (9.9056,23.3025);
\path[draw=black,line width=0.064pt,rounded corners=0.0000cm] (2.8792,13.2219)
    rectangle (15.5182,15.9945);
\path[draw=black,line width=0.064pt,rounded corners=0.0000cm] (8.4918,15.9945)
    rectangle (9.9056,18.2622);
\path[draw=black,line width=0.064pt,rounded corners=0.0000cm] (2.8792,8.1816)
    rectangle (15.5182,10.9542);
\path[draw=black,line width=0.064pt,rounded corners=0.0000cm] (8.4918,10.9542)
    rectangle (9.9056,13.2219);
\path[draw=black,line width=0.064pt,rounded corners=0.0000cm] (39.2814,38.4234)
    rectangle (51.9205,41.1961);
\path[rotate=-20.0,draw=black,line width=0.064pt,rounded corners=0.0000cm]
    (23.7908,46.6553) rectangle (36.4299,49.4280);
\path[draw=black,line width=0.064pt] (45.8691,35.9050) .. controls
    (46.0190,36.3069) and (46.1319,36.7216) .. (46.2064,37.1439) --
    (46.2064,37.1439) .. controls (46.2809,37.5663) and (46.3166,37.9946) ..
    (46.3132,38.4235);
\path[draw=black,line width=0.064pt] (44.5396,36.3890) .. controls
    (44.6614,36.7134) and (44.7529,37.0484) .. (44.8131,37.3896) --
    (44.8131,37.3896) .. controls (44.8733,37.7309) and (44.9018,38.0770) ..
    (44.8983,38.4235);
\path[rotate=-40.0,draw=black,line width=0.064pt,rounded corners=0.0000cm]
    (6.4145,49.0909) rectangle (19.0536,51.8636);
\path[draw=black,line width=0.064pt] (43.6377,31.0866)arc(319.538:330.000:7.042)
    -- (44.3784,32.1354)arc(330.000:340.462:7.042);
\path[draw=black,line width=0.064pt] (42.5538,31.9961)arc(319.422:330.000:5.627)
    -- (43.1531,32.8428)arc(330.000:340.578:5.627);
\path[draw=black,line width=0.064pt] (44.9163,33.3013) -- (43.5868,33.7853);
\path[rotate=-60.0,draw=black,line width=0.064pt,rounded corners=0.0000cm]
    (-10.7469,45.4366) rectangle (1.8922,48.2092);
\path[draw=black,line width=0.064pt] (39.8928,27.3220)arc(299.538:310.000:7.042)
    -- (40.9476,28.0542)arc(310.000:320.462:7.042);
\path[draw=black,line width=0.064pt] (39.1854,28.5473)arc(299.422:310.000:5.627)
    -- (40.0381,29.1380)arc(310.000:320.578:5.627);
\path[draw=black,line width=0.064pt] (41.8518,28.9658) -- (40.7680,29.8753);
\path[rotate=-80.0,draw=black,line width=0.064pt,rounded corners=0.0000cm]
    (-25.6235,36.1331) rectangle (-12.9845,38.9058);
\path[draw=black,line width=0.064pt] (35.0862,25.0652)arc(279.538:290.000:7.042)
    -- (36.3278,25.3925)arc(290.000:300.462:7.042);
\path[draw=black,line width=0.064pt] (34.8405,26.4586)arc(279.422:290.000:5.627)
    -- (35.8439,26.7220)arc(290.000:300.578:5.627);
\path[draw=black,line width=0.064pt] (37.4893,25.9399) -- (36.7819,27.1652);
\path[rotate=-100.0,draw=black,line width=0.064pt,rounded corners=0.0000cm]
    (-36.4210,22.3026) rectangle (-23.7819,25.0753);
\path[draw=black,line width=0.064pt] (29.7976,24.5885)arc(259.538:270.000:7.042)
    -- (31.0763,24.4714)arc(270.000:280.462:7.042);
\path[draw=black,line width=0.064pt] (30.0433,25.9818)arc(259.422:270.000:5.627)
    -- (31.0763,25.8862)arc(270.000:280.578:5.627);
\path[draw=black,line width=0.064pt] (32.3549,24.5885) -- (32.1093,25.9819);
\path[rotate=-100.0,draw=black,line width=0.064pt,rounded corners=0.0000cm]
    (-36.4210,22.3026) rectangle (-23.7819,25.0753);
\path[draw=black,line width=0.064pt] (29.7976,24.5885)arc(259.538:270.000:7.042)
    -- (31.0763,24.4714)arc(270.000:280.462:7.042);
\path[draw=black,line width=0.064pt] (30.0433,25.9818)arc(259.422:270.000:5.627)
    -- (31.0763,25.8862)arc(270.000:280.578:5.627);
\path[draw=black,line width=0.064pt] (32.3549,24.5885) -- (32.1093,25.9819);
\path[rotate=-120.0,draw=black,line width=0.064pt,rounded corners=0.0000cm]
    (-41.8370,5.6133) rectangle (-29.1979,8.3859);
\path[draw=black,line width=0.064pt] (24.6649,25.9493)arc(239.538:250.000:7.042)
    -- (25.8264,25.4020)arc(250.000:260.462:7.042);
\path[draw=black,line width=0.064pt] (25.3723,27.1746)arc(239.422:250.000:5.627)
    -- (26.3103,26.7315)arc(250.000:260.578:5.627);
\path[draw=black,line width=0.064pt] (27.0680,25.0747) -- (27.3137,26.4680);
\path[rotate=-120.0,draw=black,line width=0.064pt,rounded corners=0.0000cm]
    (-41.8370,5.6133) rectangle (-29.1979,8.3859);
\path[draw=black,line width=0.064pt] (24.6649,25.9493)arc(239.538:250.000:7.042)
    -- (25.8264,25.4020)arc(250.000:260.462:7.042);
\path[draw=black,line width=0.064pt] (25.3723,27.1746)arc(239.422:250.000:5.627)
    -- (26.3103,26.7315)arc(250.000:260.578:5.627);
\path[draw=black,line width=0.064pt] (27.0680,25.0747) -- (27.3137,26.4680);
\path[draw=c0000ff,line width=1pt] (1.8883,9.5679) -- (16.5091,9.5679) --
    (16.5091,14.6082) -- (1.8883,14.6082) -- cycle;
\path[draw=c0000ff,line width=0.1pt] (2,45) node[above right]
    (text8487) {\small Undeformed};
\path[draw=c0000ff,line width=0.1pt] (33,45) node[above right]
    (text12111) {\small Deformed};
\path[draw=c0000ff,line width=0.1pt] (20.5,11.5) node[above right]
    (text14029) {\small Repeated unit};
\path[draw=c0000ff,line width=0.1pt] (16.5091,11.6369) .. controls
    (18.2959,11.6785) and (18.1705,10.1628) .. (19.7695,10.1610);
\path[draw=c0000ff,line width=1pt] (51.5876,32.3364) .. controls
    (52.4633,34.7307) and (52.9114,37.2603) .. (52.9114,39.8097) --
    (38.3025,39.8097) .. controls (38.3025,38.9721) and (38.1552,38.1409) ..
    (37.8675,37.3543) -- cycle;
\end{tikzpicture}
\caption{Bending section with periodic structure.}
\label{fig:bending_section}
\end{figure}
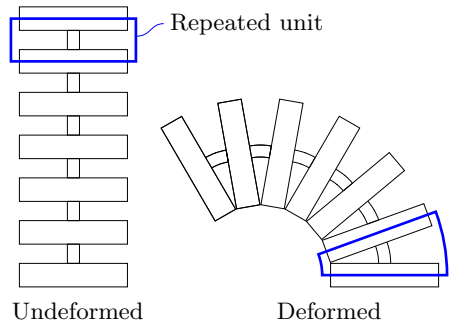

\begin{figure}
\centering

\begin{subfigure}[t]{0.48\textwidth}
\begin{tikzpicture}

\newcommand{\dx}{0.45} 
\newcommand{\LX}{5} 
\newcommand{\LY}{2} 
\newcommand{\dhatch}{0.35} 
\newcommand{\buff}{0.2} 

\draw [-, color=red,draw opacity=0] (-1.4,-0.8) -- (5.8,2.7);

\draw [fill=white, draw=black] (0,0) rectangle (\LX,\LY);
\draw [fill=gray!90, draw=black] (\dx,\dx) rectangle (\LX-\dx,\LY-\dx);
\draw [fill=black, draw=black] (\dx,0) rectangle (\LX-\dx,\dx);
\draw [fill=black, draw=black] (\dx,\LY-\dx) rectangle (\LX-\dx,\LY);
\fill [pattern=north east lines] (-\dhatch,-\dhatch) rectangle (\LX+\dhatch,0);
\draw [-] (-\dhatch,0) -- (\LX+\dhatch,0);

\node (L) at (\LX/2,-\dhatch-\buff) {\normalsize $L$};
\draw [|<-] (0,-\dhatch-\buff) -- (L);
\draw [->|] (L) -- (\LX,-\dhatch-\buff);
\node (H) at (0-2*\buff, \LY/2) {\normalsize $H$};
\draw [|<-] (0-2*\buff, 0) -- (H);
\draw [->|] (H) -- (0-2*\buff,\LY);

\node [color=white] (omega_s) at (2*\dx, \LY-0.6*\dx) {\normalsize $\Omega_s$};
\node (omega_d) at (2*\dx, \LY-1.7*\dx) {\normalsize $\Omega_d$};
\node (omega_v) at (\dx/2, \LY-1.7*\dx) {\normalsize $\Omega_\mathrm{v}$};

\node (gamma_u) [color=blue] at (\LX+2*\buff, \LY+\buff) {\normalsize $\Gamma_{\scriptscriptstyle top}$};
\draw [color=blue] (gamma_u.west) to [out=180, in=90] (\LX-1.5*\buff,\LY);
\draw [line width=0.5mm,color=blue] (0,\LY) -- (\LX,\LY);
\node (gamma_u2) [color=blue] at (\LX+2*\buff, \buff) {\normalsize $\Gamma_{\scriptscriptstyle bot}$};
\draw [color=blue] (gamma_u2.west) to [out=180, in=90] (\LX-1.5*\buff,0);
\draw [line width=0.5mm,color=blue] (0,0) -- (\LX,0);

\node (D) at (\dx/2,2*\buff) {\normalsize $t_2$};
\draw [|<->|] (0, \buff) -- (\dx,\buff);
\node [color=white] (t) at (\dx+3*\buff, \dx/2) {\normalsize $t_1$};
\draw [|<->|, color=white] (\dx+2*\buff, 0) -- (\dx+2*\buff,\dx);

\draw [<->] (-1.25,0.35) [anchor=south] node {\scriptsize $y$}  -- (-1.25,-0.15) -- (-0.75,-0.15)  [anchor=west] node {\scriptsize $x$};

\end{tikzpicture}
\caption{Undeformed configuration.}
\label{fig:bend_domain}
\end{subfigure}
\hfill
\begin{subfigure}[t]{0.48\textwidth}
\begin{tikzpicture}

\newcommand{\dx}{0.45} 
\newcommand{\LX}{5} 
\newcommand{\LY}{2} 
\newcommand{\dhatch}{0.35} 
\newcommand{\buff}{0.2} 
\newcommand{\alf}{20} 
\newcommand{\pinum}{3.14159265359} 
\newcommand{\radi}{\LY/\alf*180/\pinum} 

\draw [-, color=red,draw opacity=0] (-1.4,-0.8) -- (5.8,2.7);

\draw [fill=black, draw=black] (\dx,0) rectangle (\LX-\dx,\dx);
\draw [fill=black, draw=black,rotate around={\alf:(-\radi+\LX/2,0)}] (\dx,-\dx) rectangle (\LX-\dx,0);

\fill [pattern=north east lines] (-\dhatch,-\dhatch) rectangle (\LX+\dhatch,0);
\draw [-] (-\dhatch,0) -- (\LX+\dhatch,0);

\coordinate (k1) at (\dx,\dx);
\coordinate (k2) at ({cos(\alf)*(\dx+\radi-\LX/2)+sin(\alf)*\dx -\radi+\LX/2}, {sin(\alf)*(\dx+\radi-\LX/2)-cos(\alf)*\dx});
\coordinate (k3) at ({cos(\alf)*(\LX-\dx+\radi-\LX/2)+sin(\alf)*\dx -\radi+\LX/2}, {sin(\alf)*(\LX-\dx+\radi-\LX/2)-cos(\alf)*\dx});
\coordinate (k4) at (\LX-\dx,\dx);
\draw [color=black,fill=gray!90] (k1) to [out=90, in={\alf-90}] (k2) -- (k3) to [out={\alf-90}, in=90] (k4) -- cycle;

\node [color=blue] (omega_d) at (\LX+\buff, \LY/2+\buff) {\normalsize $\delta \Omega$};
\coordinate (c2) at (-\radi+\LX/2,0);
\draw[blue,line width=0.5mm] ($(c2) + (0:\radi-\LX/2)$) arc (0:\alf:\radi-\LX/2) -- ($(c2) + (\alf:\radi+\LX/2)$) arc (\alf:0:\radi+\LX/2)  -- cycle;

\node [color=white] (H) at (\LX/2+\buff, \LY/2+\buff) {\normalsize $H$};
\draw [|<->|,color=white,line width=0.3mm] ($(c2) + (\alf:\radi)$) arc (\alf:0:\radi);

\coordinate (k5) at ({cos(\alf)*(\LX+\radi-\LX/2) -\radi+\LX/2}, {sin(\alf)*(\LX+\radi-\LX/2)});
\coordinate (k6) at ({cos(\alf)*(\LX+\radi-\LX/2) -\radi+\LX/7}, {sin(\alf)*(\LX+\radi-\LX/2)});
\draw [-] (k5) -- (k6);
\draw [->,line width=0.3mm] ($(k5)- (0:\LX*0.3)$) arc (0:\alf:-\LX*0.3);
\node [] at ($(k5)+(-\LX*0.3+\buff,-\buff)$) {\normalsize $\alpha$};

\coordinate (p1) at ({cos(\alf)*(\radi)-sin(\alf)*(\buff+2*\dx) -\radi+\LX/2}, {sin(\alf)*(\radi)+cos(\alf)*(\buff+2*\dx)});
\coordinate (p2) at ({cos(\alf)*(\radi)-sin(\alf)*2*\buff -\radi+\LX/2}, {sin(\alf)*(\radi)+cos(\alf)*2*\buff});
\draw[<-] (p1) -- (p2);
\node [] at ($(p1)+(-\buff,-\buff)$) {\normalsize $n$};

\coordinate (p3) at ({cos(\alf)*(-\dx+\radi)-sin(\alf)*\buff -\radi+\LX/2}, {sin(\alf)*(-\dx+\radi)+cos(\alf)*\buff});
\coordinate (p4) at ({cos(\alf)*( \dx+\radi)-sin(\alf)*\buff -\radi+\LX/2}, {sin(\alf)*(\dx+\radi)+cos(\alf)*\buff});
\draw[->] (p3) -- (p4);
\node [] at ($(p4)+(-\buff,+\buff)$) {\normalsize $\hat{n}$};

\draw [<->] (-1.25,0.35) [anchor=south] node {\scriptsize $y$}  -- (-1.25,-0.15) -- (-0.75,-0.15)  [anchor=west] node {\scriptsize $x$};

\end{tikzpicture}
\caption{Deformed configuration.}
\label{fig:bend_domain_deformed}
\end{subfigure}
\caption{Domain of bending section with prescribed solid (black) and prescribed void (white). Initial guess $\rho=0.4$ in $\Omega_d$.}
\end{figure}
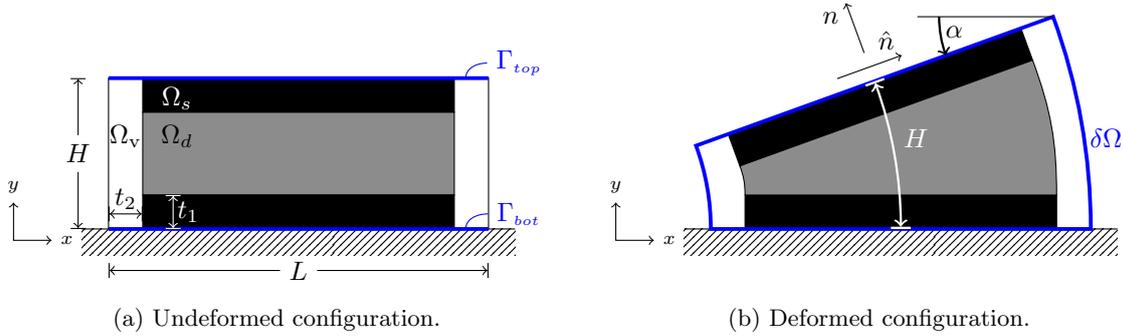

This section applies the method outlined in section \ref{sec:modelling_approach} to a bending problem which is inspired by the bending section found in endoscopes.
Endoscope bending sections, which comprise the controllable part at the distal end of an endoscope, typically comprise periodic structures. Thus, when designing such a component it is reasonable to focus on a single unit of the periodic structure. An illustration of a bending section and the repeated unit is shown in \cref{fig:bending_section}.
Let the design domain of the repeated unit from \cref{fig:bending_section} be defined as shown in \cref{fig:bend_domain} where the boundaries $\Gamma_{top}$ and $\Gamma_{bot}$ are the periodic boundaries.

During use, the endoscope bending section is deformed such that the segment in \cref{fig:bend_domain} forms a section of an annulus as shown in \cref{fig:bend_domain_deformed}, where the arc length at the center of the annular section has the length $H$. Deformations are prescribed on the outer boundary $\delta \Omega$ by specifying the angle $\alpha$ of the annular section. There are no strains along the boundaries $\Gamma_{top}$ and $\Gamma_{bot}$, i.e. they rotate as rigid bodies.

From the perspective of the endoscopist it may be desirable to have a bending section which can bend very easily at any angle below a certain threshold. This way the endoscopist may easily manoeuvre the bending section without it surpassing a certain curvature. Surpassing the threshold angle may cause damage to tools being passed through the endoscope. Further, it may be desirable for the obtained design to be subject to a state of pure bending. This may reduce interactions with other components in the endoscope bending section. Additionally, the control of the endoscope is simpler for a state of pure bending with no bending-stretching coupling.

Reaction forces and bending moment on the respective boundaries are calculated through the reaction traction $q$ by
\begin{equation}
    \begin{split}
        Q_i &= \int_{\Gamma_i} q \, d \Gamma_i \hspace{5mm} \text{and}\\
        M_i &= \int_{\Gamma_i} ((q-\Bar{q}_i) \cdot n) (p \cdot \hat{n}) \, d \Gamma_i,
    \end{split}
\end{equation}
where $n$ and $\hat{n}$ respectively are the unit normal and tangent vectors shown in \cref{fig:bend_domain_deformed}, $p$ is a point on the boundary $\Gamma_i$, and $\Bar{q}_i$ is the average surface traction.

Based on two angular control points $\alpha_0$ and $\alpha_1$ where $\alpha_0<\alpha_1$, the desired properties may be cast into an objective function which approximates the desired outcome.
Here we propose an objective function defined by
\begin{equation}
\begin{split}
    C(\Tilde{\rho}_e,u,q)  = k_q \bigg( - M_{top}|_{\alpha_1} - M_{bot}|_{\alpha_1} + k_m \left(M^2_{top}|_{\alpha_0} + \,M^2_{bot}|_{\alpha_0}\right)  + k_s \, \left(||Q_{top}||^2|_{\alpha_0} +  ||Q_{bot}||^2|_{\alpha_0} \right) \bigg).
\end{split}
\label{eq:bend_obj}
\end{equation}
The objective function in eq. (\ref{eq:bend_obj}) minimizes any positive or negative values of the bending moment at an angle $\alpha_0$, this ensures low bending resistance at low angles. Simultaneously, it maximizes the bending moment at an angle $\alpha_1>\alpha_0$, which inhibits bending at large angles. Inclusion of the terms containing the reaction forces ensures that reaction forces are minimized at the relevant boundaries. $k_s$ is a weighting factor for the reaction forces on the boundaries $\Gamma_{top}$ and $\Gamma_{bot}$. Likewise, $k_m$ is a weighting factor for the bending moment at angle $\alpha_0$. Higher values of $k_m$ increasingly penalize bending moments at lower angles.

\begin{figure}
    \begin{subfigure}{\textwidth}
    \setlength{\unitlength}{0.1\textwidth}
    \begin{picture}(10,3.9)
       \put(0.8,0){\includegraphics[width=0.92\textwidth]{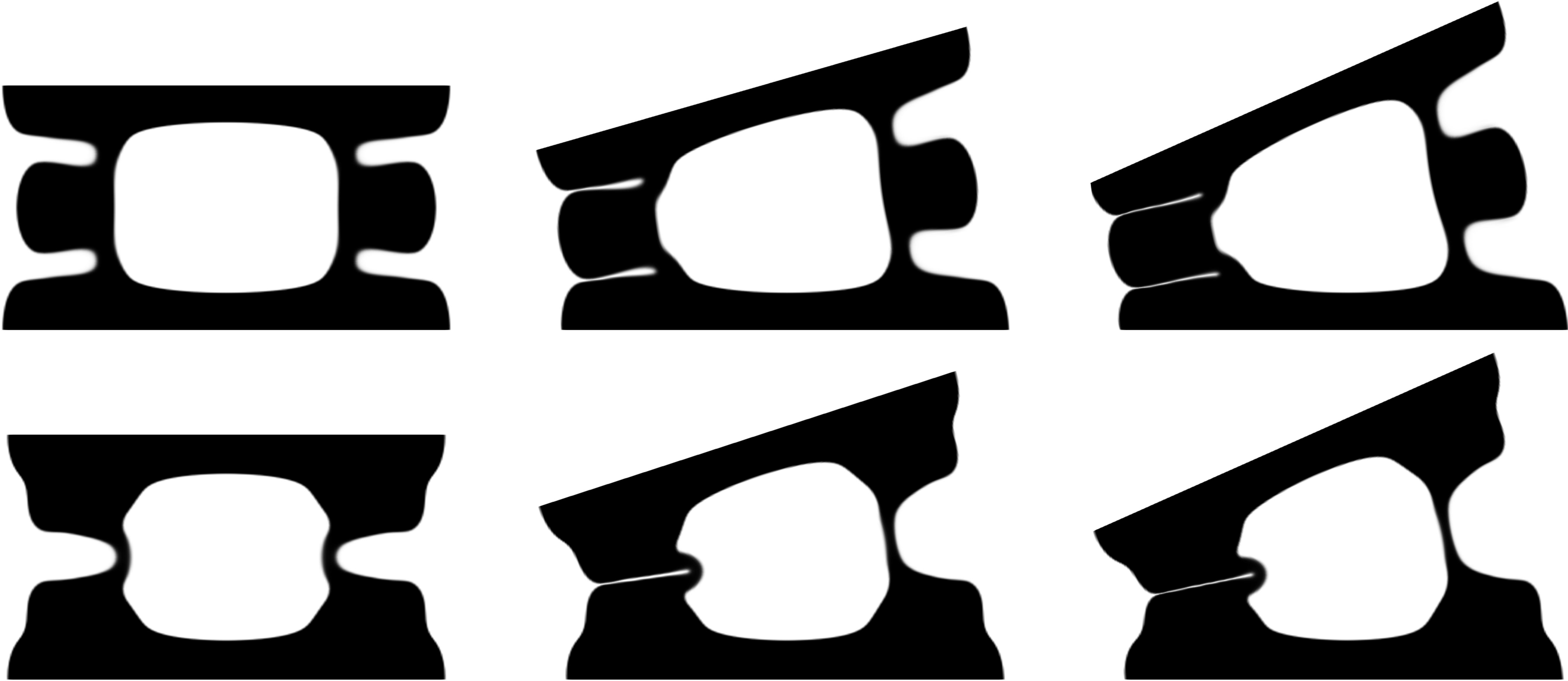}}
       \put(0.,2.8){\parbox{0.05\textwidth}{$\alpha_0=$ $14^\circ$}}
       \put(0.,0.6){\parbox{0.05\textwidth}{$\alpha_0=$ $18^\circ$}}
       \put(1.9,-.3){$\alpha=0$}
       \put(5.0,-.3){$\alpha=\alpha_0$}
       \put(8.3,-.3){$\alpha=\alpha_1$}
    \end{picture}
    \end{subfigure}
    \vspace{2mm}
    \caption{Bend design at angles $\alpha=0$, $\alpha=\alpha_0$, and $\alpha=\alpha_1$ for $\alpha_0=14^\circ$ and $\alpha_0=18^\circ$.}
    \label{fig:bend_result_deformed}
\end{figure}
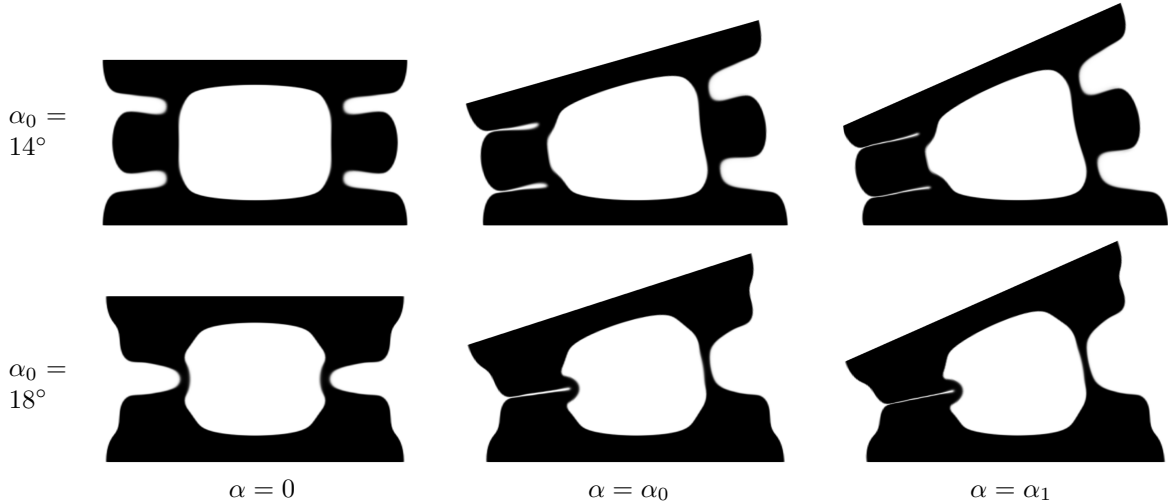

Results for two different values of the control point $\alpha_0$ and parameters listed in Table \ref{tab:bend} are shown in \cref{fig:bend_result_deformed}. 
\begin{table}[hb]
    \centering
    \begin{tabular}{llrl}
    \hline
         Domain size &$L\times H$ & $9.0\times3.5$& mm$^2$ \\
         Max angle & $\alpha_1$ & $24$ & deg \\
         Mesh size & $N_x\times N_y$ & $200\times 100$ & - \\
         Volume constraint & $V^*$ & 0.55 & - \\
         Filter radius & $r$ & 0.60 & mm \\
         Fixed region widths & $(t_1,t_2)$ & $(r,2r)$ & mm \\
         Threshold parameters & $(\beta_{ini},\beta_{max})$ & $(2,240)$ & -\\
         Void regularisation weight & $k_{r}$ & $3\cdot10^{-6}$ & - \\
         Objective weight & $k_q$ & $10^{4}$& - \\
         Objective moment weight & $k_m$ & $10$& 1/Nmm \\
         Objective force weight & $k_s$ & $10^{-2}$& mm/N \\ \hline
    \end{tabular}
    \caption{Parameters for bend problem.}
    \label{tab:bend}
\end{table}
All discretizations remain as specified in section \ref{sec:lift}. Note that the regularization scaling term $k_r$ has been increased compared to the previous two examples, this is in order to further stabilize the problem.
The deformed configurations show how the optimized designs establish contact around an angle $\alpha_0$.
A notable difference in the two designs is that the $\alpha_0=14^\circ$ design includes two void regions for contact, whereas the $\alpha_0=18^\circ$ design only includes one.
In contrast to the example in section \ref{sec:hook} the contact in \cref{fig:bend_result_deformed} does not form through a forced void region, instead the void which serves as the contact medium is a direct result of the optimization. This could also be seen in the results from \cref{sec:lift}. It thus shows that contacting surfaces can be established in design regions without any pre-specification of a void to serve as the contact medium. 

\begin{figure}
\begin{subfigure}[t]{0.48\textwidth}
    \input{hook_moment_curves_case_NX200_NY100_bend_v18_c13_v2.pgf}
\end{subfigure}
\hfill
\begin{subfigure}[t]{0.48\textwidth}
    \input{hook_force_curves_case_NX200_NY100_bend_v18_c13_v2.pgf}
\end{subfigure}
\caption{Resulting bending moment and force for bending problem.}
\label{fig:bend_results_plots}
\end{figure}
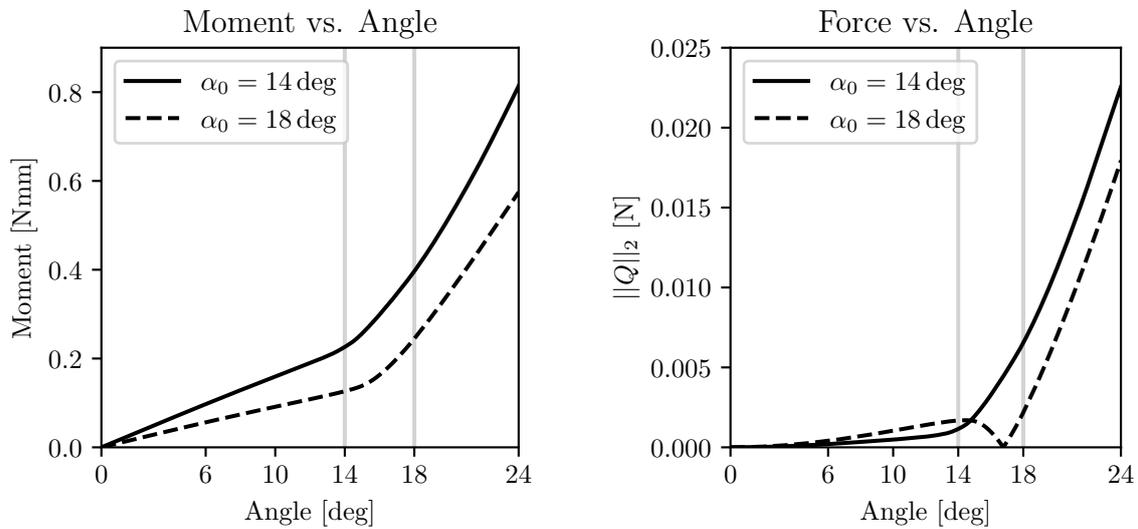

\cref{fig:bend_results_plots} shows the bending moment and the norm of the force as a function of the bending angle for the resulting designs. 
The slope of the moment-curve, which represents the bending stiffness, significantly increases around the point $\alpha_0$. Both moment-curves show a soft kink, indicating that contact is established, at an angle of approximately $14^\circ$ and $16^\circ$ respectively. From the force-curve it can be seen that the force is kept at a relatively low level, as intended by the objective function, until contact is established.

\section{Conclusion}
\label{sec:conclusion}

The present work has adapted the third medium contact method for topology optimization from \cite{Bluhm2021InternalOptimization} to standard topology optimization methods and applied it to advanced kinematic examples. This includes PDE filtering through an inhomogeneous Helmholtz equation, threshold projection, SIMP, as well as the application of MMA. 
To avoid non-physical exploitation of void regions, which serve as the third medium, it has proven useful to penalize small features by evaluating the volume constraint on a dilated design - as it is known from the robust formulation in topology optimization \cite{Wang2011OnOptimization}. This penalization also adds an indirect control of the length scale of the obtained structures. In addition, a solution algorithm which includes $\beta$-continuation as proposed in \cite{Wang2011OnOptimization} is suggested.

The application of the third medium contact has been extended to more complex applications, which e.q. require additional control of the tangential stiffness of the structure.
Three examples have been used to illustrate possible applications as well as how the approach works for problems subject to large deformations: 1) a lifting mechanism maximizing vertical reaction force upon a horizontal displacement, 2) a design of a coupling between two domains separated by a void region, resulting in two hooks, and 3) the design of a bending mechanism inspired by the bending section of an endoscope.
Results show that the optimized structures can exploit prescribed void regions to establish contact, as well as establish new void regions within the design domain to be used for contact. This inclusion of contact extends the possible solution space in topology optimization and should allow to program even more complex force displacement responses in compliant mechanism design than provided in \cite{Li2022DigitalResponses}.

\bibliographystyle{abbrv}

\bibliography{main.bib}

\end{document}

%% file: box_force_curves_pyramid_v2.pgf
\begingroup%
\makeatletter%
\begin{pgfpicture}%
\pgfpathrectangle{\pgfpointorigin}{\pgfqpoint{2.730000in}{3.730000in}}%
\pgfusepath{use as bounding box, clip}%
\begin{pgfscope}%
\pgfsetbuttcap%
\pgfsetmiterjoin%
\definecolor{currentfill}{rgb}{1.000000,1.000000,1.000000}%
\pgfsetfillcolor{currentfill}%
\pgfsetlinewidth{0.000000pt}%
\definecolor{currentstroke}{rgb}{1.000000,1.000000,1.000000}%
\pgfsetstrokecolor{currentstroke}%
\pgfsetdash{}{0pt}%
\pgfpathmoveto{\pgfqpoint{0.000000in}{0.000000in}}%
\pgfpathlineto{\pgfqpoint{2.730000in}{0.000000in}}%
\pgfpathlineto{\pgfqpoint{2.730000in}{3.730000in}}%
\pgfpathlineto{\pgfqpoint{0.000000in}{3.730000in}}%
\pgfpathlineto{\pgfqpoint{0.000000in}{0.000000in}}%
\pgfpathclose%
\pgfusepath{fill}%
\end{pgfscope}%
\begin{pgfscope}%
\pgfsetbuttcap%
\pgfsetmiterjoin%
\definecolor{currentfill}{rgb}{1.000000,1.000000,1.000000}%
\pgfsetfillcolor{currentfill}%
\pgfsetlinewidth{0.000000pt}%
\definecolor{currentstroke}{rgb}{0.000000,0.000000,0.000000}%
\pgfsetstrokecolor{currentstroke}%
\pgfsetstrokeopacity{0.000000}%
\pgfsetdash{}{0pt}%
\pgfpathmoveto{\pgfqpoint{0.506171in}{0.388889in}}%
\pgfpathlineto{\pgfqpoint{2.730000in}{0.388889in}}%
\pgfpathlineto{\pgfqpoint{2.730000in}{3.559861in}}%
\pgfpathlineto{\pgfqpoint{0.506171in}{3.559861in}}%
\pgfpathlineto{\pgfqpoint{0.506171in}{0.388889in}}%
\pgfpathclose%
\pgfusepath{fill}%
\end{pgfscope}%
\begin{pgfscope}%
\pgfsetbuttcap%
\pgfsetroundjoin%
\definecolor{currentfill}{rgb}{0.000000,0.000000,0.000000}%
\pgfsetfillcolor{currentfill}%
\pgfsetlinewidth{0.803000pt}%
\definecolor{currentstroke}{rgb}{0.000000,0.000000,0.000000}%
\pgfsetstrokecolor{currentstroke}%
\pgfsetdash{}{0pt}%
\pgfsys@defobject{currentmarker}{\pgfqpoint{0.000000in}{-0.048611in}}{\pgfqpoint{0.000000in}{0.000000in}}{%
\pgfpathmoveto{\pgfqpoint{0.000000in}{0.000000in}}%
\pgfpathlineto{\pgfqpoint{0.000000in}{-0.048611in}}%
\pgfusepath{stroke,fill}%
}%
\begin{pgfscope}%
\pgfsys@transformshift{0.506171in}{0.388889in}%
\pgfsys@useobject{currentmarker}{}%
\end{pgfscope}%
\end{pgfscope}%
\begin{pgfscope}%
\definecolor{textcolor}{rgb}{0.000000,0.000000,0.000000}%
\pgfsetstrokecolor{textcolor}%
\pgfsetfillcolor{textcolor}%
\pgftext[x=0.506171in,y=0.291667in,,top]{\color{textcolor}\rmfamily\fontsize{9.000000}{10.800000}\selectfont 0.0}%
\end{pgfscope}%
\begin{pgfscope}%
\pgfsetbuttcap%
\pgfsetroundjoin%
\definecolor{currentfill}{rgb}{0.000000,0.000000,0.000000}%
\pgfsetfillcolor{currentfill}%
\pgfsetlinewidth{0.803000pt}%
\definecolor{currentstroke}{rgb}{0.000000,0.000000,0.000000}%
\pgfsetstrokecolor{currentstroke}%
\pgfsetdash{}{0pt}%
\pgfsys@defobject{currentmarker}{\pgfqpoint{0.000000in}{-0.048611in}}{\pgfqpoint{0.000000in}{0.000000in}}{%
\pgfpathmoveto{\pgfqpoint{0.000000in}{0.000000in}}%
\pgfpathlineto{\pgfqpoint{0.000000in}{-0.048611in}}%
\pgfusepath{stroke,fill}%
}%
\begin{pgfscope}%
\pgfsys@transformshift{0.910504in}{0.388889in}%
\pgfsys@useobject{currentmarker}{}%
\end{pgfscope}%
\end{pgfscope}%
\begin{pgfscope}%
\definecolor{textcolor}{rgb}{0.000000,0.000000,0.000000}%
\pgfsetstrokecolor{textcolor}%
\pgfsetfillcolor{textcolor}%
\pgftext[x=0.910504in,y=0.291667in,,top]{\color{textcolor}\rmfamily\fontsize{9.000000}{10.800000}\selectfont 0.1}%
\end{pgfscope}%
\begin{pgfscope}%
\pgfsetbuttcap%
\pgfsetroundjoin%
\definecolor{currentfill}{rgb}{0.000000,0.000000,0.000000}%
\pgfsetfillcolor{currentfill}%
\pgfsetlinewidth{0.803000pt}%
\definecolor{currentstroke}{rgb}{0.000000,0.000000,0.000000}%
\pgfsetstrokecolor{currentstroke}%
\pgfsetdash{}{0pt}%
\pgfsys@defobject{currentmarker}{\pgfqpoint{0.000000in}{-0.048611in}}{\pgfqpoint{0.000000in}{0.000000in}}{%
\pgfpathmoveto{\pgfqpoint{0.000000in}{0.000000in}}%
\pgfpathlineto{\pgfqpoint{0.000000in}{-0.048611in}}%
\pgfusepath{stroke,fill}%
}%
\begin{pgfscope}%
\pgfsys@transformshift{1.314836in}{0.388889in}%
\pgfsys@useobject{currentmarker}{}%
\end{pgfscope}%
\end{pgfscope}%
\begin{pgfscope}%
\definecolor{textcolor}{rgb}{0.000000,0.000000,0.000000}%
\pgfsetstrokecolor{textcolor}%
\pgfsetfillcolor{textcolor}%
\pgftext[x=1.314836in,y=0.291667in,,top]{\color{textcolor}\rmfamily\fontsize{9.000000}{10.800000}\selectfont 0.2}%
\end{pgfscope}%
\begin{pgfscope}%
\pgfsetbuttcap%
\pgfsetroundjoin%
\definecolor{currentfill}{rgb}{0.000000,0.000000,0.000000}%
\pgfsetfillcolor{currentfill}%
\pgfsetlinewidth{0.803000pt}%
\definecolor{currentstroke}{rgb}{0.000000,0.000000,0.000000}%
\pgfsetstrokecolor{currentstroke}%
\pgfsetdash{}{0pt}%
\pgfsys@defobject{currentmarker}{\pgfqpoint{0.000000in}{-0.048611in}}{\pgfqpoint{0.000000in}{0.000000in}}{%
\pgfpathmoveto{\pgfqpoint{0.000000in}{0.000000in}}%
\pgfpathlineto{\pgfqpoint{0.000000in}{-0.048611in}}%
\pgfusepath{stroke,fill}%
}%
\begin{pgfscope}%
\pgfsys@transformshift{1.719169in}{0.388889in}%
\pgfsys@useobject{currentmarker}{}%
\end{pgfscope}%
\end{pgfscope}%
\begin{pgfscope}%
\definecolor{textcolor}{rgb}{0.000000,0.000000,0.000000}%
\pgfsetstrokecolor{textcolor}%
\pgfsetfillcolor{textcolor}%
\pgftext[x=1.719169in,y=0.291667in,,top]{\color{textcolor}\rmfamily\fontsize{9.000000}{10.800000}\selectfont 0.3}%
\end{pgfscope}%
\begin{pgfscope}%
\pgfsetbuttcap%
\pgfsetroundjoin%
\definecolor{currentfill}{rgb}{0.000000,0.000000,0.000000}%
\pgfsetfillcolor{currentfill}%
\pgfsetlinewidth{0.803000pt}%
\definecolor{currentstroke}{rgb}{0.000000,0.000000,0.000000}%
\pgfsetstrokecolor{currentstroke}%
\pgfsetdash{}{0pt}%
\pgfsys@defobject{currentmarker}{\pgfqpoint{0.000000in}{-0.048611in}}{\pgfqpoint{0.000000in}{0.000000in}}{%
\pgfpathmoveto{\pgfqpoint{0.000000in}{0.000000in}}%
\pgfpathlineto{\pgfqpoint{0.000000in}{-0.048611in}}%
\pgfusepath{stroke,fill}%
}%
\begin{pgfscope}%
\pgfsys@transformshift{2.123501in}{0.388889in}%
\pgfsys@useobject{currentmarker}{}%
\end{pgfscope}%
\end{pgfscope}%
\begin{pgfscope}%
\definecolor{textcolor}{rgb}{0.000000,0.000000,0.000000}%
\pgfsetstrokecolor{textcolor}%
\pgfsetfillcolor{textcolor}%
\pgftext[x=2.123501in,y=0.291667in,,top]{\color{textcolor}\rmfamily\fontsize{9.000000}{10.800000}\selectfont \(\displaystyle u_{max}\)}%
\end{pgfscope}%
\begin{pgfscope}%
\pgfsetbuttcap%
\pgfsetroundjoin%
\definecolor{currentfill}{rgb}{0.000000,0.000000,0.000000}%
\pgfsetfillcolor{currentfill}%
\pgfsetlinewidth{0.803000pt}%
\definecolor{currentstroke}{rgb}{0.000000,0.000000,0.000000}%
\pgfsetstrokecolor{currentstroke}%
\pgfsetdash{}{0pt}%
\pgfsys@defobject{currentmarker}{\pgfqpoint{0.000000in}{-0.048611in}}{\pgfqpoint{0.000000in}{0.000000in}}{%
\pgfpathmoveto{\pgfqpoint{0.000000in}{0.000000in}}%
\pgfpathlineto{\pgfqpoint{0.000000in}{-0.048611in}}%
\pgfusepath{stroke,fill}%
}%
\begin{pgfscope}%
\pgfsys@transformshift{2.527834in}{0.388889in}%
\pgfsys@useobject{currentmarker}{}%
\end{pgfscope}%
\end{pgfscope}%
\begin{pgfscope}%
\definecolor{textcolor}{rgb}{0.000000,0.000000,0.000000}%
\pgfsetstrokecolor{textcolor}%
\pgfsetfillcolor{textcolor}%
\pgftext[x=2.527834in,y=0.291667in,,top]{\color{textcolor}\rmfamily\fontsize{9.000000}{10.800000}\selectfont 0.5}%
\end{pgfscope}%
\begin{pgfscope}%
\definecolor{textcolor}{rgb}{0.000000,0.000000,0.000000}%
\pgfsetstrokecolor{textcolor}%
\pgfsetfillcolor{textcolor}%
\pgftext[x=1.618086in,y=0.125000in,,top]{\color{textcolor}\rmfamily\fontsize{9.000000}{10.800000}\selectfont \(\displaystyle u_D\,\)[mm]}%
\end{pgfscope}%
\begin{pgfscope}%
\pgfsetbuttcap%
\pgfsetroundjoin%
\definecolor{currentfill}{rgb}{0.000000,0.000000,0.000000}%
\pgfsetfillcolor{currentfill}%
\pgfsetlinewidth{0.803000pt}%
\definecolor{currentstroke}{rgb}{0.000000,0.000000,0.000000}%
\pgfsetstrokecolor{currentstroke}%
\pgfsetdash{}{0pt}%
\pgfsys@defobject{currentmarker}{\pgfqpoint{-0.048611in}{0.000000in}}{\pgfqpoint{-0.000000in}{0.000000in}}{%
\pgfpathmoveto{\pgfqpoint{-0.000000in}{0.000000in}}%
\pgfpathlineto{\pgfqpoint{-0.048611in}{0.000000in}}%
\pgfusepath{stroke,fill}%
}%
\begin{pgfscope}%
\pgfsys@transformshift{0.506171in}{0.388889in}%
\pgfsys@useobject{currentmarker}{}%
\end{pgfscope}%
\end{pgfscope}%
\begin{pgfscope}%
\definecolor{textcolor}{rgb}{0.000000,0.000000,0.000000}%
\pgfsetstrokecolor{textcolor}%
\pgfsetfillcolor{textcolor}%
\pgftext[x=0.180556in, y=0.345486in, left, base]{\color{textcolor}\rmfamily\fontsize{9.000000}{10.800000}\selectfont \(\displaystyle {0.00}\)}%
\end{pgfscope}%
\begin{pgfscope}%
\pgfsetbuttcap%
\pgfsetroundjoin%
\definecolor{currentfill}{rgb}{0.000000,0.000000,0.000000}%
\pgfsetfillcolor{currentfill}%
\pgfsetlinewidth{0.803000pt}%
\definecolor{currentstroke}{rgb}{0.000000,0.000000,0.000000}%
\pgfsetstrokecolor{currentstroke}%
\pgfsetdash{}{0pt}%
\pgfsys@defobject{currentmarker}{\pgfqpoint{-0.048611in}{0.000000in}}{\pgfqpoint{-0.000000in}{0.000000in}}{%
\pgfpathmoveto{\pgfqpoint{-0.000000in}{0.000000in}}%
\pgfpathlineto{\pgfqpoint{-0.048611in}{0.000000in}}%
\pgfusepath{stroke,fill}%
}%
\begin{pgfscope}%
\pgfsys@transformshift{0.506171in}{0.811685in}%
\pgfsys@useobject{currentmarker}{}%
\end{pgfscope}%
\end{pgfscope}%
\begin{pgfscope}%
\definecolor{textcolor}{rgb}{0.000000,0.000000,0.000000}%
\pgfsetstrokecolor{textcolor}%
\pgfsetfillcolor{textcolor}%
\pgftext[x=0.180556in, y=0.768282in, left, base]{\color{textcolor}\rmfamily\fontsize{9.000000}{10.800000}\selectfont \(\displaystyle {0.02}\)}%
\end{pgfscope}%
\begin{pgfscope}%
\pgfsetbuttcap%
\pgfsetroundjoin%
\definecolor{currentfill}{rgb}{0.000000,0.000000,0.000000}%
\pgfsetfillcolor{currentfill}%
\pgfsetlinewidth{0.803000pt}%
\definecolor{currentstroke}{rgb}{0.000000,0.000000,0.000000}%
\pgfsetstrokecolor{currentstroke}%
\pgfsetdash{}{0pt}%
\pgfsys@defobject{currentmarker}{\pgfqpoint{-0.048611in}{0.000000in}}{\pgfqpoint{-0.000000in}{0.000000in}}{%
\pgfpathmoveto{\pgfqpoint{-0.000000in}{0.000000in}}%
\pgfpathlineto{\pgfqpoint{-0.048611in}{0.000000in}}%
\pgfusepath{stroke,fill}%
}%
\begin{pgfscope}%
\pgfsys@transformshift{0.506171in}{1.234481in}%
\pgfsys@useobject{currentmarker}{}%
\end{pgfscope}%
\end{pgfscope}%
\begin{pgfscope}%
\definecolor{textcolor}{rgb}{0.000000,0.000000,0.000000}%
\pgfsetstrokecolor{textcolor}%
\pgfsetfillcolor{textcolor}%
\pgftext[x=0.180556in, y=1.191079in, left, base]{\color{textcolor}\rmfamily\fontsize{9.000000}{10.800000}\selectfont \(\displaystyle {0.04}\)}%
\end{pgfscope}%
\begin{pgfscope}%
\pgfsetbuttcap%
\pgfsetroundjoin%
\definecolor{currentfill}{rgb}{0.000000,0.000000,0.000000}%
\pgfsetfillcolor{currentfill}%
\pgfsetlinewidth{0.803000pt}%
\definecolor{currentstroke}{rgb}{0.000000,0.000000,0.000000}%
\pgfsetstrokecolor{currentstroke}%
\pgfsetdash{}{0pt}%
\pgfsys@defobject{currentmarker}{\pgfqpoint{-0.048611in}{0.000000in}}{\pgfqpoint{-0.000000in}{0.000000in}}{%
\pgfpathmoveto{\pgfqpoint{-0.000000in}{0.000000in}}%
\pgfpathlineto{\pgfqpoint{-0.048611in}{0.000000in}}%
\pgfusepath{stroke,fill}%
}%
\begin{pgfscope}%
\pgfsys@transformshift{0.506171in}{1.657278in}%
\pgfsys@useobject{currentmarker}{}%
\end{pgfscope}%
\end{pgfscope}%
\begin{pgfscope}%
\definecolor{textcolor}{rgb}{0.000000,0.000000,0.000000}%
\pgfsetstrokecolor{textcolor}%
\pgfsetfillcolor{textcolor}%
\pgftext[x=0.180556in, y=1.613875in, left, base]{\color{textcolor}\rmfamily\fontsize{9.000000}{10.800000}\selectfont \(\displaystyle {0.06}\)}%
\end{pgfscope}%
\begin{pgfscope}%
\pgfsetbuttcap%
\pgfsetroundjoin%
\definecolor{currentfill}{rgb}{0.000000,0.000000,0.000000}%
\pgfsetfillcolor{currentfill}%
\pgfsetlinewidth{0.803000pt}%
\definecolor{currentstroke}{rgb}{0.000000,0.000000,0.000000}%
\pgfsetstrokecolor{currentstroke}%
\pgfsetdash{}{0pt}%
\pgfsys@defobject{currentmarker}{\pgfqpoint{-0.048611in}{0.000000in}}{\pgfqpoint{-0.000000in}{0.000000in}}{%
\pgfpathmoveto{\pgfqpoint{-0.000000in}{0.000000in}}%
\pgfpathlineto{\pgfqpoint{-0.048611in}{0.000000in}}%
\pgfusepath{stroke,fill}%
}%
\begin{pgfscope}%
\pgfsys@transformshift{0.506171in}{2.080074in}%
\pgfsys@useobject{currentmarker}{}%
\end{pgfscope}%
\end{pgfscope}%
\begin{pgfscope}%
\definecolor{textcolor}{rgb}{0.000000,0.000000,0.000000}%
\pgfsetstrokecolor{textcolor}%
\pgfsetfillcolor{textcolor}%
\pgftext[x=0.180556in, y=2.036671in, left, base]{\color{textcolor}\rmfamily\fontsize{9.000000}{10.800000}\selectfont \(\displaystyle {0.08}\)}%
\end{pgfscope}%
\begin{pgfscope}%
\pgfsetbuttcap%
\pgfsetroundjoin%
\definecolor{currentfill}{rgb}{0.000000,0.000000,0.000000}%
\pgfsetfillcolor{currentfill}%
\pgfsetlinewidth{0.803000pt}%
\definecolor{currentstroke}{rgb}{0.000000,0.000000,0.000000}%
\pgfsetstrokecolor{currentstroke}%
\pgfsetdash{}{0pt}%
\pgfsys@defobject{currentmarker}{\pgfqpoint{-0.048611in}{0.000000in}}{\pgfqpoint{-0.000000in}{0.000000in}}{%
\pgfpathmoveto{\pgfqpoint{-0.000000in}{0.000000in}}%
\pgfpathlineto{\pgfqpoint{-0.048611in}{0.000000in}}%
\pgfusepath{stroke,fill}%
}%
\begin{pgfscope}%
\pgfsys@transformshift{0.506171in}{2.502870in}%
\pgfsys@useobject{currentmarker}{}%
\end{pgfscope}%
\end{pgfscope}%
\begin{pgfscope}%
\definecolor{textcolor}{rgb}{0.000000,0.000000,0.000000}%
\pgfsetstrokecolor{textcolor}%
\pgfsetfillcolor{textcolor}%
\pgftext[x=0.180556in, y=2.459468in, left, base]{\color{textcolor}\rmfamily\fontsize{9.000000}{10.800000}\selectfont \(\displaystyle {0.10}\)}%
\end{pgfscope}%
\begin{pgfscope}%
\pgfsetbuttcap%
\pgfsetroundjoin%
\definecolor{currentfill}{rgb}{0.000000,0.000000,0.000000}%
\pgfsetfillcolor{currentfill}%
\pgfsetlinewidth{0.803000pt}%
\definecolor{currentstroke}{rgb}{0.000000,0.000000,0.000000}%
\pgfsetstrokecolor{currentstroke}%
\pgfsetdash{}{0pt}%
\pgfsys@defobject{currentmarker}{\pgfqpoint{-0.048611in}{0.000000in}}{\pgfqpoint{-0.000000in}{0.000000in}}{%
\pgfpathmoveto{\pgfqpoint{-0.000000in}{0.000000in}}%
\pgfpathlineto{\pgfqpoint{-0.048611in}{0.000000in}}%
\pgfusepath{stroke,fill}%
}%
\begin{pgfscope}%
\pgfsys@transformshift{0.506171in}{2.925667in}%
\pgfsys@useobject{currentmarker}{}%
\end{pgfscope}%
\end{pgfscope}%
\begin{pgfscope}%
\definecolor{textcolor}{rgb}{0.000000,0.000000,0.000000}%
\pgfsetstrokecolor{textcolor}%
\pgfsetfillcolor{textcolor}%
\pgftext[x=0.180556in, y=2.882264in, left, base]{\color{textcolor}\rmfamily\fontsize{9.000000}{10.800000}\selectfont \(\displaystyle {0.12}\)}%
\end{pgfscope}%
\begin{pgfscope}%
\pgfsetbuttcap%
\pgfsetroundjoin%
\definecolor{currentfill}{rgb}{0.000000,0.000000,0.000000}%
\pgfsetfillcolor{currentfill}%
\pgfsetlinewidth{0.803000pt}%
\definecolor{currentstroke}{rgb}{0.000000,0.000000,0.000000}%
\pgfsetstrokecolor{currentstroke}%
\pgfsetdash{}{0pt}%
\pgfsys@defobject{currentmarker}{\pgfqpoint{-0.048611in}{0.000000in}}{\pgfqpoint{-0.000000in}{0.000000in}}{%
\pgfpathmoveto{\pgfqpoint{-0.000000in}{0.000000in}}%
\pgfpathlineto{\pgfqpoint{-0.048611in}{0.000000in}}%
\pgfusepath{stroke,fill}%
}%
\begin{pgfscope}%
\pgfsys@transformshift{0.506171in}{3.348463in}%
\pgfsys@useobject{currentmarker}{}%
\end{pgfscope}%
\end{pgfscope}%
\begin{pgfscope}%
\definecolor{textcolor}{rgb}{0.000000,0.000000,0.000000}%
\pgfsetstrokecolor{textcolor}%
\pgfsetfillcolor{textcolor}%
\pgftext[x=0.180556in, y=3.305060in, left, base]{\color{textcolor}\rmfamily\fontsize{9.000000}{10.800000}\selectfont \(\displaystyle {0.14}\)}%
\end{pgfscope}%
\begin{pgfscope}%
\definecolor{textcolor}{rgb}{0.000000,0.000000,0.000000}%
\pgfsetstrokecolor{textcolor}%
\pgfsetfillcolor{textcolor}%
\pgftext[x=0.125000in,y=1.974375in,,bottom,rotate=90.000000]{\color{textcolor}\rmfamily\fontsize{9.000000}{10.800000}\selectfont \(\displaystyle |Q\cdot n|\) [N]}%
\end{pgfscope}%
\begin{pgfscope}%
\pgfpathrectangle{\pgfqpoint{0.506171in}{0.388889in}}{\pgfqpoint{2.223829in}{3.170972in}}%
\pgfusepath{clip}%
\pgfsetrectcap%
\pgfsetroundjoin%
\pgfsetlinewidth{1.505625pt}%
\definecolor{currentstroke}{rgb}{0.000000,0.000000,1.000000}%
\pgfsetstrokecolor{currentstroke}%
\pgfsetdash{}{0pt}%
\pgfpathmoveto{\pgfqpoint{0.506171in}{0.388889in}}%
\pgfpathlineto{\pgfqpoint{0.682497in}{0.389820in}}%
\pgfpathlineto{\pgfqpoint{0.706818in}{0.392681in}}%
\pgfpathlineto{\pgfqpoint{0.725058in}{0.396789in}}%
\pgfpathlineto{\pgfqpoint{0.743299in}{0.403141in}}%
\pgfpathlineto{\pgfqpoint{0.767620in}{0.414707in}}%
\pgfpathlineto{\pgfqpoint{0.791940in}{0.428789in}}%
\pgfpathlineto{\pgfqpoint{0.822341in}{0.449499in}}%
\pgfpathlineto{\pgfqpoint{0.858822in}{0.477981in}}%
\pgfpathlineto{\pgfqpoint{0.901384in}{0.514570in}}%
\pgfpathlineto{\pgfqpoint{0.943946in}{0.554149in}}%
\pgfpathlineto{\pgfqpoint{0.986506in}{0.596497in}}%
\pgfpathlineto{\pgfqpoint{1.035147in}{0.647997in}}%
\pgfpathlineto{\pgfqpoint{1.089870in}{0.709521in}}%
\pgfpathlineto{\pgfqpoint{1.144592in}{0.774745in}}%
\pgfpathlineto{\pgfqpoint{1.199315in}{0.843879in}}%
\pgfpathlineto{\pgfqpoint{1.254033in}{0.916974in}}%
\pgfpathlineto{\pgfqpoint{1.308755in}{0.994088in}}%
\pgfpathlineto{\pgfqpoint{1.363478in}{1.075406in}}%
\pgfpathlineto{\pgfqpoint{1.418200in}{1.160989in}}%
\pgfpathlineto{\pgfqpoint{1.472922in}{1.250998in}}%
\pgfpathlineto{\pgfqpoint{1.521563in}{1.335166in}}%
\pgfpathlineto{\pgfqpoint{1.570205in}{1.423655in}}%
\pgfpathlineto{\pgfqpoint{1.618846in}{1.517013in}}%
\pgfpathlineto{\pgfqpoint{1.667487in}{1.615620in}}%
\pgfpathlineto{\pgfqpoint{1.758688in}{1.807781in}}%
\pgfpathlineto{\pgfqpoint{1.831654in}{1.965152in}}%
\pgfpathlineto{\pgfqpoint{2.038377in}{2.416290in}}%
\pgfpathlineto{\pgfqpoint{2.093100in}{2.528619in}}%
\pgfpathlineto{\pgfqpoint{2.147822in}{2.635650in}}%
\pgfpathlineto{\pgfqpoint{2.196463in}{2.725980in}}%
\pgfpathlineto{\pgfqpoint{2.245104in}{2.811829in}}%
\pgfpathlineto{\pgfqpoint{2.299827in}{2.903470in}}%
\pgfpathlineto{\pgfqpoint{2.354549in}{2.990756in}}%
\pgfpathlineto{\pgfqpoint{2.415348in}{3.083750in}}%
\pgfpathlineto{\pgfqpoint{2.482233in}{3.181776in}}%
\pgfpathlineto{\pgfqpoint{2.555195in}{3.284388in}}%
\pgfpathlineto{\pgfqpoint{2.634238in}{3.391335in}}%
\pgfpathlineto{\pgfqpoint{2.719358in}{3.502974in}}%
\pgfpathlineto{\pgfqpoint{2.740000in}{3.529611in}}%
\pgfpathlineto{\pgfqpoint{2.740000in}{3.529611in}}%
\pgfusepath{stroke}%
\end{pgfscope}%
\begin{pgfscope}%
\pgfpathrectangle{\pgfqpoint{0.506171in}{0.388889in}}{\pgfqpoint{2.223829in}{3.170972in}}%
\pgfusepath{clip}%
\pgfsetrectcap%
\pgfsetroundjoin%
\pgfsetlinewidth{1.505625pt}%
\definecolor{currentstroke}{rgb}{0.000000,0.000000,0.000000}%
\pgfsetstrokecolor{currentstroke}%
\pgfsetdash{}{0pt}%
\pgfpathmoveto{\pgfqpoint{0.506171in}{0.388889in}}%
\pgfpathlineto{\pgfqpoint{0.718978in}{0.391861in}}%
\pgfpathlineto{\pgfqpoint{0.731138in}{0.394363in}}%
\pgfpathlineto{\pgfqpoint{0.749379in}{0.401748in}}%
\pgfpathlineto{\pgfqpoint{0.767620in}{0.411333in}}%
\pgfpathlineto{\pgfqpoint{0.791940in}{0.426759in}}%
\pgfpathlineto{\pgfqpoint{0.822341in}{0.449154in}}%
\pgfpathlineto{\pgfqpoint{0.852742in}{0.474402in}}%
\pgfpathlineto{\pgfqpoint{0.883143in}{0.502526in}}%
\pgfpathlineto{\pgfqpoint{0.913544in}{0.533625in}}%
\pgfpathlineto{\pgfqpoint{0.974344in}{0.599742in}}%
\pgfpathlineto{\pgfqpoint{1.029066in}{0.662451in}}%
\pgfpathlineto{\pgfqpoint{1.083789in}{0.728777in}}%
\pgfpathlineto{\pgfqpoint{1.150669in}{0.813421in}}%
\pgfpathlineto{\pgfqpoint{1.217554in}{0.901840in}}%
\pgfpathlineto{\pgfqpoint{1.296597in}{1.009801in}}%
\pgfpathlineto{\pgfqpoint{1.387798in}{1.138160in}}%
\pgfpathlineto{\pgfqpoint{1.485081in}{1.279234in}}%
\pgfpathlineto{\pgfqpoint{1.588444in}{1.433266in}}%
\pgfpathlineto{\pgfqpoint{1.691808in}{1.591241in}}%
\pgfpathlineto{\pgfqpoint{1.807334in}{1.771775in}}%
\pgfpathlineto{\pgfqpoint{1.935014in}{1.975303in}}%
\pgfpathlineto{\pgfqpoint{2.062698in}{2.182917in}}%
\pgfpathlineto{\pgfqpoint{2.232946in}{2.463889in}}%
\pgfpathlineto{\pgfqpoint{2.451831in}{2.824555in}}%
\pgfpathlineto{\pgfqpoint{2.597755in}{3.066585in}}%
\pgfpathlineto{\pgfqpoint{2.740000in}{3.304548in}}%
\pgfpathlineto{\pgfqpoint{2.740000in}{3.304548in}}%
\pgfusepath{stroke}%
\end{pgfscope}%
\begin{pgfscope}%
\pgfpathrectangle{\pgfqpoint{0.506171in}{0.388889in}}{\pgfqpoint{2.223829in}{3.170972in}}%
\pgfusepath{clip}%
\pgfsetbuttcap%
\pgfsetroundjoin%
\pgfsetlinewidth{0.702625pt}%
\definecolor{currentstroke}{rgb}{0.000000,0.000000,0.000000}%
\pgfsetstrokecolor{currentstroke}%
\pgfsetdash{{0.700000pt}{1.155000pt}}{0.000000pt}%
\pgfpathmoveto{\pgfqpoint{2.123501in}{0.388889in}}%
\pgfpathlineto{\pgfqpoint{2.123501in}{3.569861in}}%
\pgfusepath{stroke}%
\end{pgfscope}%
\begin{pgfscope}%
\pgfsetrectcap%
\pgfsetmiterjoin%
\pgfsetlinewidth{0.803000pt}%
\definecolor{currentstroke}{rgb}{0.000000,0.000000,0.000000}%
\pgfsetstrokecolor{currentstroke}%
\pgfsetdash{}{0pt}%
\pgfpathmoveto{\pgfqpoint{0.506171in}{0.388889in}}%
\pgfpathlineto{\pgfqpoint{0.506171in}{3.559861in}}%
\pgfusepath{stroke}%
\end{pgfscope}%
\begin{pgfscope}%
\pgfsetrectcap%
\pgfsetmiterjoin%
\pgfsetlinewidth{0.803000pt}%
\definecolor{currentstroke}{rgb}{0.000000,0.000000,0.000000}%
\pgfsetstrokecolor{currentstroke}%
\pgfsetdash{}{0pt}%
\pgfpathmoveto{\pgfqpoint{2.730000in}{0.388889in}}%
\pgfpathlineto{\pgfqpoint{2.730000in}{3.559861in}}%
\pgfusepath{stroke}%
\end{pgfscope}%
\begin{pgfscope}%
\pgfsetrectcap%
\pgfsetmiterjoin%
\pgfsetlinewidth{0.803000pt}%
\definecolor{currentstroke}{rgb}{0.000000,0.000000,0.000000}%
\pgfsetstrokecolor{currentstroke}%
\pgfsetdash{}{0pt}%
\pgfpathmoveto{\pgfqpoint{0.506171in}{0.388889in}}%
\pgfpathlineto{\pgfqpoint{2.730000in}{0.388889in}}%
\pgfusepath{stroke}%
\end{pgfscope}%
\begin{pgfscope}%
\pgfsetrectcap%
\pgfsetmiterjoin%
\pgfsetlinewidth{0.803000pt}%
\definecolor{currentstroke}{rgb}{0.000000,0.000000,0.000000}%
\pgfsetstrokecolor{currentstroke}%
\pgfsetdash{}{0pt}%
\pgfpathmoveto{\pgfqpoint{0.506171in}{3.559861in}}%
\pgfpathlineto{\pgfqpoint{2.730000in}{3.559861in}}%
\pgfusepath{stroke}%
\end{pgfscope}%
\begin{pgfscope}%
\definecolor{textcolor}{rgb}{0.000000,0.000000,0.000000}%
\pgfsetstrokecolor{textcolor}%
\pgfsetfillcolor{textcolor}%
\pgftext[x=1.618086in,y=3.643194in,,base]{\color{textcolor}\rmfamily\fontsize{9.000000}{10.800000}\selectfont Reaction force vs. displacement}%
\end{pgfscope}%
\begin{pgfscope}%
\pgfsetbuttcap%
\pgfsetmiterjoin%
\definecolor{currentfill}{rgb}{1.000000,1.000000,1.000000}%
\pgfsetfillcolor{currentfill}%
\pgfsetlinewidth{1.003750pt}%
\definecolor{currentstroke}{rgb}{0.800000,0.800000,0.800000}%
\pgfsetstrokecolor{currentstroke}%
\pgfsetdash{}{0pt}%
\pgfpathmoveto{\pgfqpoint{0.583949in}{3.153943in}}%
\pgfpathlineto{\pgfqpoint{2.309136in}{3.153943in}}%
\pgfpathquadraticcurveto{\pgfqpoint{2.331358in}{3.153943in}}{\pgfqpoint{2.331358in}{3.176165in}}%
\pgfpathlineto{\pgfqpoint{2.331358in}{3.482083in}}%
\pgfpathquadraticcurveto{\pgfqpoint{2.331358in}{3.504306in}}{\pgfqpoint{2.309136in}{3.504306in}}%
\pgfpathlineto{\pgfqpoint{0.583949in}{3.504306in}}%
\pgfpathquadraticcurveto{\pgfqpoint{0.561727in}{3.504306in}}{\pgfqpoint{0.561727in}{3.482083in}}%
\pgfpathlineto{\pgfqpoint{0.561727in}{3.176165in}}%
\pgfpathquadraticcurveto{\pgfqpoint{0.561727in}{3.153943in}}{\pgfqpoint{0.583949in}{3.153943in}}%
\pgfpathlineto{\pgfqpoint{0.583949in}{3.153943in}}%
\pgfpathclose%
\pgfusepath{stroke,fill}%
\end{pgfscope}%
\begin{pgfscope}%
\pgfsetrectcap%
\pgfsetroundjoin%
\pgfsetlinewidth{1.505625pt}%
\definecolor{currentstroke}{rgb}{0.000000,0.000000,1.000000}%
\pgfsetstrokecolor{currentstroke}%
\pgfsetdash{}{0pt}%
\pgfpathmoveto{\pgfqpoint{0.606171in}{3.413819in}}%
\pgfpathlineto{\pgfqpoint{0.717283in}{3.413819in}}%
\pgfpathlineto{\pgfqpoint{0.828394in}{3.413819in}}%
\pgfusepath{stroke}%
\end{pgfscope}%
\begin{pgfscope}%
\definecolor{textcolor}{rgb}{0.000000,0.000000,0.000000}%
\pgfsetstrokecolor{textcolor}%
\pgfsetfillcolor{textcolor}%
\pgftext[x=0.917283in,y=3.374931in,left,base]{\color{textcolor}\rmfamily\fontsize{8.000000}{9.600000}\selectfont \(\displaystyle C_0\) optimized for \(\displaystyle V^*= 1.0\)}%
\end{pgfscope}%
\begin{pgfscope}%
\pgfsetrectcap%
\pgfsetroundjoin%
\pgfsetlinewidth{1.505625pt}%
\definecolor{currentstroke}{rgb}{0.000000,0.000000,0.000000}%
\pgfsetstrokecolor{currentstroke}%
\pgfsetdash{}{0pt}%
\pgfpathmoveto{\pgfqpoint{0.606171in}{3.258881in}}%
\pgfpathlineto{\pgfqpoint{0.717283in}{3.258881in}}%
\pgfpathlineto{\pgfqpoint{0.828394in}{3.258881in}}%
\pgfusepath{stroke}%
\end{pgfscope}%
\begin{pgfscope}%
\definecolor{textcolor}{rgb}{0.000000,0.000000,0.000000}%
\pgfsetstrokecolor{textcolor}%
\pgfsetfillcolor{textcolor}%
\pgftext[x=0.917283in,y=3.219992in,left,base]{\color{textcolor}\rmfamily\fontsize{8.000000}{9.600000}\selectfont Reference design}%
\end{pgfscope}%
\end{pgfpicture}%
\makeatother%
\endgroup%

%% file: box_force_curves_v4.pgf
\begingroup%
\makeatletter%
\begin{pgfpicture}%
\pgfpathrectangle{\pgfpointorigin}{\pgfqpoint{2.752825in}{3.730000in}}%
\pgfusepath{use as bounding box, clip}%
\begin{pgfscope}%
\pgfsetbuttcap%
\pgfsetmiterjoin%
\definecolor{currentfill}{rgb}{1.000000,1.000000,1.000000}%
\pgfsetfillcolor{currentfill}%
\pgfsetlinewidth{0.000000pt}%
\definecolor{currentstroke}{rgb}{1.000000,1.000000,1.000000}%
\pgfsetstrokecolor{currentstroke}%
\pgfsetdash{}{0pt}%
\pgfpathmoveto{\pgfqpoint{0.000000in}{0.000000in}}%
\pgfpathlineto{\pgfqpoint{2.752825in}{0.000000in}}%
\pgfpathlineto{\pgfqpoint{2.752825in}{3.730000in}}%
\pgfpathlineto{\pgfqpoint{0.000000in}{3.730000in}}%
\pgfpathlineto{\pgfqpoint{0.000000in}{0.000000in}}%
\pgfpathclose%
\pgfusepath{fill}%
\end{pgfscope}%
\begin{pgfscope}%
\pgfsetbuttcap%
\pgfsetmiterjoin%
\definecolor{currentfill}{rgb}{1.000000,1.000000,1.000000}%
\pgfsetfillcolor{currentfill}%
\pgfsetlinewidth{0.000000pt}%
\definecolor{currentstroke}{rgb}{0.000000,0.000000,0.000000}%
\pgfsetstrokecolor{currentstroke}%
\pgfsetstrokeopacity{0.000000}%
\pgfsetdash{}{0pt}%
\pgfpathmoveto{\pgfqpoint{0.570407in}{0.533078in}}%
\pgfpathlineto{\pgfqpoint{2.689982in}{0.533078in}}%
\pgfpathlineto{\pgfqpoint{2.689982in}{3.559861in}}%
\pgfpathlineto{\pgfqpoint{0.570407in}{3.559861in}}%
\pgfpathlineto{\pgfqpoint{0.570407in}{0.533078in}}%
\pgfpathclose%
\pgfusepath{fill}%
\end{pgfscope}%
\begin{pgfscope}%
\pgfsetbuttcap%
\pgfsetroundjoin%
\definecolor{currentfill}{rgb}{0.000000,0.000000,0.000000}%
\pgfsetfillcolor{currentfill}%
\pgfsetlinewidth{0.803000pt}%
\definecolor{currentstroke}{rgb}{0.000000,0.000000,0.000000}%
\pgfsetstrokecolor{currentstroke}%
\pgfsetdash{}{0pt}%
\pgfsys@defobject{currentmarker}{\pgfqpoint{0.000000in}{-0.048611in}}{\pgfqpoint{0.000000in}{0.000000in}}{%
\pgfpathmoveto{\pgfqpoint{0.000000in}{0.000000in}}%
\pgfpathlineto{\pgfqpoint{0.000000in}{-0.048611in}}%
\pgfusepath{stroke,fill}%
}%
\begin{pgfscope}%
\pgfsys@transformshift{0.570407in}{0.533078in}%
\pgfsys@useobject{currentmarker}{}%
\end{pgfscope}%
\end{pgfscope}%
\begin{pgfscope}%
\definecolor{textcolor}{rgb}{0.000000,0.000000,0.000000}%
\pgfsetstrokecolor{textcolor}%
\pgfsetfillcolor{textcolor}%
\pgftext[x=0.570407in,y=0.435855in,,top]{\color{textcolor}\rmfamily\fontsize{9.000000}{10.800000}\selectfont 0.0}%
\end{pgfscope}%
\begin{pgfscope}%
\pgfsetbuttcap%
\pgfsetroundjoin%
\definecolor{currentfill}{rgb}{0.000000,0.000000,0.000000}%
\pgfsetfillcolor{currentfill}%
\pgfsetlinewidth{0.803000pt}%
\definecolor{currentstroke}{rgb}{0.000000,0.000000,0.000000}%
\pgfsetstrokecolor{currentstroke}%
\pgfsetdash{}{0pt}%
\pgfsys@defobject{currentmarker}{\pgfqpoint{0.000000in}{-0.048611in}}{\pgfqpoint{0.000000in}{0.000000in}}{%
\pgfpathmoveto{\pgfqpoint{0.000000in}{0.000000in}}%
\pgfpathlineto{\pgfqpoint{0.000000in}{-0.048611in}}%
\pgfusepath{stroke,fill}%
}%
\begin{pgfscope}%
\pgfsys@transformshift{1.041424in}{0.533078in}%
\pgfsys@useobject{currentmarker}{}%
\end{pgfscope}%
\end{pgfscope}%
\begin{pgfscope}%
\definecolor{textcolor}{rgb}{0.000000,0.000000,0.000000}%
\pgfsetstrokecolor{textcolor}%
\pgfsetfillcolor{textcolor}%
\pgftext[x=1.041424in,y=0.435855in,,top]{\color{textcolor}\rmfamily\fontsize{9.000000}{10.800000}\selectfont 0.1}%
\end{pgfscope}%
\begin{pgfscope}%
\pgfsetbuttcap%
\pgfsetroundjoin%
\definecolor{currentfill}{rgb}{0.000000,0.000000,0.000000}%
\pgfsetfillcolor{currentfill}%
\pgfsetlinewidth{0.803000pt}%
\definecolor{currentstroke}{rgb}{0.000000,0.000000,0.000000}%
\pgfsetstrokecolor{currentstroke}%
\pgfsetdash{}{0pt}%
\pgfsys@defobject{currentmarker}{\pgfqpoint{0.000000in}{-0.048611in}}{\pgfqpoint{0.000000in}{0.000000in}}{%
\pgfpathmoveto{\pgfqpoint{0.000000in}{0.000000in}}%
\pgfpathlineto{\pgfqpoint{0.000000in}{-0.048611in}}%
\pgfusepath{stroke,fill}%
}%
\begin{pgfscope}%
\pgfsys@transformshift{1.512440in}{0.533078in}%
\pgfsys@useobject{currentmarker}{}%
\end{pgfscope}%
\end{pgfscope}%
\begin{pgfscope}%
\definecolor{textcolor}{rgb}{0.000000,0.000000,0.000000}%
\pgfsetstrokecolor{textcolor}%
\pgfsetfillcolor{textcolor}%
\pgftext[x=1.512440in,y=0.435855in,,top]{\color{textcolor}\rmfamily\fontsize{9.000000}{10.800000}\selectfont 0.2}%
\end{pgfscope}%
\begin{pgfscope}%
\pgfsetbuttcap%
\pgfsetroundjoin%
\definecolor{currentfill}{rgb}{0.000000,0.000000,0.000000}%
\pgfsetfillcolor{currentfill}%
\pgfsetlinewidth{0.803000pt}%
\definecolor{currentstroke}{rgb}{0.000000,0.000000,0.000000}%
\pgfsetstrokecolor{currentstroke}%
\pgfsetdash{}{0pt}%
\pgfsys@defobject{currentmarker}{\pgfqpoint{0.000000in}{-0.048611in}}{\pgfqpoint{0.000000in}{0.000000in}}{%
\pgfpathmoveto{\pgfqpoint{0.000000in}{0.000000in}}%
\pgfpathlineto{\pgfqpoint{0.000000in}{-0.048611in}}%
\pgfusepath{stroke,fill}%
}%
\begin{pgfscope}%
\pgfsys@transformshift{1.983457in}{0.533078in}%
\pgfsys@useobject{currentmarker}{}%
\end{pgfscope}%
\end{pgfscope}%
\begin{pgfscope}%
\definecolor{textcolor}{rgb}{0.000000,0.000000,0.000000}%
\pgfsetstrokecolor{textcolor}%
\pgfsetfillcolor{textcolor}%
\pgftext[x=1.983457in,y=0.435855in,,top]{\color{textcolor}\rmfamily\fontsize{9.000000}{10.800000}\selectfont 0.3}%
\end{pgfscope}%
\begin{pgfscope}%
\pgfsetbuttcap%
\pgfsetroundjoin%
\definecolor{currentfill}{rgb}{0.000000,0.000000,0.000000}%
\pgfsetfillcolor{currentfill}%
\pgfsetlinewidth{0.803000pt}%
\definecolor{currentstroke}{rgb}{0.000000,0.000000,0.000000}%
\pgfsetstrokecolor{currentstroke}%
\pgfsetdash{}{0pt}%
\pgfsys@defobject{currentmarker}{\pgfqpoint{0.000000in}{-0.048611in}}{\pgfqpoint{0.000000in}{0.000000in}}{%
\pgfpathmoveto{\pgfqpoint{0.000000in}{0.000000in}}%
\pgfpathlineto{\pgfqpoint{0.000000in}{-0.048611in}}%
\pgfusepath{stroke,fill}%
}%
\begin{pgfscope}%
\pgfsys@transformshift{2.218965in}{0.533078in}%
\pgfsys@useobject{currentmarker}{}%
\end{pgfscope}%
\end{pgfscope}%
\begin{pgfscope}%
\definecolor{textcolor}{rgb}{0.000000,0.000000,0.000000}%
\pgfsetstrokecolor{textcolor}%
\pgfsetfillcolor{textcolor}%
\pgftext[x=2.232907in, y=0.364749in, left, base,rotate=325.000000]{\color{textcolor}\rmfamily\fontsize{9.000000}{10.800000}\selectfont \(\displaystyle u_{min}\)}%
\end{pgfscope}%
\begin{pgfscope}%
\pgfsetbuttcap%
\pgfsetroundjoin%
\definecolor{currentfill}{rgb}{0.000000,0.000000,0.000000}%
\pgfsetfillcolor{currentfill}%
\pgfsetlinewidth{0.803000pt}%
\definecolor{currentstroke}{rgb}{0.000000,0.000000,0.000000}%
\pgfsetstrokecolor{currentstroke}%
\pgfsetdash{}{0pt}%
\pgfsys@defobject{currentmarker}{\pgfqpoint{0.000000in}{-0.048611in}}{\pgfqpoint{0.000000in}{0.000000in}}{%
\pgfpathmoveto{\pgfqpoint{0.000000in}{0.000000in}}%
\pgfpathlineto{\pgfqpoint{0.000000in}{-0.048611in}}%
\pgfusepath{stroke,fill}%
}%
\begin{pgfscope}%
\pgfsys@transformshift{2.454474in}{0.533078in}%
\pgfsys@useobject{currentmarker}{}%
\end{pgfscope}%
\end{pgfscope}%
\begin{pgfscope}%
\definecolor{textcolor}{rgb}{0.000000,0.000000,0.000000}%
\pgfsetstrokecolor{textcolor}%
\pgfsetfillcolor{textcolor}%
\pgftext[x=2.468415in, y=0.364749in, left, base,rotate=325.000000]{\color{textcolor}\rmfamily\fontsize{9.000000}{10.800000}\selectfont \(\displaystyle u_{max}\)}%
\end{pgfscope}%
\begin{pgfscope}%
\definecolor{textcolor}{rgb}{0.000000,0.000000,0.000000}%
\pgfsetstrokecolor{textcolor}%
\pgfsetfillcolor{textcolor}%
\pgftext[x=1.630195in,y=0.125000in,,top]{\color{textcolor}\rmfamily\fontsize{9.000000}{10.800000}\selectfont \(\displaystyle u_D\,\)[mm]}%
\end{pgfscope}%
\begin{pgfscope}%
\pgfsetbuttcap%
\pgfsetroundjoin%
\definecolor{currentfill}{rgb}{0.000000,0.000000,0.000000}%
\pgfsetfillcolor{currentfill}%
\pgfsetlinewidth{0.803000pt}%
\definecolor{currentstroke}{rgb}{0.000000,0.000000,0.000000}%
\pgfsetstrokecolor{currentstroke}%
\pgfsetdash{}{0pt}%
\pgfsys@defobject{currentmarker}{\pgfqpoint{-0.048611in}{0.000000in}}{\pgfqpoint{-0.000000in}{0.000000in}}{%
\pgfpathmoveto{\pgfqpoint{-0.000000in}{0.000000in}}%
\pgfpathlineto{\pgfqpoint{-0.048611in}{0.000000in}}%
\pgfusepath{stroke,fill}%
}%
\begin{pgfscope}%
\pgfsys@transformshift{0.570407in}{0.533078in}%
\pgfsys@useobject{currentmarker}{}%
\end{pgfscope}%
\end{pgfscope}%
\begin{pgfscope}%
\definecolor{textcolor}{rgb}{0.000000,0.000000,0.000000}%
\pgfsetstrokecolor{textcolor}%
\pgfsetfillcolor{textcolor}%
\pgftext[x=0.180556in, y=0.489675in, left, base]{\color{textcolor}\rmfamily\fontsize{9.000000}{10.800000}\selectfont \(\displaystyle {0.000}\)}%
\end{pgfscope}%
\begin{pgfscope}%
\pgfsetbuttcap%
\pgfsetroundjoin%
\definecolor{currentfill}{rgb}{0.000000,0.000000,0.000000}%
\pgfsetfillcolor{currentfill}%
\pgfsetlinewidth{0.803000pt}%
\definecolor{currentstroke}{rgb}{0.000000,0.000000,0.000000}%
\pgfsetstrokecolor{currentstroke}%
\pgfsetdash{}{0pt}%
\pgfsys@defobject{currentmarker}{\pgfqpoint{-0.048611in}{0.000000in}}{\pgfqpoint{-0.000000in}{0.000000in}}{%
\pgfpathmoveto{\pgfqpoint{-0.000000in}{0.000000in}}%
\pgfpathlineto{\pgfqpoint{-0.048611in}{0.000000in}}%
\pgfusepath{stroke,fill}%
}%
\begin{pgfscope}%
\pgfsys@transformshift{0.570407in}{0.998737in}%
\pgfsys@useobject{currentmarker}{}%
\end{pgfscope}%
\end{pgfscope}%
\begin{pgfscope}%
\definecolor{textcolor}{rgb}{0.000000,0.000000,0.000000}%
\pgfsetstrokecolor{textcolor}%
\pgfsetfillcolor{textcolor}%
\pgftext[x=0.180556in, y=0.955334in, left, base]{\color{textcolor}\rmfamily\fontsize{9.000000}{10.800000}\selectfont \(\displaystyle {0.002}\)}%
\end{pgfscope}%
\begin{pgfscope}%
\pgfsetbuttcap%
\pgfsetroundjoin%
\definecolor{currentfill}{rgb}{0.000000,0.000000,0.000000}%
\pgfsetfillcolor{currentfill}%
\pgfsetlinewidth{0.803000pt}%
\definecolor{currentstroke}{rgb}{0.000000,0.000000,0.000000}%
\pgfsetstrokecolor{currentstroke}%
\pgfsetdash{}{0pt}%
\pgfsys@defobject{currentmarker}{\pgfqpoint{-0.048611in}{0.000000in}}{\pgfqpoint{-0.000000in}{0.000000in}}{%
\pgfpathmoveto{\pgfqpoint{-0.000000in}{0.000000in}}%
\pgfpathlineto{\pgfqpoint{-0.048611in}{0.000000in}}%
\pgfusepath{stroke,fill}%
}%
\begin{pgfscope}%
\pgfsys@transformshift{0.570407in}{1.464396in}%
\pgfsys@useobject{currentmarker}{}%
\end{pgfscope}%
\end{pgfscope}%
\begin{pgfscope}%
\definecolor{textcolor}{rgb}{0.000000,0.000000,0.000000}%
\pgfsetstrokecolor{textcolor}%
\pgfsetfillcolor{textcolor}%
\pgftext[x=0.180556in, y=1.420993in, left, base]{\color{textcolor}\rmfamily\fontsize{9.000000}{10.800000}\selectfont \(\displaystyle {0.004}\)}%
\end{pgfscope}%
\begin{pgfscope}%
\pgfsetbuttcap%
\pgfsetroundjoin%
\definecolor{currentfill}{rgb}{0.000000,0.000000,0.000000}%
\pgfsetfillcolor{currentfill}%
\pgfsetlinewidth{0.803000pt}%
\definecolor{currentstroke}{rgb}{0.000000,0.000000,0.000000}%
\pgfsetstrokecolor{currentstroke}%
\pgfsetdash{}{0pt}%
\pgfsys@defobject{currentmarker}{\pgfqpoint{-0.048611in}{0.000000in}}{\pgfqpoint{-0.000000in}{0.000000in}}{%
\pgfpathmoveto{\pgfqpoint{-0.000000in}{0.000000in}}%
\pgfpathlineto{\pgfqpoint{-0.048611in}{0.000000in}}%
\pgfusepath{stroke,fill}%
}%
\begin{pgfscope}%
\pgfsys@transformshift{0.570407in}{1.930055in}%
\pgfsys@useobject{currentmarker}{}%
\end{pgfscope}%
\end{pgfscope}%
\begin{pgfscope}%
\definecolor{textcolor}{rgb}{0.000000,0.000000,0.000000}%
\pgfsetstrokecolor{textcolor}%
\pgfsetfillcolor{textcolor}%
\pgftext[x=0.180556in, y=1.886652in, left, base]{\color{textcolor}\rmfamily\fontsize{9.000000}{10.800000}\selectfont \(\displaystyle {0.006}\)}%
\end{pgfscope}%
\begin{pgfscope}%
\pgfsetbuttcap%
\pgfsetroundjoin%
\definecolor{currentfill}{rgb}{0.000000,0.000000,0.000000}%
\pgfsetfillcolor{currentfill}%
\pgfsetlinewidth{0.803000pt}%
\definecolor{currentstroke}{rgb}{0.000000,0.000000,0.000000}%
\pgfsetstrokecolor{currentstroke}%
\pgfsetdash{}{0pt}%
\pgfsys@defobject{currentmarker}{\pgfqpoint{-0.048611in}{0.000000in}}{\pgfqpoint{-0.000000in}{0.000000in}}{%
\pgfpathmoveto{\pgfqpoint{-0.000000in}{0.000000in}}%
\pgfpathlineto{\pgfqpoint{-0.048611in}{0.000000in}}%
\pgfusepath{stroke,fill}%
}%
\begin{pgfscope}%
\pgfsys@transformshift{0.570407in}{2.395714in}%
\pgfsys@useobject{currentmarker}{}%
\end{pgfscope}%
\end{pgfscope}%
\begin{pgfscope}%
\definecolor{textcolor}{rgb}{0.000000,0.000000,0.000000}%
\pgfsetstrokecolor{textcolor}%
\pgfsetfillcolor{textcolor}%
\pgftext[x=0.180556in, y=2.352311in, left, base]{\color{textcolor}\rmfamily\fontsize{9.000000}{10.800000}\selectfont \(\displaystyle {0.008}\)}%
\end{pgfscope}%
\begin{pgfscope}%
\pgfsetbuttcap%
\pgfsetroundjoin%
\definecolor{currentfill}{rgb}{0.000000,0.000000,0.000000}%
\pgfsetfillcolor{currentfill}%
\pgfsetlinewidth{0.803000pt}%
\definecolor{currentstroke}{rgb}{0.000000,0.000000,0.000000}%
\pgfsetstrokecolor{currentstroke}%
\pgfsetdash{}{0pt}%
\pgfsys@defobject{currentmarker}{\pgfqpoint{-0.048611in}{0.000000in}}{\pgfqpoint{-0.000000in}{0.000000in}}{%
\pgfpathmoveto{\pgfqpoint{-0.000000in}{0.000000in}}%
\pgfpathlineto{\pgfqpoint{-0.048611in}{0.000000in}}%
\pgfusepath{stroke,fill}%
}%
\begin{pgfscope}%
\pgfsys@transformshift{0.570407in}{2.861373in}%
\pgfsys@useobject{currentmarker}{}%
\end{pgfscope}%
\end{pgfscope}%
\begin{pgfscope}%
\definecolor{textcolor}{rgb}{0.000000,0.000000,0.000000}%
\pgfsetstrokecolor{textcolor}%
\pgfsetfillcolor{textcolor}%
\pgftext[x=0.180556in, y=2.817970in, left, base]{\color{textcolor}\rmfamily\fontsize{9.000000}{10.800000}\selectfont \(\displaystyle {0.010}\)}%
\end{pgfscope}%
\begin{pgfscope}%
\pgfsetbuttcap%
\pgfsetroundjoin%
\definecolor{currentfill}{rgb}{0.000000,0.000000,0.000000}%
\pgfsetfillcolor{currentfill}%
\pgfsetlinewidth{0.803000pt}%
\definecolor{currentstroke}{rgb}{0.000000,0.000000,0.000000}%
\pgfsetstrokecolor{currentstroke}%
\pgfsetdash{}{0pt}%
\pgfsys@defobject{currentmarker}{\pgfqpoint{-0.048611in}{0.000000in}}{\pgfqpoint{-0.000000in}{0.000000in}}{%
\pgfpathmoveto{\pgfqpoint{-0.000000in}{0.000000in}}%
\pgfpathlineto{\pgfqpoint{-0.048611in}{0.000000in}}%
\pgfusepath{stroke,fill}%
}%
\begin{pgfscope}%
\pgfsys@transformshift{0.570407in}{3.327032in}%
\pgfsys@useobject{currentmarker}{}%
\end{pgfscope}%
\end{pgfscope}%
\begin{pgfscope}%
\definecolor{textcolor}{rgb}{0.000000,0.000000,0.000000}%
\pgfsetstrokecolor{textcolor}%
\pgfsetfillcolor{textcolor}%
\pgftext[x=0.180556in, y=3.283629in, left, base]{\color{textcolor}\rmfamily\fontsize{9.000000}{10.800000}\selectfont \(\displaystyle {0.012}\)}%
\end{pgfscope}%
\begin{pgfscope}%
\definecolor{textcolor}{rgb}{0.000000,0.000000,0.000000}%
\pgfsetstrokecolor{textcolor}%
\pgfsetfillcolor{textcolor}%
\pgftext[x=0.125000in,y=2.046469in,,bottom,rotate=90.000000]{\color{textcolor}\rmfamily\fontsize{9.000000}{10.800000}\selectfont \(\displaystyle |Q\cdot n|\) [N]}%
\end{pgfscope}%
\begin{pgfscope}%
\pgfpathrectangle{\pgfqpoint{0.570407in}{0.533078in}}{\pgfqpoint{2.119575in}{3.026783in}}%
\pgfusepath{clip}%
\pgfsetrectcap%
\pgfsetroundjoin%
\pgfsetlinewidth{1.505625pt}%
\definecolor{currentstroke}{rgb}{0.000000,0.000000,1.000000}%
\pgfsetstrokecolor{currentstroke}%
\pgfsetdash{}{0pt}%
\pgfpathmoveto{\pgfqpoint{0.570407in}{0.533078in}}%
\pgfpathlineto{\pgfqpoint{1.639935in}{0.534183in}}%
\pgfpathlineto{\pgfqpoint{1.788678in}{0.536407in}}%
\pgfpathlineto{\pgfqpoint{1.880752in}{0.539817in}}%
\pgfpathlineto{\pgfqpoint{1.944499in}{0.544157in}}%
\pgfpathlineto{\pgfqpoint{1.994083in}{0.549498in}}%
\pgfpathlineto{\pgfqpoint{2.029494in}{0.555055in}}%
\pgfpathlineto{\pgfqpoint{2.057826in}{0.561347in}}%
\pgfpathlineto{\pgfqpoint{2.079078in}{0.567720in}}%
\pgfpathlineto{\pgfqpoint{2.100326in}{0.576181in}}%
\pgfpathlineto{\pgfqpoint{2.121573in}{0.587453in}}%
\pgfpathlineto{\pgfqpoint{2.135741in}{0.597251in}}%
\pgfpathlineto{\pgfqpoint{2.149905in}{0.609379in}}%
\pgfpathlineto{\pgfqpoint{2.164073in}{0.624335in}}%
\pgfpathlineto{\pgfqpoint{2.178237in}{0.643104in}}%
\pgfpathlineto{\pgfqpoint{2.192405in}{0.667075in}}%
\pgfpathlineto{\pgfqpoint{2.206568in}{0.695610in}}%
\pgfpathlineto{\pgfqpoint{2.220736in}{0.729501in}}%
\pgfpathlineto{\pgfqpoint{2.234900in}{0.770082in}}%
\pgfpathlineto{\pgfqpoint{2.249068in}{0.818000in}}%
\pgfpathlineto{\pgfqpoint{2.270316in}{0.898711in}}%
\pgfpathlineto{\pgfqpoint{2.298647in}{1.019847in}}%
\pgfpathlineto{\pgfqpoint{2.326979in}{1.153901in}}%
\pgfpathlineto{\pgfqpoint{2.348231in}{1.264670in}}%
\pgfpathlineto{\pgfqpoint{2.383642in}{1.464284in}}%
\pgfpathlineto{\pgfqpoint{2.426142in}{1.717405in}}%
\pgfpathlineto{\pgfqpoint{2.475721in}{2.027049in}}%
\pgfpathlineto{\pgfqpoint{2.525305in}{2.347825in}}%
\pgfpathlineto{\pgfqpoint{2.581969in}{2.727691in}}%
\pgfpathlineto{\pgfqpoint{2.659879in}{3.266892in}}%
\pgfpathlineto{\pgfqpoint{2.699982in}{3.548884in}}%
\pgfpathlineto{\pgfqpoint{2.699982in}{3.548884in}}%
\pgfusepath{stroke}%
\end{pgfscope}%
\begin{pgfscope}%
\pgfpathrectangle{\pgfqpoint{0.570407in}{0.533078in}}{\pgfqpoint{2.119575in}{3.026783in}}%
\pgfusepath{clip}%
\pgfsetrectcap%
\pgfsetroundjoin%
\pgfsetlinewidth{1.505625pt}%
\definecolor{currentstroke}{rgb}{0.000000,0.000000,0.000000}%
\pgfsetstrokecolor{currentstroke}%
\pgfsetdash{}{0pt}%
\pgfpathmoveto{\pgfqpoint{0.570407in}{0.533078in}}%
\pgfpathlineto{\pgfqpoint{1.533688in}{0.534175in}}%
\pgfpathlineto{\pgfqpoint{1.703683in}{0.536571in}}%
\pgfpathlineto{\pgfqpoint{1.838257in}{0.540720in}}%
\pgfpathlineto{\pgfqpoint{1.916168in}{0.545051in}}%
\pgfpathlineto{\pgfqpoint{1.965752in}{0.549683in}}%
\pgfpathlineto{\pgfqpoint{2.008247in}{0.555812in}}%
\pgfpathlineto{\pgfqpoint{2.043662in}{0.563445in}}%
\pgfpathlineto{\pgfqpoint{2.071994in}{0.572092in}}%
\pgfpathlineto{\pgfqpoint{2.093242in}{0.580835in}}%
\pgfpathlineto{\pgfqpoint{2.114494in}{0.592204in}}%
\pgfpathlineto{\pgfqpoint{2.128657in}{0.601629in}}%
\pgfpathlineto{\pgfqpoint{2.142826in}{0.613104in}}%
\pgfpathlineto{\pgfqpoint{2.156989in}{0.627139in}}%
\pgfpathlineto{\pgfqpoint{2.171157in}{0.644009in}}%
\pgfpathlineto{\pgfqpoint{2.185321in}{0.664511in}}%
\pgfpathlineto{\pgfqpoint{2.199489in}{0.689383in}}%
\pgfpathlineto{\pgfqpoint{2.213652in}{0.718062in}}%
\pgfpathlineto{\pgfqpoint{2.227821in}{0.751814in}}%
\pgfpathlineto{\pgfqpoint{2.241984in}{0.791886in}}%
\pgfpathlineto{\pgfqpoint{2.256152in}{0.837395in}}%
\pgfpathlineto{\pgfqpoint{2.277400in}{0.914699in}}%
\pgfpathlineto{\pgfqpoint{2.305731in}{1.028211in}}%
\pgfpathlineto{\pgfqpoint{2.334063in}{1.154127in}}%
\pgfpathlineto{\pgfqpoint{2.362395in}{1.289962in}}%
\pgfpathlineto{\pgfqpoint{2.390726in}{1.434957in}}%
\pgfpathlineto{\pgfqpoint{2.433226in}{1.664550in}}%
\pgfpathlineto{\pgfqpoint{2.518221in}{2.139315in}}%
\pgfpathlineto{\pgfqpoint{2.581969in}{2.491628in}}%
\pgfpathlineto{\pgfqpoint{2.631548in}{2.754997in}}%
\pgfpathlineto{\pgfqpoint{2.681127in}{3.005820in}}%
\pgfpathlineto{\pgfqpoint{2.699982in}{3.097190in}}%
\pgfpathlineto{\pgfqpoint{2.699982in}{3.097190in}}%
\pgfusepath{stroke}%
\end{pgfscope}%
\begin{pgfscope}%
\pgfpathrectangle{\pgfqpoint{0.570407in}{0.533078in}}{\pgfqpoint{2.119575in}{3.026783in}}%
\pgfusepath{clip}%
\pgfsetbuttcap%
\pgfsetroundjoin%
\pgfsetlinewidth{0.702625pt}%
\definecolor{currentstroke}{rgb}{0.000000,0.000000,0.000000}%
\pgfsetstrokecolor{currentstroke}%
\pgfsetdash{{0.700000pt}{1.155000pt}}{0.000000pt}%
\pgfpathmoveto{\pgfqpoint{2.218965in}{0.533078in}}%
\pgfpathlineto{\pgfqpoint{2.218965in}{3.569861in}}%
\pgfusepath{stroke}%
\end{pgfscope}%
\begin{pgfscope}%
\pgfpathrectangle{\pgfqpoint{0.570407in}{0.533078in}}{\pgfqpoint{2.119575in}{3.026783in}}%
\pgfusepath{clip}%
\pgfsetbuttcap%
\pgfsetroundjoin%
\pgfsetlinewidth{0.702625pt}%
\definecolor{currentstroke}{rgb}{0.000000,0.000000,0.000000}%
\pgfsetstrokecolor{currentstroke}%
\pgfsetdash{{0.700000pt}{1.155000pt}}{0.000000pt}%
\pgfpathmoveto{\pgfqpoint{2.454474in}{0.533078in}}%
\pgfpathlineto{\pgfqpoint{2.454474in}{3.569861in}}%
\pgfusepath{stroke}%
\end{pgfscope}%
\begin{pgfscope}%
\pgfsetrectcap%
\pgfsetmiterjoin%
\pgfsetlinewidth{0.803000pt}%
\definecolor{currentstroke}{rgb}{0.000000,0.000000,0.000000}%
\pgfsetstrokecolor{currentstroke}%
\pgfsetdash{}{0pt}%
\pgfpathmoveto{\pgfqpoint{0.570407in}{0.533078in}}%
\pgfpathlineto{\pgfqpoint{0.570407in}{3.559861in}}%
\pgfusepath{stroke}%
\end{pgfscope}%
\begin{pgfscope}%
\pgfsetrectcap%
\pgfsetmiterjoin%
\pgfsetlinewidth{0.803000pt}%
\definecolor{currentstroke}{rgb}{0.000000,0.000000,0.000000}%
\pgfsetstrokecolor{currentstroke}%
\pgfsetdash{}{0pt}%
\pgfpathmoveto{\pgfqpoint{2.689982in}{0.533078in}}%
\pgfpathlineto{\pgfqpoint{2.689982in}{3.559861in}}%
\pgfusepath{stroke}%
\end{pgfscope}%
\begin{pgfscope}%
\pgfsetrectcap%
\pgfsetmiterjoin%
\pgfsetlinewidth{0.803000pt}%
\definecolor{currentstroke}{rgb}{0.000000,0.000000,0.000000}%
\pgfsetstrokecolor{currentstroke}%
\pgfsetdash{}{0pt}%
\pgfpathmoveto{\pgfqpoint{0.570407in}{0.533078in}}%
\pgfpathlineto{\pgfqpoint{2.689982in}{0.533078in}}%
\pgfusepath{stroke}%
\end{pgfscope}%
\begin{pgfscope}%
\pgfsetrectcap%
\pgfsetmiterjoin%
\pgfsetlinewidth{0.803000pt}%
\definecolor{currentstroke}{rgb}{0.000000,0.000000,0.000000}%
\pgfsetstrokecolor{currentstroke}%
\pgfsetdash{}{0pt}%
\pgfpathmoveto{\pgfqpoint{0.570407in}{3.559861in}}%
\pgfpathlineto{\pgfqpoint{2.689982in}{3.559861in}}%
\pgfusepath{stroke}%
\end{pgfscope}%
\begin{pgfscope}%
\definecolor{textcolor}{rgb}{0.000000,0.000000,0.000000}%
\pgfsetstrokecolor{textcolor}%
\pgfsetfillcolor{textcolor}%
\pgftext[x=1.630195in,y=3.643194in,,base]{\color{textcolor}\rmfamily\fontsize{9.000000}{10.800000}\selectfont Reaction force vs. displacement}%
\end{pgfscope}%
\begin{pgfscope}%
\pgfsetbuttcap%
\pgfsetmiterjoin%
\definecolor{currentfill}{rgb}{1.000000,1.000000,1.000000}%
\pgfsetfillcolor{currentfill}%
\pgfsetlinewidth{1.003750pt}%
\definecolor{currentstroke}{rgb}{0.800000,0.800000,0.800000}%
\pgfsetstrokecolor{currentstroke}%
\pgfsetdash{}{0pt}%
\pgfpathmoveto{\pgfqpoint{0.648185in}{3.146790in}}%
\pgfpathlineto{\pgfqpoint{2.373371in}{3.146790in}}%
\pgfpathquadraticcurveto{\pgfqpoint{2.395593in}{3.146790in}}{\pgfqpoint{2.395593in}{3.169012in}}%
\pgfpathlineto{\pgfqpoint{2.395593in}{3.482083in}}%
\pgfpathquadraticcurveto{\pgfqpoint{2.395593in}{3.504306in}}{\pgfqpoint{2.373371in}{3.504306in}}%
\pgfpathlineto{\pgfqpoint{0.648185in}{3.504306in}}%
\pgfpathquadraticcurveto{\pgfqpoint{0.625962in}{3.504306in}}{\pgfqpoint{0.625962in}{3.482083in}}%
\pgfpathlineto{\pgfqpoint{0.625962in}{3.169012in}}%
\pgfpathquadraticcurveto{\pgfqpoint{0.625962in}{3.146790in}}{\pgfqpoint{0.648185in}{3.146790in}}%
\pgfpathlineto{\pgfqpoint{0.648185in}{3.146790in}}%
\pgfpathclose%
\pgfusepath{stroke,fill}%
\end{pgfscope}%
\begin{pgfscope}%
\pgfsetrectcap%
\pgfsetroundjoin%
\pgfsetlinewidth{1.505625pt}%
\definecolor{currentstroke}{rgb}{0.000000,0.000000,1.000000}%
\pgfsetstrokecolor{currentstroke}%
\pgfsetdash{}{0pt}%
\pgfpathmoveto{\pgfqpoint{0.670407in}{3.413819in}}%
\pgfpathlineto{\pgfqpoint{0.781518in}{3.413819in}}%
\pgfpathlineto{\pgfqpoint{0.892629in}{3.413819in}}%
\pgfusepath{stroke}%
\end{pgfscope}%
\begin{pgfscope}%
\definecolor{textcolor}{rgb}{0.000000,0.000000,0.000000}%
\pgfsetstrokecolor{textcolor}%
\pgfsetfillcolor{textcolor}%
\pgftext[x=0.981518in,y=3.374931in,left,base]{\color{textcolor}\rmfamily\fontsize{8.000000}{9.600000}\selectfont \(\displaystyle C_1\) optimized for \(\displaystyle V^*= 1.0\)}%
\end{pgfscope}%
\begin{pgfscope}%
\pgfsetrectcap%
\pgfsetroundjoin%
\pgfsetlinewidth{1.505625pt}%
\definecolor{currentstroke}{rgb}{0.000000,0.000000,0.000000}%
\pgfsetstrokecolor{currentstroke}%
\pgfsetdash{}{0pt}%
\pgfpathmoveto{\pgfqpoint{0.670407in}{3.251728in}}%
\pgfpathlineto{\pgfqpoint{0.781518in}{3.251728in}}%
\pgfpathlineto{\pgfqpoint{0.892629in}{3.251728in}}%
\pgfusepath{stroke}%
\end{pgfscope}%
\begin{pgfscope}%
\definecolor{textcolor}{rgb}{0.000000,0.000000,0.000000}%
\pgfsetstrokecolor{textcolor}%
\pgfsetfillcolor{textcolor}%
\pgftext[x=0.981518in,y=3.212839in,left,base]{\color{textcolor}\rmfamily\fontsize{8.000000}{9.600000}\selectfont \(\displaystyle C_1\) optimized for \(\displaystyle V^*= 0.6\)}%
\end{pgfscope}%
\end{pgfpicture}%
\makeatother%
\endgroup%

%% file: hook_moment_curves_case_NX200_NY100_bend_v18_c13_v2.pgf
\begingroup%
\makeatletter%
\begin{pgfpicture}%
\pgfpathrectangle{\pgfpointorigin}{\pgfqpoint{3.000000in}{3.000000in}}%
\pgfusepath{use as bounding box, clip}%
\begin{pgfscope}%
\pgfsetbuttcap%
\pgfsetmiterjoin%
\definecolor{currentfill}{rgb}{1.000000,1.000000,1.000000}%
\pgfsetfillcolor{currentfill}%
\pgfsetlinewidth{0.000000pt}%
\definecolor{currentstroke}{rgb}{1.000000,1.000000,1.000000}%
\pgfsetstrokecolor{currentstroke}%
\pgfsetdash{}{0pt}%
\pgfpathmoveto{\pgfqpoint{0.000000in}{0.000000in}}%
\pgfpathlineto{\pgfqpoint{3.000000in}{0.000000in}}%
\pgfpathlineto{\pgfqpoint{3.000000in}{3.000000in}}%
\pgfpathlineto{\pgfqpoint{0.000000in}{3.000000in}}%
\pgfpathlineto{\pgfqpoint{0.000000in}{0.000000in}}%
\pgfpathclose%
\pgfusepath{fill}%
\end{pgfscope}%
\begin{pgfscope}%
\pgfsetbuttcap%
\pgfsetmiterjoin%
\definecolor{currentfill}{rgb}{1.000000,1.000000,1.000000}%
\pgfsetfillcolor{currentfill}%
\pgfsetlinewidth{0.000000pt}%
\definecolor{currentstroke}{rgb}{0.000000,0.000000,0.000000}%
\pgfsetstrokecolor{currentstroke}%
\pgfsetstrokeopacity{0.000000}%
\pgfsetdash{}{0pt}%
\pgfpathmoveto{\pgfqpoint{0.619136in}{0.565123in}}%
\pgfpathlineto{\pgfqpoint{2.780555in}{0.565123in}}%
\pgfpathlineto{\pgfqpoint{2.780555in}{2.650926in}}%
\pgfpathlineto{\pgfqpoint{0.619136in}{2.650926in}}%
\pgfpathlineto{\pgfqpoint{0.619136in}{0.565123in}}%
\pgfpathclose%
\pgfusepath{fill}%
\end{pgfscope}%
\begin{pgfscope}%
\pgfsetbuttcap%
\pgfsetroundjoin%
\definecolor{currentfill}{rgb}{0.000000,0.000000,0.000000}%
\pgfsetfillcolor{currentfill}%
\pgfsetlinewidth{0.803000pt}%
\definecolor{currentstroke}{rgb}{0.000000,0.000000,0.000000}%
\pgfsetstrokecolor{currentstroke}%
\pgfsetdash{}{0pt}%
\pgfsys@defobject{currentmarker}{\pgfqpoint{0.000000in}{-0.048611in}}{\pgfqpoint{0.000000in}{0.000000in}}{%
\pgfpathmoveto{\pgfqpoint{0.000000in}{0.000000in}}%
\pgfpathlineto{\pgfqpoint{0.000000in}{-0.048611in}}%
\pgfusepath{stroke,fill}%
}%
\begin{pgfscope}%
\pgfsys@transformshift{0.619136in}{0.565123in}%
\pgfsys@useobject{currentmarker}{}%
\end{pgfscope}%
\end{pgfscope}%
\begin{pgfscope}%
\definecolor{textcolor}{rgb}{0.000000,0.000000,0.000000}%
\pgfsetstrokecolor{textcolor}%
\pgfsetfillcolor{textcolor}%
\pgftext[x=0.619136in,y=0.467901in,,top]{\color{textcolor}\rmfamily\fontsize{10.000000}{12.000000}\selectfont 0}%
\end{pgfscope}%
\begin{pgfscope}%
\pgfsetbuttcap%
\pgfsetroundjoin%
\definecolor{currentfill}{rgb}{0.000000,0.000000,0.000000}%
\pgfsetfillcolor{currentfill}%
\pgfsetlinewidth{0.803000pt}%
\definecolor{currentstroke}{rgb}{0.000000,0.000000,0.000000}%
\pgfsetstrokecolor{currentstroke}%
\pgfsetdash{}{0pt}%
\pgfsys@defobject{currentmarker}{\pgfqpoint{0.000000in}{-0.048611in}}{\pgfqpoint{0.000000in}{0.000000in}}{%
\pgfpathmoveto{\pgfqpoint{0.000000in}{0.000000in}}%
\pgfpathlineto{\pgfqpoint{0.000000in}{-0.048611in}}%
\pgfusepath{stroke,fill}%
}%
\begin{pgfscope}%
\pgfsys@transformshift{1.159491in}{0.565123in}%
\pgfsys@useobject{currentmarker}{}%
\end{pgfscope}%
\end{pgfscope}%
\begin{pgfscope}%
\definecolor{textcolor}{rgb}{0.000000,0.000000,0.000000}%
\pgfsetstrokecolor{textcolor}%
\pgfsetfillcolor{textcolor}%
\pgftext[x=1.159491in,y=0.467901in,,top]{\color{textcolor}\rmfamily\fontsize{10.000000}{12.000000}\selectfont 6}%
\end{pgfscope}%
\begin{pgfscope}%
\pgfsetbuttcap%
\pgfsetroundjoin%
\definecolor{currentfill}{rgb}{0.000000,0.000000,0.000000}%
\pgfsetfillcolor{currentfill}%
\pgfsetlinewidth{0.803000pt}%
\definecolor{currentstroke}{rgb}{0.000000,0.000000,0.000000}%
\pgfsetstrokecolor{currentstroke}%
\pgfsetdash{}{0pt}%
\pgfsys@defobject{currentmarker}{\pgfqpoint{0.000000in}{-0.048611in}}{\pgfqpoint{0.000000in}{0.000000in}}{%
\pgfpathmoveto{\pgfqpoint{0.000000in}{0.000000in}}%
\pgfpathlineto{\pgfqpoint{0.000000in}{-0.048611in}}%
\pgfusepath{stroke,fill}%
}%
\begin{pgfscope}%
\pgfsys@transformshift{1.519728in}{0.565123in}%
\pgfsys@useobject{currentmarker}{}%
\end{pgfscope}%
\end{pgfscope}%
\begin{pgfscope}%
\definecolor{textcolor}{rgb}{0.000000,0.000000,0.000000}%
\pgfsetstrokecolor{textcolor}%
\pgfsetfillcolor{textcolor}%
\pgftext[x=1.519728in,y=0.467901in,,top]{\color{textcolor}\rmfamily\fontsize{10.000000}{12.000000}\selectfont 10}%
\end{pgfscope}%
\begin{pgfscope}%
\pgfsetbuttcap%
\pgfsetroundjoin%
\definecolor{currentfill}{rgb}{0.000000,0.000000,0.000000}%
\pgfsetfillcolor{currentfill}%
\pgfsetlinewidth{0.803000pt}%
\definecolor{currentstroke}{rgb}{0.000000,0.000000,0.000000}%
\pgfsetstrokecolor{currentstroke}%
\pgfsetdash{}{0pt}%
\pgfsys@defobject{currentmarker}{\pgfqpoint{0.000000in}{-0.048611in}}{\pgfqpoint{0.000000in}{0.000000in}}{%
\pgfpathmoveto{\pgfqpoint{0.000000in}{0.000000in}}%
\pgfpathlineto{\pgfqpoint{0.000000in}{-0.048611in}}%
\pgfusepath{stroke,fill}%
}%
\begin{pgfscope}%
\pgfsys@transformshift{1.879964in}{0.565123in}%
\pgfsys@useobject{currentmarker}{}%
\end{pgfscope}%
\end{pgfscope}%
\begin{pgfscope}%
\definecolor{textcolor}{rgb}{0.000000,0.000000,0.000000}%
\pgfsetstrokecolor{textcolor}%
\pgfsetfillcolor{textcolor}%
\pgftext[x=1.879964in,y=0.467901in,,top]{\color{textcolor}\rmfamily\fontsize{10.000000}{12.000000}\selectfont 14}%
\end{pgfscope}%
\begin{pgfscope}%
\pgfsetbuttcap%
\pgfsetroundjoin%
\definecolor{currentfill}{rgb}{0.000000,0.000000,0.000000}%
\pgfsetfillcolor{currentfill}%
\pgfsetlinewidth{0.803000pt}%
\definecolor{currentstroke}{rgb}{0.000000,0.000000,0.000000}%
\pgfsetstrokecolor{currentstroke}%
\pgfsetdash{}{0pt}%
\pgfsys@defobject{currentmarker}{\pgfqpoint{0.000000in}{-0.048611in}}{\pgfqpoint{0.000000in}{0.000000in}}{%
\pgfpathmoveto{\pgfqpoint{0.000000in}{0.000000in}}%
\pgfpathlineto{\pgfqpoint{0.000000in}{-0.048611in}}%
\pgfusepath{stroke,fill}%
}%
\begin{pgfscope}%
\pgfsys@transformshift{2.240201in}{0.565123in}%
\pgfsys@useobject{currentmarker}{}%
\end{pgfscope}%
\end{pgfscope}%
\begin{pgfscope}%
\definecolor{textcolor}{rgb}{0.000000,0.000000,0.000000}%
\pgfsetstrokecolor{textcolor}%
\pgfsetfillcolor{textcolor}%
\pgftext[x=2.240201in,y=0.467901in,,top]{\color{textcolor}\rmfamily\fontsize{10.000000}{12.000000}\selectfont 18}%
\end{pgfscope}%
\begin{pgfscope}%
\pgfsetbuttcap%
\pgfsetroundjoin%
\definecolor{currentfill}{rgb}{0.000000,0.000000,0.000000}%
\pgfsetfillcolor{currentfill}%
\pgfsetlinewidth{0.803000pt}%
\definecolor{currentstroke}{rgb}{0.000000,0.000000,0.000000}%
\pgfsetstrokecolor{currentstroke}%
\pgfsetdash{}{0pt}%
\pgfsys@defobject{currentmarker}{\pgfqpoint{0.000000in}{-0.048611in}}{\pgfqpoint{0.000000in}{0.000000in}}{%
\pgfpathmoveto{\pgfqpoint{0.000000in}{0.000000in}}%
\pgfpathlineto{\pgfqpoint{0.000000in}{-0.048611in}}%
\pgfusepath{stroke,fill}%
}%
\begin{pgfscope}%
\pgfsys@transformshift{2.780555in}{0.565123in}%
\pgfsys@useobject{currentmarker}{}%
\end{pgfscope}%
\end{pgfscope}%
\begin{pgfscope}%
\definecolor{textcolor}{rgb}{0.000000,0.000000,0.000000}%
\pgfsetstrokecolor{textcolor}%
\pgfsetfillcolor{textcolor}%
\pgftext[x=2.780555in,y=0.467901in,,top]{\color{textcolor}\rmfamily\fontsize{10.000000}{12.000000}\selectfont 24}%
\end{pgfscope}%
\begin{pgfscope}%
\definecolor{textcolor}{rgb}{0.000000,0.000000,0.000000}%
\pgfsetstrokecolor{textcolor}%
\pgfsetfillcolor{textcolor}%
\pgftext[x=1.699846in,y=0.288889in,,top]{\color{textcolor}\rmfamily\fontsize{10.000000}{12.000000}\selectfont Angle [deg]}%
\end{pgfscope}%
\begin{pgfscope}%
\pgfsetbuttcap%
\pgfsetroundjoin%
\definecolor{currentfill}{rgb}{0.000000,0.000000,0.000000}%
\pgfsetfillcolor{currentfill}%
\pgfsetlinewidth{0.803000pt}%
\definecolor{currentstroke}{rgb}{0.000000,0.000000,0.000000}%
\pgfsetstrokecolor{currentstroke}%
\pgfsetdash{}{0pt}%
\pgfsys@defobject{currentmarker}{\pgfqpoint{-0.048611in}{0.000000in}}{\pgfqpoint{-0.000000in}{0.000000in}}{%
\pgfpathmoveto{\pgfqpoint{-0.000000in}{0.000000in}}%
\pgfpathlineto{\pgfqpoint{-0.048611in}{0.000000in}}%
\pgfusepath{stroke,fill}%
}%
\begin{pgfscope}%
\pgfsys@transformshift{0.619136in}{0.565123in}%
\pgfsys@useobject{currentmarker}{}%
\end{pgfscope}%
\end{pgfscope}%
\begin{pgfscope}%
\definecolor{textcolor}{rgb}{0.000000,0.000000,0.000000}%
\pgfsetstrokecolor{textcolor}%
\pgfsetfillcolor{textcolor}%
\pgftext[x=0.344444in, y=0.516898in, left, base]{\color{textcolor}\rmfamily\fontsize{10.000000}{12.000000}\selectfont \(\displaystyle {0.0}\)}%
\end{pgfscope}%
\begin{pgfscope}%
\pgfsetbuttcap%
\pgfsetroundjoin%
\definecolor{currentfill}{rgb}{0.000000,0.000000,0.000000}%
\pgfsetfillcolor{currentfill}%
\pgfsetlinewidth{0.803000pt}%
\definecolor{currentstroke}{rgb}{0.000000,0.000000,0.000000}%
\pgfsetstrokecolor{currentstroke}%
\pgfsetdash{}{0pt}%
\pgfsys@defobject{currentmarker}{\pgfqpoint{-0.048611in}{0.000000in}}{\pgfqpoint{-0.000000in}{0.000000in}}{%
\pgfpathmoveto{\pgfqpoint{-0.000000in}{0.000000in}}%
\pgfpathlineto{\pgfqpoint{-0.048611in}{0.000000in}}%
\pgfusepath{stroke,fill}%
}%
\begin{pgfscope}%
\pgfsys@transformshift{0.619136in}{1.028635in}%
\pgfsys@useobject{currentmarker}{}%
\end{pgfscope}%
\end{pgfscope}%
\begin{pgfscope}%
\definecolor{textcolor}{rgb}{0.000000,0.000000,0.000000}%
\pgfsetstrokecolor{textcolor}%
\pgfsetfillcolor{textcolor}%
\pgftext[x=0.344444in, y=0.980410in, left, base]{\color{textcolor}\rmfamily\fontsize{10.000000}{12.000000}\selectfont \(\displaystyle {0.2}\)}%
\end{pgfscope}%
\begin{pgfscope}%
\pgfsetbuttcap%
\pgfsetroundjoin%
\definecolor{currentfill}{rgb}{0.000000,0.000000,0.000000}%
\pgfsetfillcolor{currentfill}%
\pgfsetlinewidth{0.803000pt}%
\definecolor{currentstroke}{rgb}{0.000000,0.000000,0.000000}%
\pgfsetstrokecolor{currentstroke}%
\pgfsetdash{}{0pt}%
\pgfsys@defobject{currentmarker}{\pgfqpoint{-0.048611in}{0.000000in}}{\pgfqpoint{-0.000000in}{0.000000in}}{%
\pgfpathmoveto{\pgfqpoint{-0.000000in}{0.000000in}}%
\pgfpathlineto{\pgfqpoint{-0.048611in}{0.000000in}}%
\pgfusepath{stroke,fill}%
}%
\begin{pgfscope}%
\pgfsys@transformshift{0.619136in}{1.492147in}%
\pgfsys@useobject{currentmarker}{}%
\end{pgfscope}%
\end{pgfscope}%
\begin{pgfscope}%
\definecolor{textcolor}{rgb}{0.000000,0.000000,0.000000}%
\pgfsetstrokecolor{textcolor}%
\pgfsetfillcolor{textcolor}%
\pgftext[x=0.344444in, y=1.443922in, left, base]{\color{textcolor}\rmfamily\fontsize{10.000000}{12.000000}\selectfont \(\displaystyle {0.4}\)}%
\end{pgfscope}%
\begin{pgfscope}%
\pgfsetbuttcap%
\pgfsetroundjoin%
\definecolor{currentfill}{rgb}{0.000000,0.000000,0.000000}%
\pgfsetfillcolor{currentfill}%
\pgfsetlinewidth{0.803000pt}%
\definecolor{currentstroke}{rgb}{0.000000,0.000000,0.000000}%
\pgfsetstrokecolor{currentstroke}%
\pgfsetdash{}{0pt}%
\pgfsys@defobject{currentmarker}{\pgfqpoint{-0.048611in}{0.000000in}}{\pgfqpoint{-0.000000in}{0.000000in}}{%
\pgfpathmoveto{\pgfqpoint{-0.000000in}{0.000000in}}%
\pgfpathlineto{\pgfqpoint{-0.048611in}{0.000000in}}%
\pgfusepath{stroke,fill}%
}%
\begin{pgfscope}%
\pgfsys@transformshift{0.619136in}{1.955659in}%
\pgfsys@useobject{currentmarker}{}%
\end{pgfscope}%
\end{pgfscope}%
\begin{pgfscope}%
\definecolor{textcolor}{rgb}{0.000000,0.000000,0.000000}%
\pgfsetstrokecolor{textcolor}%
\pgfsetfillcolor{textcolor}%
\pgftext[x=0.344444in, y=1.907433in, left, base]{\color{textcolor}\rmfamily\fontsize{10.000000}{12.000000}\selectfont \(\displaystyle {0.6}\)}%
\end{pgfscope}%
\begin{pgfscope}%
\pgfsetbuttcap%
\pgfsetroundjoin%
\definecolor{currentfill}{rgb}{0.000000,0.000000,0.000000}%
\pgfsetfillcolor{currentfill}%
\pgfsetlinewidth{0.803000pt}%
\definecolor{currentstroke}{rgb}{0.000000,0.000000,0.000000}%
\pgfsetstrokecolor{currentstroke}%
\pgfsetdash{}{0pt}%
\pgfsys@defobject{currentmarker}{\pgfqpoint{-0.048611in}{0.000000in}}{\pgfqpoint{-0.000000in}{0.000000in}}{%
\pgfpathmoveto{\pgfqpoint{-0.000000in}{0.000000in}}%
\pgfpathlineto{\pgfqpoint{-0.048611in}{0.000000in}}%
\pgfusepath{stroke,fill}%
}%
\begin{pgfscope}%
\pgfsys@transformshift{0.619136in}{2.419170in}%
\pgfsys@useobject{currentmarker}{}%
\end{pgfscope}%
\end{pgfscope}%
\begin{pgfscope}%
\definecolor{textcolor}{rgb}{0.000000,0.000000,0.000000}%
\pgfsetstrokecolor{textcolor}%
\pgfsetfillcolor{textcolor}%
\pgftext[x=0.344444in, y=2.370945in, left, base]{\color{textcolor}\rmfamily\fontsize{10.000000}{12.000000}\selectfont \(\displaystyle {0.8}\)}%
\end{pgfscope}%
\begin{pgfscope}%
\definecolor{textcolor}{rgb}{0.000000,0.000000,0.000000}%
\pgfsetstrokecolor{textcolor}%
\pgfsetfillcolor{textcolor}%
\pgftext[x=0.288889in,y=1.608025in,,bottom,rotate=90.000000]{\color{textcolor}\rmfamily\fontsize{10.000000}{12.000000}\selectfont Moment [Nmm]}%
\end{pgfscope}%
\begin{pgfscope}%
\pgfpathrectangle{\pgfqpoint{0.619136in}{0.565123in}}{\pgfqpoint{2.161419in}{2.085803in}}%
\pgfusepath{clip}%
\pgfsetbuttcap%
\pgfsetroundjoin%
\pgfsetlinewidth{1.505625pt}%
\definecolor{currentstroke}{rgb}{0.827451,0.827451,0.827451}%
\pgfsetstrokecolor{currentstroke}%
\pgfsetdash{}{0pt}%
\pgfpathmoveto{\pgfqpoint{1.879964in}{0.565123in}}%
\pgfpathlineto{\pgfqpoint{1.879964in}{2.650926in}}%
\pgfusepath{stroke}%
\end{pgfscope}%
\begin{pgfscope}%
\pgfpathrectangle{\pgfqpoint{0.619136in}{0.565123in}}{\pgfqpoint{2.161419in}{2.085803in}}%
\pgfusepath{clip}%
\pgfsetbuttcap%
\pgfsetroundjoin%
\pgfsetlinewidth{1.505625pt}%
\definecolor{currentstroke}{rgb}{0.827451,0.827451,0.827451}%
\pgfsetstrokecolor{currentstroke}%
\pgfsetdash{}{0pt}%
\pgfpathmoveto{\pgfqpoint{2.240201in}{0.565123in}}%
\pgfpathlineto{\pgfqpoint{2.240201in}{2.650926in}}%
\pgfusepath{stroke}%
\end{pgfscope}%
\begin{pgfscope}%
\pgfpathrectangle{\pgfqpoint{0.619136in}{0.565123in}}{\pgfqpoint{2.161419in}{2.085803in}}%
\pgfusepath{clip}%
\pgfsetrectcap%
\pgfsetroundjoin%
\pgfsetlinewidth{1.505625pt}%
\definecolor{currentstroke}{rgb}{0.000000,0.000000,0.000000}%
\pgfsetstrokecolor{currentstroke}%
\pgfsetdash{}{0pt}%
\pgfpathmoveto{\pgfqpoint{0.619136in}{0.565123in}}%
\pgfpathlineto{\pgfqpoint{1.005886in}{0.727188in}}%
\pgfpathlineto{\pgfqpoint{1.155282in}{0.788808in}}%
\pgfpathlineto{\pgfqpoint{1.334561in}{0.861370in}}%
\pgfpathlineto{\pgfqpoint{1.764823in}{1.031300in}}%
\pgfpathlineto{\pgfqpoint{1.818606in}{1.055032in}}%
\pgfpathlineto{\pgfqpoint{1.845500in}{1.068924in}}%
\pgfpathlineto{\pgfqpoint{1.885836in}{1.092565in}}%
\pgfpathlineto{\pgfqpoint{1.919453in}{1.114793in}}%
\pgfpathlineto{\pgfqpoint{1.939618in}{1.130995in}}%
\pgfpathlineto{\pgfqpoint{1.966378in}{1.155980in}}%
\pgfpathlineto{\pgfqpoint{2.003551in}{1.194350in}}%
\pgfpathlineto{\pgfqpoint{2.052065in}{1.248372in}}%
\pgfpathlineto{\pgfqpoint{2.060084in}{1.257635in}}%
\pgfpathlineto{\pgfqpoint{2.132128in}{1.344062in}}%
\pgfpathlineto{\pgfqpoint{2.204177in}{1.435635in}}%
\pgfpathlineto{\pgfqpoint{2.245602in}{1.491681in}}%
\pgfpathlineto{\pgfqpoint{2.296227in}{1.565284in}}%
\pgfpathlineto{\pgfqpoint{2.342909in}{1.637254in}}%
\pgfpathlineto{\pgfqpoint{2.398937in}{1.728445in}}%
\pgfpathlineto{\pgfqpoint{2.454959in}{1.824042in}}%
\pgfpathlineto{\pgfqpoint{2.516585in}{1.933403in}}%
\pgfpathlineto{\pgfqpoint{2.578211in}{2.047042in}}%
\pgfpathlineto{\pgfqpoint{2.636699in}{2.160401in}}%
\pgfpathlineto{\pgfqpoint{2.709306in}{2.305981in}}%
\pgfpathlineto{\pgfqpoint{2.780555in}{2.451398in}}%
\pgfpathlineto{\pgfqpoint{2.780555in}{2.451398in}}%
\pgfusepath{stroke}%
\end{pgfscope}%
\begin{pgfscope}%
\pgfpathrectangle{\pgfqpoint{0.619136in}{0.565123in}}{\pgfqpoint{2.161419in}{2.085803in}}%
\pgfusepath{clip}%
\pgfsetbuttcap%
\pgfsetroundjoin%
\pgfsetlinewidth{1.505625pt}%
\definecolor{currentstroke}{rgb}{0.000000,0.000000,0.000000}%
\pgfsetstrokecolor{currentstroke}%
\pgfsetdash{{5.550000pt}{2.400000pt}}{0.000000pt}%
\pgfpathmoveto{\pgfqpoint{0.619136in}{0.565123in}}%
\pgfpathlineto{\pgfqpoint{0.700190in}{0.584986in}}%
\pgfpathlineto{\pgfqpoint{0.797453in}{0.608763in}}%
\pgfpathlineto{\pgfqpoint{0.914170in}{0.637116in}}%
\pgfpathlineto{\pgfqpoint{1.054230in}{0.670715in}}%
\pgfpathlineto{\pgfqpoint{1.222300in}{0.710155in}}%
\pgfpathlineto{\pgfqpoint{1.423989in}{0.755842in}}%
\pgfpathlineto{\pgfqpoint{1.625673in}{0.799992in}}%
\pgfpathlineto{\pgfqpoint{1.726515in}{0.821988in}}%
\pgfpathlineto{\pgfqpoint{1.827363in}{0.844841in}}%
\pgfpathlineto{\pgfqpoint{1.877781in}{0.857252in}}%
\pgfpathlineto{\pgfqpoint{1.928205in}{0.871382in}}%
\pgfpathlineto{\pgfqpoint{1.953416in}{0.879776in}}%
\pgfpathlineto{\pgfqpoint{1.966022in}{0.884685in}}%
\pgfpathlineto{\pgfqpoint{1.978628in}{0.890321in}}%
\pgfpathlineto{\pgfqpoint{1.984928in}{0.893380in}}%
\pgfpathlineto{\pgfqpoint{1.991229in}{0.896611in}}%
\pgfpathlineto{\pgfqpoint{1.997534in}{0.900015in}}%
\pgfpathlineto{\pgfqpoint{2.000687in}{0.901777in}}%
\pgfpathlineto{\pgfqpoint{2.004464in}{0.903946in}}%
\pgfpathlineto{\pgfqpoint{2.009005in}{0.906641in}}%
\pgfpathlineto{\pgfqpoint{2.013541in}{0.909445in}}%
\pgfpathlineto{\pgfqpoint{2.018082in}{0.912347in}}%
\pgfpathlineto{\pgfqpoint{2.022617in}{0.915327in}}%
\pgfpathlineto{\pgfqpoint{2.027158in}{0.918396in}}%
\pgfpathlineto{\pgfqpoint{2.031694in}{0.921564in}}%
\pgfpathlineto{\pgfqpoint{2.036234in}{0.924836in}}%
\pgfpathlineto{\pgfqpoint{2.040770in}{0.928201in}}%
\pgfpathlineto{\pgfqpoint{2.045306in}{0.931648in}}%
\pgfpathlineto{\pgfqpoint{2.049847in}{0.935163in}}%
\pgfpathlineto{\pgfqpoint{2.055290in}{0.939469in}}%
\pgfpathlineto{\pgfqpoint{2.060739in}{0.943863in}}%
\pgfpathlineto{\pgfqpoint{2.067272in}{0.949256in}}%
\pgfpathlineto{\pgfqpoint{2.073804in}{0.954791in}}%
\pgfpathlineto{\pgfqpoint{2.080342in}{0.960473in}}%
\pgfpathlineto{\pgfqpoint{2.086875in}{0.966300in}}%
\pgfpathlineto{\pgfqpoint{2.093413in}{0.972260in}}%
\pgfpathlineto{\pgfqpoint{2.099945in}{0.978356in}}%
\pgfpathlineto{\pgfqpoint{2.106478in}{0.984581in}}%
\pgfpathlineto{\pgfqpoint{2.113015in}{0.990917in}}%
\pgfpathlineto{\pgfqpoint{2.119548in}{0.997355in}}%
\pgfpathlineto{\pgfqpoint{2.126086in}{1.003891in}}%
\pgfpathlineto{\pgfqpoint{2.132618in}{1.010528in}}%
\pgfpathlineto{\pgfqpoint{2.139151in}{1.017272in}}%
\pgfpathlineto{\pgfqpoint{2.145689in}{1.024118in}}%
\pgfpathlineto{\pgfqpoint{2.152221in}{1.031062in}}%
\pgfpathlineto{\pgfqpoint{2.158759in}{1.038091in}}%
\pgfpathlineto{\pgfqpoint{2.165291in}{1.045201in}}%
\pgfpathlineto{\pgfqpoint{2.171824in}{1.052390in}}%
\pgfpathlineto{\pgfqpoint{2.178362in}{1.059656in}}%
\pgfpathlineto{\pgfqpoint{2.184894in}{1.066998in}}%
\pgfpathlineto{\pgfqpoint{2.191432in}{1.074418in}}%
\pgfpathlineto{\pgfqpoint{2.197965in}{1.081911in}}%
\pgfpathlineto{\pgfqpoint{2.204497in}{1.089473in}}%
\pgfpathlineto{\pgfqpoint{2.211035in}{1.097107in}}%
\pgfpathlineto{\pgfqpoint{2.217567in}{1.104816in}}%
\pgfpathlineto{\pgfqpoint{2.224105in}{1.112610in}}%
\pgfpathlineto{\pgfqpoint{2.230638in}{1.120492in}}%
\pgfpathlineto{\pgfqpoint{2.237170in}{1.128466in}}%
\pgfpathlineto{\pgfqpoint{2.240199in}{1.132188in}}%
\pgfpathlineto{\pgfqpoint{2.321253in}{1.236295in}}%
\pgfpathlineto{\pgfqpoint{2.361779in}{1.290814in}}%
\pgfpathlineto{\pgfqpoint{2.382043in}{1.318580in}}%
\pgfpathlineto{\pgfqpoint{2.392177in}{1.332564in}}%
\pgfpathlineto{\pgfqpoint{2.402306in}{1.346602in}}%
\pgfpathlineto{\pgfqpoint{2.412440in}{1.360683in}}%
\pgfpathlineto{\pgfqpoint{2.422569in}{1.374802in}}%
\pgfpathlineto{\pgfqpoint{2.427637in}{1.381875in}}%
\pgfpathlineto{\pgfqpoint{2.433715in}{1.390369in}}%
\pgfpathlineto{\pgfqpoint{2.441011in}{1.400575in}}%
\pgfpathlineto{\pgfqpoint{2.449763in}{1.412842in}}%
\pgfpathlineto{\pgfqpoint{2.458519in}{1.425128in}}%
\pgfpathlineto{\pgfqpoint{2.467271in}{1.437436in}}%
\pgfpathlineto{\pgfqpoint{2.476027in}{1.449766in}}%
\pgfpathlineto{\pgfqpoint{2.484778in}{1.462116in}}%
\pgfpathlineto{\pgfqpoint{2.495284in}{1.476969in}}%
\pgfpathlineto{\pgfqpoint{2.507890in}{1.494889in}}%
\pgfpathlineto{\pgfqpoint{2.520496in}{1.512975in}}%
\pgfpathlineto{\pgfqpoint{2.526796in}{1.522078in}}%
\pgfpathlineto{\pgfqpoint{2.533102in}{1.531214in}}%
\pgfpathlineto{\pgfqpoint{2.539402in}{1.540371in}}%
\pgfpathlineto{\pgfqpoint{2.545703in}{1.549537in}}%
\pgfpathlineto{\pgfqpoint{2.552008in}{1.558716in}}%
\pgfpathlineto{\pgfqpoint{2.559573in}{1.569746in}}%
\pgfpathlineto{\pgfqpoint{2.568644in}{1.583021in}}%
\pgfpathlineto{\pgfqpoint{2.577721in}{1.596349in}}%
\pgfpathlineto{\pgfqpoint{2.582261in}{1.603026in}}%
\pgfpathlineto{\pgfqpoint{2.587705in}{1.611045in}}%
\pgfpathlineto{\pgfqpoint{2.594243in}{1.620674in}}%
\pgfpathlineto{\pgfqpoint{2.602081in}{1.632236in}}%
\pgfpathlineto{\pgfqpoint{2.611493in}{1.646119in}}%
\pgfpathlineto{\pgfqpoint{2.622783in}{1.662784in}}%
\pgfpathlineto{\pgfqpoint{2.636333in}{1.682794in}}%
\pgfpathlineto{\pgfqpoint{2.649883in}{1.702834in}}%
\pgfpathlineto{\pgfqpoint{2.663433in}{1.722980in}}%
\pgfpathlineto{\pgfqpoint{2.676984in}{1.743189in}}%
\pgfpathlineto{\pgfqpoint{2.690534in}{1.763406in}}%
\pgfpathlineto{\pgfqpoint{2.704084in}{1.783589in}}%
\pgfpathlineto{\pgfqpoint{2.717634in}{1.803733in}}%
\pgfpathlineto{\pgfqpoint{2.731184in}{1.823843in}}%
\pgfpathlineto{\pgfqpoint{2.737959in}{1.833896in}}%
\pgfpathlineto{\pgfqpoint{2.746092in}{1.845976in}}%
\pgfpathlineto{\pgfqpoint{2.754219in}{1.858080in}}%
\pgfpathlineto{\pgfqpoint{2.762351in}{1.870217in}}%
\pgfpathlineto{\pgfqpoint{2.770483in}{1.882398in}}%
\pgfpathlineto{\pgfqpoint{2.778610in}{1.894623in}}%
\pgfpathlineto{\pgfqpoint{2.780555in}{1.897548in}}%
\pgfusepath{stroke}%
\end{pgfscope}%
\begin{pgfscope}%
\pgfsetrectcap%
\pgfsetmiterjoin%
\pgfsetlinewidth{0.803000pt}%
\definecolor{currentstroke}{rgb}{0.000000,0.000000,0.000000}%
\pgfsetstrokecolor{currentstroke}%
\pgfsetdash{}{0pt}%
\pgfpathmoveto{\pgfqpoint{0.619136in}{0.565123in}}%
\pgfpathlineto{\pgfqpoint{0.619136in}{2.650926in}}%
\pgfusepath{stroke}%
\end{pgfscope}%
\begin{pgfscope}%
\pgfsetrectcap%
\pgfsetmiterjoin%
\pgfsetlinewidth{0.803000pt}%
\definecolor{currentstroke}{rgb}{0.000000,0.000000,0.000000}%
\pgfsetstrokecolor{currentstroke}%
\pgfsetdash{}{0pt}%
\pgfpathmoveto{\pgfqpoint{2.780555in}{0.565123in}}%
\pgfpathlineto{\pgfqpoint{2.780555in}{2.650926in}}%
\pgfusepath{stroke}%
\end{pgfscope}%
\begin{pgfscope}%
\pgfsetrectcap%
\pgfsetmiterjoin%
\pgfsetlinewidth{0.803000pt}%
\definecolor{currentstroke}{rgb}{0.000000,0.000000,0.000000}%
\pgfsetstrokecolor{currentstroke}%
\pgfsetdash{}{0pt}%
\pgfpathmoveto{\pgfqpoint{0.619136in}{0.565123in}}%
\pgfpathlineto{\pgfqpoint{2.780555in}{0.565123in}}%
\pgfusepath{stroke}%
\end{pgfscope}%
\begin{pgfscope}%
\pgfsetrectcap%
\pgfsetmiterjoin%
\pgfsetlinewidth{0.803000pt}%
\definecolor{currentstroke}{rgb}{0.000000,0.000000,0.000000}%
\pgfsetstrokecolor{currentstroke}%
\pgfsetdash{}{0pt}%
\pgfpathmoveto{\pgfqpoint{0.619136in}{2.650926in}}%
\pgfpathlineto{\pgfqpoint{2.780555in}{2.650926in}}%
\pgfusepath{stroke}%
\end{pgfscope}%
\begin{pgfscope}%
\definecolor{textcolor}{rgb}{0.000000,0.000000,0.000000}%
\pgfsetstrokecolor{textcolor}%
\pgfsetfillcolor{textcolor}%
\pgftext[x=1.699846in,y=2.734260in,,base]{\color{textcolor}\rmfamily\fontsize{12.000000}{14.400000}\selectfont Moment vs. Angle}%
\end{pgfscope}%
\begin{pgfscope}%
\pgfsetbuttcap%
\pgfsetmiterjoin%
\definecolor{currentfill}{rgb}{1.000000,1.000000,1.000000}%
\pgfsetfillcolor{currentfill}%
\pgfsetfillopacity{0.800000}%
\pgfsetlinewidth{1.003750pt}%
\definecolor{currentstroke}{rgb}{0.800000,0.800000,0.800000}%
\pgfsetstrokecolor{currentstroke}%
\pgfsetstrokeopacity{0.800000}%
\pgfsetdash{}{0pt}%
\pgfpathmoveto{\pgfqpoint{0.716359in}{2.152470in}}%
\pgfpathlineto{\pgfqpoint{1.867513in}{2.152470in}}%
\pgfpathquadraticcurveto{\pgfqpoint{1.895291in}{2.152470in}}{\pgfqpoint{1.895291in}{2.180247in}}%
\pgfpathlineto{\pgfqpoint{1.895291in}{2.553704in}}%
\pgfpathquadraticcurveto{\pgfqpoint{1.895291in}{2.581482in}}{\pgfqpoint{1.867513in}{2.581482in}}%
\pgfpathlineto{\pgfqpoint{0.716359in}{2.581482in}}%
\pgfpathquadraticcurveto{\pgfqpoint{0.688581in}{2.581482in}}{\pgfqpoint{0.688581in}{2.553704in}}%
\pgfpathlineto{\pgfqpoint{0.688581in}{2.180247in}}%
\pgfpathquadraticcurveto{\pgfqpoint{0.688581in}{2.152470in}}{\pgfqpoint{0.716359in}{2.152470in}}%
\pgfpathlineto{\pgfqpoint{0.716359in}{2.152470in}}%
\pgfpathclose%
\pgfusepath{stroke,fill}%
\end{pgfscope}%
\begin{pgfscope}%
\pgfsetrectcap%
\pgfsetroundjoin%
\pgfsetlinewidth{1.505625pt}%
\definecolor{currentstroke}{rgb}{0.000000,0.000000,0.000000}%
\pgfsetstrokecolor{currentstroke}%
\pgfsetdash{}{0pt}%
\pgfpathmoveto{\pgfqpoint{0.744136in}{2.477315in}}%
\pgfpathlineto{\pgfqpoint{0.883025in}{2.477315in}}%
\pgfpathlineto{\pgfqpoint{1.021914in}{2.477315in}}%
\pgfusepath{stroke}%
\end{pgfscope}%
\begin{pgfscope}%
\definecolor{textcolor}{rgb}{0.000000,0.000000,0.000000}%
\pgfsetstrokecolor{textcolor}%
\pgfsetfillcolor{textcolor}%
\pgftext[x=1.133025in,y=2.428704in,left,base]{\color{textcolor}\rmfamily\fontsize{10.000000}{12.000000}\selectfont \(\displaystyle \alpha_0=14\)\,deg}%
\end{pgfscope}%
\begin{pgfscope}%
\pgfsetbuttcap%
\pgfsetroundjoin%
\pgfsetlinewidth{1.505625pt}%
\definecolor{currentstroke}{rgb}{0.000000,0.000000,0.000000}%
\pgfsetstrokecolor{currentstroke}%
\pgfsetdash{{5.550000pt}{2.400000pt}}{0.000000pt}%
\pgfpathmoveto{\pgfqpoint{0.744136in}{2.283642in}}%
\pgfpathlineto{\pgfqpoint{0.883025in}{2.283642in}}%
\pgfpathlineto{\pgfqpoint{1.021914in}{2.283642in}}%
\pgfusepath{stroke}%
\end{pgfscope}%
\begin{pgfscope}%
\definecolor{textcolor}{rgb}{0.000000,0.000000,0.000000}%
\pgfsetstrokecolor{textcolor}%
\pgfsetfillcolor{textcolor}%
\pgftext[x=1.133025in,y=2.235031in,left,base]{\color{textcolor}\rmfamily\fontsize{10.000000}{12.000000}\selectfont \(\displaystyle \alpha_0=18\)\,deg}%
\end{pgfscope}%
\end{pgfpicture}%
\makeatother%
\endgroup%

%% file: hook_force_curves_case_NX200_NY100_bend_v18_c13_v2.pgf
\begingroup%
\makeatletter%
\begin{pgfpicture}%
\pgfpathrectangle{\pgfpointorigin}{\pgfqpoint{3.000000in}{3.000000in}}%
\pgfusepath{use as bounding box, clip}%
\begin{pgfscope}%
\pgfsetbuttcap%
\pgfsetmiterjoin%
\definecolor{currentfill}{rgb}{1.000000,1.000000,1.000000}%
\pgfsetfillcolor{currentfill}%
\pgfsetlinewidth{0.000000pt}%
\definecolor{currentstroke}{rgb}{1.000000,1.000000,1.000000}%
\pgfsetstrokecolor{currentstroke}%
\pgfsetdash{}{0pt}%
\pgfpathmoveto{\pgfqpoint{0.000000in}{0.000000in}}%
\pgfpathlineto{\pgfqpoint{3.000000in}{0.000000in}}%
\pgfpathlineto{\pgfqpoint{3.000000in}{3.000000in}}%
\pgfpathlineto{\pgfqpoint{0.000000in}{3.000000in}}%
\pgfpathlineto{\pgfqpoint{0.000000in}{0.000000in}}%
\pgfpathclose%
\pgfusepath{fill}%
\end{pgfscope}%
\begin{pgfscope}%
\pgfsetbuttcap%
\pgfsetmiterjoin%
\definecolor{currentfill}{rgb}{1.000000,1.000000,1.000000}%
\pgfsetfillcolor{currentfill}%
\pgfsetlinewidth{0.000000pt}%
\definecolor{currentstroke}{rgb}{0.000000,0.000000,0.000000}%
\pgfsetstrokecolor{currentstroke}%
\pgfsetstrokeopacity{0.000000}%
\pgfsetdash{}{0pt}%
\pgfpathmoveto{\pgfqpoint{0.758026in}{0.565123in}}%
\pgfpathlineto{\pgfqpoint{2.780555in}{0.565123in}}%
\pgfpathlineto{\pgfqpoint{2.780555in}{2.650926in}}%
\pgfpathlineto{\pgfqpoint{0.758026in}{2.650926in}}%
\pgfpathlineto{\pgfqpoint{0.758026in}{0.565123in}}%
\pgfpathclose%
\pgfusepath{fill}%
\end{pgfscope}%
\begin{pgfscope}%
\pgfsetbuttcap%
\pgfsetroundjoin%
\definecolor{currentfill}{rgb}{0.000000,0.000000,0.000000}%
\pgfsetfillcolor{currentfill}%
\pgfsetlinewidth{0.803000pt}%
\definecolor{currentstroke}{rgb}{0.000000,0.000000,0.000000}%
\pgfsetstrokecolor{currentstroke}%
\pgfsetdash{}{0pt}%
\pgfsys@defobject{currentmarker}{\pgfqpoint{0.000000in}{-0.048611in}}{\pgfqpoint{0.000000in}{0.000000in}}{%
\pgfpathmoveto{\pgfqpoint{0.000000in}{0.000000in}}%
\pgfpathlineto{\pgfqpoint{0.000000in}{-0.048611in}}%
\pgfusepath{stroke,fill}%
}%
\begin{pgfscope}%
\pgfsys@transformshift{0.758026in}{0.565123in}%
\pgfsys@useobject{currentmarker}{}%
\end{pgfscope}%
\end{pgfscope}%
\begin{pgfscope}%
\definecolor{textcolor}{rgb}{0.000000,0.000000,0.000000}%
\pgfsetstrokecolor{textcolor}%
\pgfsetfillcolor{textcolor}%
\pgftext[x=0.758026in,y=0.467901in,,top]{\color{textcolor}\rmfamily\fontsize{10.000000}{12.000000}\selectfont 0}%
\end{pgfscope}%
\begin{pgfscope}%
\pgfsetbuttcap%
\pgfsetroundjoin%
\definecolor{currentfill}{rgb}{0.000000,0.000000,0.000000}%
\pgfsetfillcolor{currentfill}%
\pgfsetlinewidth{0.803000pt}%
\definecolor{currentstroke}{rgb}{0.000000,0.000000,0.000000}%
\pgfsetstrokecolor{currentstroke}%
\pgfsetdash{}{0pt}%
\pgfsys@defobject{currentmarker}{\pgfqpoint{0.000000in}{-0.048611in}}{\pgfqpoint{0.000000in}{0.000000in}}{%
\pgfpathmoveto{\pgfqpoint{0.000000in}{0.000000in}}%
\pgfpathlineto{\pgfqpoint{0.000000in}{-0.048611in}}%
\pgfusepath{stroke,fill}%
}%
\begin{pgfscope}%
\pgfsys@transformshift{1.263658in}{0.565123in}%
\pgfsys@useobject{currentmarker}{}%
\end{pgfscope}%
\end{pgfscope}%
\begin{pgfscope}%
\definecolor{textcolor}{rgb}{0.000000,0.000000,0.000000}%
\pgfsetstrokecolor{textcolor}%
\pgfsetfillcolor{textcolor}%
\pgftext[x=1.263658in,y=0.467901in,,top]{\color{textcolor}\rmfamily\fontsize{10.000000}{12.000000}\selectfont 6}%
\end{pgfscope}%
\begin{pgfscope}%
\pgfsetbuttcap%
\pgfsetroundjoin%
\definecolor{currentfill}{rgb}{0.000000,0.000000,0.000000}%
\pgfsetfillcolor{currentfill}%
\pgfsetlinewidth{0.803000pt}%
\definecolor{currentstroke}{rgb}{0.000000,0.000000,0.000000}%
\pgfsetstrokecolor{currentstroke}%
\pgfsetdash{}{0pt}%
\pgfsys@defobject{currentmarker}{\pgfqpoint{0.000000in}{-0.048611in}}{\pgfqpoint{0.000000in}{0.000000in}}{%
\pgfpathmoveto{\pgfqpoint{0.000000in}{0.000000in}}%
\pgfpathlineto{\pgfqpoint{0.000000in}{-0.048611in}}%
\pgfusepath{stroke,fill}%
}%
\begin{pgfscope}%
\pgfsys@transformshift{1.600746in}{0.565123in}%
\pgfsys@useobject{currentmarker}{}%
\end{pgfscope}%
\end{pgfscope}%
\begin{pgfscope}%
\definecolor{textcolor}{rgb}{0.000000,0.000000,0.000000}%
\pgfsetstrokecolor{textcolor}%
\pgfsetfillcolor{textcolor}%
\pgftext[x=1.600746in,y=0.467901in,,top]{\color{textcolor}\rmfamily\fontsize{10.000000}{12.000000}\selectfont 10}%
\end{pgfscope}%
\begin{pgfscope}%
\pgfsetbuttcap%
\pgfsetroundjoin%
\definecolor{currentfill}{rgb}{0.000000,0.000000,0.000000}%
\pgfsetfillcolor{currentfill}%
\pgfsetlinewidth{0.803000pt}%
\definecolor{currentstroke}{rgb}{0.000000,0.000000,0.000000}%
\pgfsetstrokecolor{currentstroke}%
\pgfsetdash{}{0pt}%
\pgfsys@defobject{currentmarker}{\pgfqpoint{0.000000in}{-0.048611in}}{\pgfqpoint{0.000000in}{0.000000in}}{%
\pgfpathmoveto{\pgfqpoint{0.000000in}{0.000000in}}%
\pgfpathlineto{\pgfqpoint{0.000000in}{-0.048611in}}%
\pgfusepath{stroke,fill}%
}%
\begin{pgfscope}%
\pgfsys@transformshift{1.937835in}{0.565123in}%
\pgfsys@useobject{currentmarker}{}%
\end{pgfscope}%
\end{pgfscope}%
\begin{pgfscope}%
\definecolor{textcolor}{rgb}{0.000000,0.000000,0.000000}%
\pgfsetstrokecolor{textcolor}%
\pgfsetfillcolor{textcolor}%
\pgftext[x=1.937835in,y=0.467901in,,top]{\color{textcolor}\rmfamily\fontsize{10.000000}{12.000000}\selectfont 14}%
\end{pgfscope}%
\begin{pgfscope}%
\pgfsetbuttcap%
\pgfsetroundjoin%
\definecolor{currentfill}{rgb}{0.000000,0.000000,0.000000}%
\pgfsetfillcolor{currentfill}%
\pgfsetlinewidth{0.803000pt}%
\definecolor{currentstroke}{rgb}{0.000000,0.000000,0.000000}%
\pgfsetstrokecolor{currentstroke}%
\pgfsetdash{}{0pt}%
\pgfsys@defobject{currentmarker}{\pgfqpoint{0.000000in}{-0.048611in}}{\pgfqpoint{0.000000in}{0.000000in}}{%
\pgfpathmoveto{\pgfqpoint{0.000000in}{0.000000in}}%
\pgfpathlineto{\pgfqpoint{0.000000in}{-0.048611in}}%
\pgfusepath{stroke,fill}%
}%
\begin{pgfscope}%
\pgfsys@transformshift{2.274923in}{0.565123in}%
\pgfsys@useobject{currentmarker}{}%
\end{pgfscope}%
\end{pgfscope}%
\begin{pgfscope}%
\definecolor{textcolor}{rgb}{0.000000,0.000000,0.000000}%
\pgfsetstrokecolor{textcolor}%
\pgfsetfillcolor{textcolor}%
\pgftext[x=2.274923in,y=0.467901in,,top]{\color{textcolor}\rmfamily\fontsize{10.000000}{12.000000}\selectfont 18}%
\end{pgfscope}%
\begin{pgfscope}%
\pgfsetbuttcap%
\pgfsetroundjoin%
\definecolor{currentfill}{rgb}{0.000000,0.000000,0.000000}%
\pgfsetfillcolor{currentfill}%
\pgfsetlinewidth{0.803000pt}%
\definecolor{currentstroke}{rgb}{0.000000,0.000000,0.000000}%
\pgfsetstrokecolor{currentstroke}%
\pgfsetdash{}{0pt}%
\pgfsys@defobject{currentmarker}{\pgfqpoint{0.000000in}{-0.048611in}}{\pgfqpoint{0.000000in}{0.000000in}}{%
\pgfpathmoveto{\pgfqpoint{0.000000in}{0.000000in}}%
\pgfpathlineto{\pgfqpoint{0.000000in}{-0.048611in}}%
\pgfusepath{stroke,fill}%
}%
\begin{pgfscope}%
\pgfsys@transformshift{2.780555in}{0.565123in}%
\pgfsys@useobject{currentmarker}{}%
\end{pgfscope}%
\end{pgfscope}%
\begin{pgfscope}%
\definecolor{textcolor}{rgb}{0.000000,0.000000,0.000000}%
\pgfsetstrokecolor{textcolor}%
\pgfsetfillcolor{textcolor}%
\pgftext[x=2.780555in,y=0.467901in,,top]{\color{textcolor}\rmfamily\fontsize{10.000000}{12.000000}\selectfont 24}%
\end{pgfscope}%
\begin{pgfscope}%
\definecolor{textcolor}{rgb}{0.000000,0.000000,0.000000}%
\pgfsetstrokecolor{textcolor}%
\pgfsetfillcolor{textcolor}%
\pgftext[x=1.769291in,y=0.288889in,,top]{\color{textcolor}\rmfamily\fontsize{10.000000}{12.000000}\selectfont Angle [deg]}%
\end{pgfscope}%
\begin{pgfscope}%
\pgfsetbuttcap%
\pgfsetroundjoin%
\definecolor{currentfill}{rgb}{0.000000,0.000000,0.000000}%
\pgfsetfillcolor{currentfill}%
\pgfsetlinewidth{0.803000pt}%
\definecolor{currentstroke}{rgb}{0.000000,0.000000,0.000000}%
\pgfsetstrokecolor{currentstroke}%
\pgfsetdash{}{0pt}%
\pgfsys@defobject{currentmarker}{\pgfqpoint{-0.048611in}{0.000000in}}{\pgfqpoint{-0.000000in}{0.000000in}}{%
\pgfpathmoveto{\pgfqpoint{-0.000000in}{0.000000in}}%
\pgfpathlineto{\pgfqpoint{-0.048611in}{0.000000in}}%
\pgfusepath{stroke,fill}%
}%
\begin{pgfscope}%
\pgfsys@transformshift{0.758026in}{0.565123in}%
\pgfsys@useobject{currentmarker}{}%
\end{pgfscope}%
\end{pgfscope}%
\begin{pgfscope}%
\definecolor{textcolor}{rgb}{0.000000,0.000000,0.000000}%
\pgfsetstrokecolor{textcolor}%
\pgfsetfillcolor{textcolor}%
\pgftext[x=0.344444in, y=0.516898in, left, base]{\color{textcolor}\rmfamily\fontsize{10.000000}{12.000000}\selectfont \(\displaystyle {0.000}\)}%
\end{pgfscope}%
\begin{pgfscope}%
\pgfsetbuttcap%
\pgfsetroundjoin%
\definecolor{currentfill}{rgb}{0.000000,0.000000,0.000000}%
\pgfsetfillcolor{currentfill}%
\pgfsetlinewidth{0.803000pt}%
\definecolor{currentstroke}{rgb}{0.000000,0.000000,0.000000}%
\pgfsetstrokecolor{currentstroke}%
\pgfsetdash{}{0pt}%
\pgfsys@defobject{currentmarker}{\pgfqpoint{-0.048611in}{0.000000in}}{\pgfqpoint{-0.000000in}{0.000000in}}{%
\pgfpathmoveto{\pgfqpoint{-0.000000in}{0.000000in}}%
\pgfpathlineto{\pgfqpoint{-0.048611in}{0.000000in}}%
\pgfusepath{stroke,fill}%
}%
\begin{pgfscope}%
\pgfsys@transformshift{0.758026in}{0.982284in}%
\pgfsys@useobject{currentmarker}{}%
\end{pgfscope}%
\end{pgfscope}%
\begin{pgfscope}%
\definecolor{textcolor}{rgb}{0.000000,0.000000,0.000000}%
\pgfsetstrokecolor{textcolor}%
\pgfsetfillcolor{textcolor}%
\pgftext[x=0.344444in, y=0.934059in, left, base]{\color{textcolor}\rmfamily\fontsize{10.000000}{12.000000}\selectfont \(\displaystyle {0.005}\)}%
\end{pgfscope}%
\begin{pgfscope}%
\pgfsetbuttcap%
\pgfsetroundjoin%
\definecolor{currentfill}{rgb}{0.000000,0.000000,0.000000}%
\pgfsetfillcolor{currentfill}%
\pgfsetlinewidth{0.803000pt}%
\definecolor{currentstroke}{rgb}{0.000000,0.000000,0.000000}%
\pgfsetstrokecolor{currentstroke}%
\pgfsetdash{}{0pt}%
\pgfsys@defobject{currentmarker}{\pgfqpoint{-0.048611in}{0.000000in}}{\pgfqpoint{-0.000000in}{0.000000in}}{%
\pgfpathmoveto{\pgfqpoint{-0.000000in}{0.000000in}}%
\pgfpathlineto{\pgfqpoint{-0.048611in}{0.000000in}}%
\pgfusepath{stroke,fill}%
}%
\begin{pgfscope}%
\pgfsys@transformshift{0.758026in}{1.399445in}%
\pgfsys@useobject{currentmarker}{}%
\end{pgfscope}%
\end{pgfscope}%
\begin{pgfscope}%
\definecolor{textcolor}{rgb}{0.000000,0.000000,0.000000}%
\pgfsetstrokecolor{textcolor}%
\pgfsetfillcolor{textcolor}%
\pgftext[x=0.344444in, y=1.351219in, left, base]{\color{textcolor}\rmfamily\fontsize{10.000000}{12.000000}\selectfont \(\displaystyle {0.010}\)}%
\end{pgfscope}%
\begin{pgfscope}%
\pgfsetbuttcap%
\pgfsetroundjoin%
\definecolor{currentfill}{rgb}{0.000000,0.000000,0.000000}%
\pgfsetfillcolor{currentfill}%
\pgfsetlinewidth{0.803000pt}%
\definecolor{currentstroke}{rgb}{0.000000,0.000000,0.000000}%
\pgfsetstrokecolor{currentstroke}%
\pgfsetdash{}{0pt}%
\pgfsys@defobject{currentmarker}{\pgfqpoint{-0.048611in}{0.000000in}}{\pgfqpoint{-0.000000in}{0.000000in}}{%
\pgfpathmoveto{\pgfqpoint{-0.000000in}{0.000000in}}%
\pgfpathlineto{\pgfqpoint{-0.048611in}{0.000000in}}%
\pgfusepath{stroke,fill}%
}%
\begin{pgfscope}%
\pgfsys@transformshift{0.758026in}{1.816605in}%
\pgfsys@useobject{currentmarker}{}%
\end{pgfscope}%
\end{pgfscope}%
\begin{pgfscope}%
\definecolor{textcolor}{rgb}{0.000000,0.000000,0.000000}%
\pgfsetstrokecolor{textcolor}%
\pgfsetfillcolor{textcolor}%
\pgftext[x=0.344444in, y=1.768380in, left, base]{\color{textcolor}\rmfamily\fontsize{10.000000}{12.000000}\selectfont \(\displaystyle {0.015}\)}%
\end{pgfscope}%
\begin{pgfscope}%
\pgfsetbuttcap%
\pgfsetroundjoin%
\definecolor{currentfill}{rgb}{0.000000,0.000000,0.000000}%
\pgfsetfillcolor{currentfill}%
\pgfsetlinewidth{0.803000pt}%
\definecolor{currentstroke}{rgb}{0.000000,0.000000,0.000000}%
\pgfsetstrokecolor{currentstroke}%
\pgfsetdash{}{0pt}%
\pgfsys@defobject{currentmarker}{\pgfqpoint{-0.048611in}{0.000000in}}{\pgfqpoint{-0.000000in}{0.000000in}}{%
\pgfpathmoveto{\pgfqpoint{-0.000000in}{0.000000in}}%
\pgfpathlineto{\pgfqpoint{-0.048611in}{0.000000in}}%
\pgfusepath{stroke,fill}%
}%
\begin{pgfscope}%
\pgfsys@transformshift{0.758026in}{2.233766in}%
\pgfsys@useobject{currentmarker}{}%
\end{pgfscope}%
\end{pgfscope}%
\begin{pgfscope}%
\definecolor{textcolor}{rgb}{0.000000,0.000000,0.000000}%
\pgfsetstrokecolor{textcolor}%
\pgfsetfillcolor{textcolor}%
\pgftext[x=0.344444in, y=2.185540in, left, base]{\color{textcolor}\rmfamily\fontsize{10.000000}{12.000000}\selectfont \(\displaystyle {0.020}\)}%
\end{pgfscope}%
\begin{pgfscope}%
\pgfsetbuttcap%
\pgfsetroundjoin%
\definecolor{currentfill}{rgb}{0.000000,0.000000,0.000000}%
\pgfsetfillcolor{currentfill}%
\pgfsetlinewidth{0.803000pt}%
\definecolor{currentstroke}{rgb}{0.000000,0.000000,0.000000}%
\pgfsetstrokecolor{currentstroke}%
\pgfsetdash{}{0pt}%
\pgfsys@defobject{currentmarker}{\pgfqpoint{-0.048611in}{0.000000in}}{\pgfqpoint{-0.000000in}{0.000000in}}{%
\pgfpathmoveto{\pgfqpoint{-0.000000in}{0.000000in}}%
\pgfpathlineto{\pgfqpoint{-0.048611in}{0.000000in}}%
\pgfusepath{stroke,fill}%
}%
\begin{pgfscope}%
\pgfsys@transformshift{0.758026in}{2.650926in}%
\pgfsys@useobject{currentmarker}{}%
\end{pgfscope}%
\end{pgfscope}%
\begin{pgfscope}%
\definecolor{textcolor}{rgb}{0.000000,0.000000,0.000000}%
\pgfsetstrokecolor{textcolor}%
\pgfsetfillcolor{textcolor}%
\pgftext[x=0.344444in, y=2.602701in, left, base]{\color{textcolor}\rmfamily\fontsize{10.000000}{12.000000}\selectfont \(\displaystyle {0.025}\)}%
\end{pgfscope}%
\begin{pgfscope}%
\definecolor{textcolor}{rgb}{0.000000,0.000000,0.000000}%
\pgfsetstrokecolor{textcolor}%
\pgfsetfillcolor{textcolor}%
\pgftext[x=0.288889in,y=1.608025in,,bottom,rotate=90.000000]{\color{textcolor}\rmfamily\fontsize{10.000000}{12.000000}\selectfont \(\displaystyle ||Q||_2\) [N]}%
\end{pgfscope}%
\begin{pgfscope}%
\pgfpathrectangle{\pgfqpoint{0.758026in}{0.565123in}}{\pgfqpoint{2.022530in}{2.085803in}}%
\pgfusepath{clip}%
\pgfsetbuttcap%
\pgfsetroundjoin%
\pgfsetlinewidth{1.505625pt}%
\definecolor{currentstroke}{rgb}{0.827451,0.827451,0.827451}%
\pgfsetstrokecolor{currentstroke}%
\pgfsetdash{}{0pt}%
\pgfpathmoveto{\pgfqpoint{1.937835in}{0.565123in}}%
\pgfpathlineto{\pgfqpoint{1.937835in}{2.650926in}}%
\pgfusepath{stroke}%
\end{pgfscope}%
\begin{pgfscope}%
\pgfpathrectangle{\pgfqpoint{0.758026in}{0.565123in}}{\pgfqpoint{2.022530in}{2.085803in}}%
\pgfusepath{clip}%
\pgfsetbuttcap%
\pgfsetroundjoin%
\pgfsetlinewidth{1.505625pt}%
\definecolor{currentstroke}{rgb}{0.827451,0.827451,0.827451}%
\pgfsetstrokecolor{currentstroke}%
\pgfsetdash{}{0pt}%
\pgfpathmoveto{\pgfqpoint{2.274923in}{0.565123in}}%
\pgfpathlineto{\pgfqpoint{2.274923in}{2.650926in}}%
\pgfusepath{stroke}%
\end{pgfscope}%
\begin{pgfscope}%
\pgfpathrectangle{\pgfqpoint{0.758026in}{0.565123in}}{\pgfqpoint{2.022530in}{2.085803in}}%
\pgfusepath{clip}%
\pgfsetrectcap%
\pgfsetroundjoin%
\pgfsetlinewidth{1.505625pt}%
\definecolor{currentstroke}{rgb}{0.000000,0.000000,0.000000}%
\pgfsetstrokecolor{currentstroke}%
\pgfsetdash{}{0pt}%
\pgfpathmoveto{\pgfqpoint{0.758026in}{0.565123in}}%
\pgfpathlineto{\pgfqpoint{0.906345in}{0.566677in}}%
\pgfpathlineto{\pgfqpoint{1.003426in}{0.569330in}}%
\pgfpathlineto{\pgfqpoint{1.119924in}{0.574091in}}%
\pgfpathlineto{\pgfqpoint{1.259719in}{0.581756in}}%
\pgfpathlineto{\pgfqpoint{1.427478in}{0.592953in}}%
\pgfpathlineto{\pgfqpoint{1.628785in}{0.607701in}}%
\pgfpathlineto{\pgfqpoint{1.779766in}{0.621837in}}%
\pgfpathlineto{\pgfqpoint{1.830093in}{0.629030in}}%
\pgfpathlineto{\pgfqpoint{1.880419in}{0.638752in}}%
\pgfpathlineto{\pgfqpoint{1.905585in}{0.646877in}}%
\pgfpathlineto{\pgfqpoint{1.937038in}{0.660480in}}%
\pgfpathlineto{\pgfqpoint{1.962203in}{0.673854in}}%
\pgfpathlineto{\pgfqpoint{1.981073in}{0.686077in}}%
\pgfpathlineto{\pgfqpoint{1.996804in}{0.700041in}}%
\pgfpathlineto{\pgfqpoint{2.018696in}{0.723064in}}%
\pgfpathlineto{\pgfqpoint{2.046957in}{0.756941in}}%
\pgfpathlineto{\pgfqpoint{2.080090in}{0.800672in}}%
\pgfpathlineto{\pgfqpoint{2.106380in}{0.837724in}}%
\pgfpathlineto{\pgfqpoint{2.173795in}{0.939839in}}%
\pgfpathlineto{\pgfqpoint{2.224358in}{1.022078in}}%
\pgfpathlineto{\pgfqpoint{2.259756in}{1.082959in}}%
\pgfpathlineto{\pgfqpoint{2.290093in}{1.140033in}}%
\pgfpathlineto{\pgfqpoint{2.327349in}{1.215870in}}%
\pgfpathlineto{\pgfqpoint{2.362297in}{1.292055in}}%
\pgfpathlineto{\pgfqpoint{2.405985in}{1.393507in}}%
\pgfpathlineto{\pgfqpoint{2.449668in}{1.501040in}}%
\pgfpathlineto{\pgfqpoint{2.502095in}{1.636984in}}%
\pgfpathlineto{\pgfqpoint{2.570243in}{1.821803in}}%
\pgfpathlineto{\pgfqpoint{2.606939in}{1.927386in}}%
\pgfpathlineto{\pgfqpoint{2.661042in}{2.090321in}}%
\pgfpathlineto{\pgfqpoint{2.755404in}{2.373748in}}%
\pgfpathlineto{\pgfqpoint{2.780555in}{2.447068in}}%
\pgfpathlineto{\pgfqpoint{2.780555in}{2.447068in}}%
\pgfusepath{stroke}%
\end{pgfscope}%
\begin{pgfscope}%
\pgfpathrectangle{\pgfqpoint{0.758026in}{0.565123in}}{\pgfqpoint{2.022530in}{2.085803in}}%
\pgfusepath{clip}%
\pgfsetbuttcap%
\pgfsetroundjoin%
\pgfsetlinewidth{1.505625pt}%
\definecolor{currentstroke}{rgb}{0.000000,0.000000,0.000000}%
\pgfsetstrokecolor{currentstroke}%
\pgfsetdash{{5.550000pt}{2.400000pt}}{0.000000pt}%
\pgfpathmoveto{\pgfqpoint{0.758026in}{0.565123in}}%
\pgfpathlineto{\pgfqpoint{0.833871in}{0.565933in}}%
\pgfpathlineto{\pgfqpoint{0.924884in}{0.568982in}}%
\pgfpathlineto{\pgfqpoint{1.034101in}{0.575569in}}%
\pgfpathlineto{\pgfqpoint{1.165161in}{0.587453in}}%
\pgfpathlineto{\pgfqpoint{1.322431in}{0.606872in}}%
\pgfpathlineto{\pgfqpoint{1.511160in}{0.636327in}}%
\pgfpathlineto{\pgfqpoint{1.699884in}{0.669675in}}%
\pgfpathlineto{\pgfqpoint{1.794246in}{0.686132in}}%
\pgfpathlineto{\pgfqpoint{1.888613in}{0.700320in}}%
\pgfpathlineto{\pgfqpoint{1.935792in}{0.705228in}}%
\pgfpathlineto{\pgfqpoint{1.982975in}{0.706690in}}%
\pgfpathlineto{\pgfqpoint{2.006567in}{0.704869in}}%
\pgfpathlineto{\pgfqpoint{2.018363in}{0.702675in}}%
\pgfpathlineto{\pgfqpoint{2.030159in}{0.699204in}}%
\pgfpathlineto{\pgfqpoint{2.036054in}{0.697038in}}%
\pgfpathlineto{\pgfqpoint{2.041950in}{0.694561in}}%
\pgfpathlineto{\pgfqpoint{2.047850in}{0.691767in}}%
\pgfpathlineto{\pgfqpoint{2.050800in}{0.690261in}}%
\pgfpathlineto{\pgfqpoint{2.054335in}{0.688356in}}%
\pgfpathlineto{\pgfqpoint{2.058584in}{0.685898in}}%
\pgfpathlineto{\pgfqpoint{2.062828in}{0.683244in}}%
\pgfpathlineto{\pgfqpoint{2.067077in}{0.680415in}}%
\pgfpathlineto{\pgfqpoint{2.071321in}{0.677433in}}%
\pgfpathlineto{\pgfqpoint{2.075570in}{0.674292in}}%
\pgfpathlineto{\pgfqpoint{2.079814in}{0.670965in}}%
\pgfpathlineto{\pgfqpoint{2.084063in}{0.667447in}}%
\pgfpathlineto{\pgfqpoint{2.088308in}{0.663758in}}%
\pgfpathlineto{\pgfqpoint{2.092552in}{0.659918in}}%
\pgfpathlineto{\pgfqpoint{2.096801in}{0.655941in}}%
\pgfpathlineto{\pgfqpoint{2.101895in}{0.651002in}}%
\pgfpathlineto{\pgfqpoint{2.106994in}{0.645890in}}%
\pgfpathlineto{\pgfqpoint{2.113106in}{0.639522in}}%
\pgfpathlineto{\pgfqpoint{2.119219in}{0.632875in}}%
\pgfpathlineto{\pgfqpoint{2.125337in}{0.625939in}}%
\pgfpathlineto{\pgfqpoint{2.131450in}{0.618724in}}%
\pgfpathlineto{\pgfqpoint{2.137567in}{0.611238in}}%
\pgfpathlineto{\pgfqpoint{2.143680in}{0.603477in}}%
\pgfpathlineto{\pgfqpoint{2.149793in}{0.595451in}}%
\pgfpathlineto{\pgfqpoint{2.155910in}{0.587209in}}%
\pgfpathlineto{\pgfqpoint{2.162023in}{0.578837in}}%
\pgfpathlineto{\pgfqpoint{2.168141in}{0.570797in}}%
\pgfpathlineto{\pgfqpoint{2.174254in}{0.570774in}}%
\pgfpathlineto{\pgfqpoint{2.180366in}{0.579263in}}%
\pgfpathlineto{\pgfqpoint{2.186484in}{0.588559in}}%
\pgfpathlineto{\pgfqpoint{2.192597in}{0.598171in}}%
\pgfpathlineto{\pgfqpoint{2.198714in}{0.608019in}}%
\pgfpathlineto{\pgfqpoint{2.204827in}{0.618074in}}%
\pgfpathlineto{\pgfqpoint{2.210940in}{0.628328in}}%
\pgfpathlineto{\pgfqpoint{2.217058in}{0.638779in}}%
\pgfpathlineto{\pgfqpoint{2.223170in}{0.649426in}}%
\pgfpathlineto{\pgfqpoint{2.229288in}{0.660268in}}%
\pgfpathlineto{\pgfqpoint{2.235401in}{0.671297in}}%
\pgfpathlineto{\pgfqpoint{2.241514in}{0.682506in}}%
\pgfpathlineto{\pgfqpoint{2.247631in}{0.693892in}}%
\pgfpathlineto{\pgfqpoint{2.253744in}{0.705459in}}%
\pgfpathlineto{\pgfqpoint{2.259862in}{0.717211in}}%
\pgfpathlineto{\pgfqpoint{2.265975in}{0.729153in}}%
\pgfpathlineto{\pgfqpoint{2.272087in}{0.741279in}}%
\pgfpathlineto{\pgfqpoint{2.274922in}{0.746958in}}%
\pgfpathlineto{\pgfqpoint{2.350767in}{0.910014in}}%
\pgfpathlineto{\pgfqpoint{2.388689in}{0.998031in}}%
\pgfpathlineto{\pgfqpoint{2.407650in}{1.043280in}}%
\pgfpathlineto{\pgfqpoint{2.417134in}{1.066196in}}%
\pgfpathlineto{\pgfqpoint{2.426612in}{1.089313in}}%
\pgfpathlineto{\pgfqpoint{2.436095in}{1.112620in}}%
\pgfpathlineto{\pgfqpoint{2.445573in}{1.136099in}}%
\pgfpathlineto{\pgfqpoint{2.450315in}{1.147895in}}%
\pgfpathlineto{\pgfqpoint{2.456002in}{1.162095in}}%
\pgfpathlineto{\pgfqpoint{2.462830in}{1.179203in}}%
\pgfpathlineto{\pgfqpoint{2.471019in}{1.199835in}}%
\pgfpathlineto{\pgfqpoint{2.479213in}{1.220590in}}%
\pgfpathlineto{\pgfqpoint{2.487402in}{1.241472in}}%
\pgfpathlineto{\pgfqpoint{2.495596in}{1.262471in}}%
\pgfpathlineto{\pgfqpoint{2.503785in}{1.283581in}}%
\pgfpathlineto{\pgfqpoint{2.513615in}{1.309059in}}%
\pgfpathlineto{\pgfqpoint{2.525411in}{1.339877in}}%
\pgfpathlineto{\pgfqpoint{2.537207in}{1.371029in}}%
\pgfpathlineto{\pgfqpoint{2.543103in}{1.386740in}}%
\pgfpathlineto{\pgfqpoint{2.549003in}{1.402531in}}%
\pgfpathlineto{\pgfqpoint{2.554898in}{1.418400in}}%
\pgfpathlineto{\pgfqpoint{2.560794in}{1.434319in}}%
\pgfpathlineto{\pgfqpoint{2.566694in}{1.450288in}}%
\pgfpathlineto{\pgfqpoint{2.573773in}{1.469527in}}%
\pgfpathlineto{\pgfqpoint{2.582261in}{1.492713in}}%
\pgfpathlineto{\pgfqpoint{2.590754in}{1.516008in}}%
\pgfpathlineto{\pgfqpoint{2.595003in}{1.527688in}}%
\pgfpathlineto{\pgfqpoint{2.600097in}{1.541746in}}%
\pgfpathlineto{\pgfqpoint{2.606215in}{1.558641in}}%
\pgfpathlineto{\pgfqpoint{2.613549in}{1.578974in}}%
\pgfpathlineto{\pgfqpoint{2.622356in}{1.603444in}}%
\pgfpathlineto{\pgfqpoint{2.632921in}{1.632913in}}%
\pgfpathlineto{\pgfqpoint{2.645601in}{1.668438in}}%
\pgfpathlineto{\pgfqpoint{2.658280in}{1.704189in}}%
\pgfpathlineto{\pgfqpoint{2.670959in}{1.740273in}}%
\pgfpathlineto{\pgfqpoint{2.683639in}{1.776658in}}%
\pgfpathlineto{\pgfqpoint{2.696318in}{1.813335in}}%
\pgfpathlineto{\pgfqpoint{2.708998in}{1.850220in}}%
\pgfpathlineto{\pgfqpoint{2.721677in}{1.887264in}}%
\pgfpathlineto{\pgfqpoint{2.734357in}{1.924458in}}%
\pgfpathlineto{\pgfqpoint{2.740697in}{1.943130in}}%
\pgfpathlineto{\pgfqpoint{2.748306in}{1.965615in}}%
\pgfpathlineto{\pgfqpoint{2.755911in}{1.988208in}}%
\pgfpathlineto{\pgfqpoint{2.763521in}{2.010910in}}%
\pgfpathlineto{\pgfqpoint{2.771130in}{2.033695in}}%
\pgfpathlineto{\pgfqpoint{2.778735in}{2.056564in}}%
\pgfpathlineto{\pgfqpoint{2.780555in}{2.062037in}}%
\pgfusepath{stroke}%
\end{pgfscope}%
\begin{pgfscope}%
\pgfsetrectcap%
\pgfsetmiterjoin%
\pgfsetlinewidth{0.803000pt}%
\definecolor{currentstroke}{rgb}{0.000000,0.000000,0.000000}%
\pgfsetstrokecolor{currentstroke}%
\pgfsetdash{}{0pt}%
\pgfpathmoveto{\pgfqpoint{0.758026in}{0.565123in}}%
\pgfpathlineto{\pgfqpoint{0.758026in}{2.650926in}}%
\pgfusepath{stroke}%
\end{pgfscope}%
\begin{pgfscope}%
\pgfsetrectcap%
\pgfsetmiterjoin%
\pgfsetlinewidth{0.803000pt}%
\definecolor{currentstroke}{rgb}{0.000000,0.000000,0.000000}%
\pgfsetstrokecolor{currentstroke}%
\pgfsetdash{}{0pt}%
\pgfpathmoveto{\pgfqpoint{2.780555in}{0.565123in}}%
\pgfpathlineto{\pgfqpoint{2.780555in}{2.650926in}}%
\pgfusepath{stroke}%
\end{pgfscope}%
\begin{pgfscope}%
\pgfsetrectcap%
\pgfsetmiterjoin%
\pgfsetlinewidth{0.803000pt}%
\definecolor{currentstroke}{rgb}{0.000000,0.000000,0.000000}%
\pgfsetstrokecolor{currentstroke}%
\pgfsetdash{}{0pt}%
\pgfpathmoveto{\pgfqpoint{0.758026in}{0.565123in}}%
\pgfpathlineto{\pgfqpoint{2.780555in}{0.565123in}}%
\pgfusepath{stroke}%
\end{pgfscope}%
\begin{pgfscope}%
\pgfsetrectcap%
\pgfsetmiterjoin%
\pgfsetlinewidth{0.803000pt}%
\definecolor{currentstroke}{rgb}{0.000000,0.000000,0.000000}%
\pgfsetstrokecolor{currentstroke}%
\pgfsetdash{}{0pt}%
\pgfpathmoveto{\pgfqpoint{0.758026in}{2.650926in}}%
\pgfpathlineto{\pgfqpoint{2.780555in}{2.650926in}}%
\pgfusepath{stroke}%
\end{pgfscope}%
\begin{pgfscope}%
\definecolor{textcolor}{rgb}{0.000000,0.000000,0.000000}%
\pgfsetstrokecolor{textcolor}%
\pgfsetfillcolor{textcolor}%
\pgftext[x=1.769291in,y=2.734260in,,base]{\color{textcolor}\rmfamily\fontsize{12.000000}{14.400000}\selectfont Force vs. Angle}%
\end{pgfscope}%
\begin{pgfscope}%
\pgfsetbuttcap%
\pgfsetmiterjoin%
\definecolor{currentfill}{rgb}{1.000000,1.000000,1.000000}%
\pgfsetfillcolor{currentfill}%
\pgfsetfillopacity{0.800000}%
\pgfsetlinewidth{1.003750pt}%
\definecolor{currentstroke}{rgb}{0.800000,0.800000,0.800000}%
\pgfsetstrokecolor{currentstroke}%
\pgfsetstrokeopacity{0.800000}%
\pgfsetdash{}{0pt}%
\pgfpathmoveto{\pgfqpoint{0.855248in}{2.152470in}}%
\pgfpathlineto{\pgfqpoint{2.006403in}{2.152470in}}%
\pgfpathquadraticcurveto{\pgfqpoint{2.034181in}{2.152470in}}{\pgfqpoint{2.034181in}{2.180247in}}%
\pgfpathlineto{\pgfqpoint{2.034181in}{2.553704in}}%
\pgfpathquadraticcurveto{\pgfqpoint{2.034181in}{2.581482in}}{\pgfqpoint{2.006403in}{2.581482in}}%
\pgfpathlineto{\pgfqpoint{0.855248in}{2.581482in}}%
\pgfpathquadraticcurveto{\pgfqpoint{0.827470in}{2.581482in}}{\pgfqpoint{0.827470in}{2.553704in}}%
\pgfpathlineto{\pgfqpoint{0.827470in}{2.180247in}}%
\pgfpathquadraticcurveto{\pgfqpoint{0.827470in}{2.152470in}}{\pgfqpoint{0.855248in}{2.152470in}}%
\pgfpathlineto{\pgfqpoint{0.855248in}{2.152470in}}%
\pgfpathclose%
\pgfusepath{stroke,fill}%
\end{pgfscope}%
\begin{pgfscope}%
\pgfsetrectcap%
\pgfsetroundjoin%
\pgfsetlinewidth{1.505625pt}%
\definecolor{currentstroke}{rgb}{0.000000,0.000000,0.000000}%
\pgfsetstrokecolor{currentstroke}%
\pgfsetdash{}{0pt}%
\pgfpathmoveto{\pgfqpoint{0.883026in}{2.477315in}}%
\pgfpathlineto{\pgfqpoint{1.021915in}{2.477315in}}%
\pgfpathlineto{\pgfqpoint{1.160803in}{2.477315in}}%
\pgfusepath{stroke}%
\end{pgfscope}%
\begin{pgfscope}%
\definecolor{textcolor}{rgb}{0.000000,0.000000,0.000000}%
\pgfsetstrokecolor{textcolor}%
\pgfsetfillcolor{textcolor}%
\pgftext[x=1.271915in,y=2.428704in,left,base]{\color{textcolor}\rmfamily\fontsize{10.000000}{12.000000}\selectfont \(\displaystyle \alpha_0=14\)\,deg}%
\end{pgfscope}%
\begin{pgfscope}%
\pgfsetbuttcap%
\pgfsetroundjoin%
\pgfsetlinewidth{1.505625pt}%
\definecolor{currentstroke}{rgb}{0.000000,0.000000,0.000000}%
\pgfsetstrokecolor{currentstroke}%
\pgfsetdash{{5.550000pt}{2.400000pt}}{0.000000pt}%
\pgfpathmoveto{\pgfqpoint{0.883026in}{2.283642in}}%
\pgfpathlineto{\pgfqpoint{1.021915in}{2.283642in}}%
\pgfpathlineto{\pgfqpoint{1.160803in}{2.283642in}}%
\pgfusepath{stroke}%
\end{pgfscope}%
\begin{pgfscope}%
\definecolor{textcolor}{rgb}{0.000000,0.000000,0.000000}%
\pgfsetstrokecolor{textcolor}%
\pgfsetfillcolor{textcolor}%
\pgftext[x=1.271915in,y=2.235031in,left,base]{\color{textcolor}\rmfamily\fontsize{10.000000}{12.000000}\selectfont \(\displaystyle \alpha_0=18\)\,deg}%
\end{pgfscope}%
\end{pgfpicture}%
\makeatother%
\endgroup%